\documentclass[aps,prd,preprint,showpacs,eqsecnum,nofootinbib]{revtex4}
\usepackage{graphicx,epsf,amssymb,feynarts}
\newcommand{\be}{\begin{equation}}
\newcommand{\ee}{\end{equation}}

\newcommand{\bea}{\begin{eqnarray}}
\newcommand{\eea}{\end{eqnarray}}
\newcommand{\Tr}{{\rm Tr}}
\newcommand{\nn}{\nonumber}

\newif\ifdraft
\drafttrue
\newif\ifpreprint
\preprinttrue

\def\Sect#1{Section~{\ref{#1}}}
\def\sect#1{section~{\ref{#1}}}
\def\app#1{appendix~{\ref{#1}}}

\def\fig#1{fig.~{\ref{#1}}}
\def\figs#1#2{figs.~{\ref{#1}} and {\ref{#2}}}

\def\tree{{\rm tree}}

\def\Tr{\, {\rm Tr}}
\def\tr{\, {\rm tr}}

\def\NeqFour{{\cal N}=4}
\def\NeqOne{{\cal N}=1}

\def\spa#1.#2{\left\langle#1\,#2\right\rangle}
\def\spb#1.#2{\left[#1\,#2\right]}
\def\sand#1.#2.#3{%
\left\langle\smash{#1}{\vphantom1}^{-}\right|{#2}%
\left|\smash{#3}{\vphantom1}^{-}\right\rangle}
\def\sandp#1.#2.#3{%
\left\langle\smash{#1}{\vphantom1}^{-}\right|{#2}%
\left|\smash{#3}{\vphantom1}^{+}\right\rangle}
\def\sandpp#1.#2.#3{%
\left\langle\smash{#1}{\vphantom1}^{+}\right|{#2}%
\left|\smash{#3}{\vphantom1}^{+}\right\rangle}
\def\sandpm#1.#2.#3{%
\left\langle\smash{#1}{\vphantom1}^{+}\right|{#2}%
\left|\smash{#3}{\vphantom1}^{-}\right\rangle}
\def\sandmp#1.#2.#3{%
\left\langle\smash{#1}{\vphantom1}^{-}\right|{#2}%
\left|\smash{#3}{\vphantom1}^{+}\right\rangle}
\def\sandmm#1.#2.#3{%
\left\langle\smash{#1}{\vphantom1}^{-}\right|{#2}%
\left|\smash{#3}{\vphantom1}^{-}\right\rangle}
\def\spab#1.#2.#3{\sandmm#1.#2.#3}
\def\spba#1.#2.#3{\sandpp#1.#2.#3}
\def\spaa#1.#2.#3.#4{\sandmp#1.{#2#3}.#4}
\def\spbb#1.#2.#3.#4{\sandpm#1.{#2#3}.#4}
\def\spash#1.#2{\spa{\smash{#1}}.{\smash{#2}}}

\newbox\charbox
\newbox\slabox
\def\s#1{{      
        \setbox\charbox=\hbox{$#1$}
        \setbox\slabox=\hbox{$/$}
        \dimen\charbox=\ht\slabox
        \advance\dimen\charbox by -\dp\slabox
        \advance\dimen\charbox by -\ht\charbox
        \advance\dimen\charbox by \dp\charbox
        \divide\dimen\charbox by 2
        \raise-\dimen\charbox\hbox to \wd\charbox{\hss/\hss}
        \llap{$#1$}
}}

\def\eqn#1{eq.~(\ref{#1})}
\def\Eqn#1{Equation~(\ref{#1})}
\def\eqns#1#2{eqs.~(\ref{#1}) and~(\ref{#2})}

\def\qb{{\overline {\kern-0.7pt q\kern -0.7pt}}}

\def\e{\epsilon}
\def\ep{\epsilon}
\def\eps{\epsilon}

\def\Li{\mathop{\rm Li}\nolimits}
\def\Split{\mathop{\rm Split}\nolimits}
\def\tree{{(0)}}
\def\Lloop{{(L)}}
\def\lloop{{(l)}}
\def\oneloop{{(1)}}
\def\twoloop{{(2)}}

\def\Ord{{\cal O}}
\def\mod{\mathop{\rm mod}\nolimits}

\def\Remainder{R}
\def\fstring{f}
\def\gstring{g}
\def\kstring{k}
\def\Cstring{C}

\begin{document}
\hfuzz 15 pt


\ifpreprint
\noindent
UCLA/08/TEP/5
\hfill SLAC--PUB--13150 \\
Saclay/IPhT--T08/045
\hfill Brown-HET-1495
\fi

\vskip 1 cm 

\title{The Two-Loop Six-Gluon MHV Amplitude 
in Maximally Supersymmetric Yang-Mills Theory}%

\author{Z.~Bern${}^a$, L.~J.~Dixon${}^{b}$,  
D.~A.~Kosower${}^{c}$, R.~Roiban${}^d$, 
M. Spradlin${}^e$, C. Vergu${}^c$ and A. Volovich${}^e$ }

\affiliation{
${}^a$\hbox{Department of Physics and Astronomy, UCLA, Los Angeles, CA
90095-1547, USA}  \\
${}^b$\hbox{Stanford Linear Accelerator Center,
              Stanford University,
             Stanford, CA 94309, USA} \\
${}^c$\hbox{Institut de Physique Th\'eorique, CEA--Saclay,
          F--91191 Gif-sur-Yvette cedex, France}\\
${}^d$\hbox{Department of Physics, Pennsylvania State University,
           University Park, PA 16802, USA}\\
${}^e$\hbox{Department of Physics, Brown University,
         Box 1843, Providence, RI 02912, USA}\\
}

\vskip 1 cm 

\begin{abstract}
 We give a representation of the parity-even part of the planar
 two-loop six-gluon MHV amplitude of $\NeqFour$ super-Yang-Mills
 theory, in terms of loop-momentum integrals with simple dual
 conformal properties. We evaluate the integrals numerically in order
 to test directly the ABDK/BDS all-loop ansatz for planar MHV
 amplitudes.  We find that the ansatz requires an additive remainder
 function, in accord with previous indications from strong-coupling
 and Regge limits.  The planar six-gluon amplitude can also be
 compared with the hexagonal Wilson loop computed by Drummond, Henn,
 Korchemsky and Sokatchev in arXiv:0803.1466 [hep-th].  After
 accounting for differing singularities and other constants
 independent of the kinematics, we find that the Wilson loop and
 MHV-amplitude remainders are {\it identical}, to within our numerical
 precision. This result provides non-trivial confirmation of a
 proposed $n$-point equivalence between Wilson loops and planar MHV
 amplitudes, and suggests that an additional mechanism besides dual
 conformal symmetry fixes their form at six points and beyond.
\end{abstract}

\pacs{11.15.Bt, 11.15.Pg, 11.25.Db, 11.25.Tq, 12.60.Jv  \hspace{1cm}}

\maketitle


\section{Introduction}

Gauge theories play a central role in modern particle physics.  
How far can we go in understanding their properties quantitatively?
Long ago~\cite{tHooft}, 
't~Hooft suggested that the theories simplify dramatically
in the so-called planar limit, that of a large number of colors $N_c$.
In this limit, he suggested that the theory is expressible as a string
theory.  This idea was given a concrete realization by 
Maldacena's~\cite{Maldacena} conjecture of
the anti-de Sitter/conformal field theory (AdS/CFT) duality
between weakly-coupled type-IIB string theory on an AdS background and the
maximally supersymmetric ($\NeqFour$) gauge theory at strong coupling.  

The duality requires that the perturbative series for a variety of
quantities, including scattering amplitudes, sum up to a simple
quantity.  In order for this to be possible, it would appear almost
essential for the terms in the perturbative series to be related to
each other in a simple way.  Such a relation is guaranteed for the
infrared-singular terms in amplitudes by the requirement that physical
quantities be infrared-finite.  However, the finite terms are not
required to obey such an iterative relation.  The $\NeqFour$ theory is
special, as was shown by Anastasiou and three of the authors
(ABDK)~\cite{ABDK}, in that a simple relation does hold between the
finite terms of the planar one- and two-loop four-gluon amplitudes.
(Using the Ward identities of $\NeqFour$ supersymmetry,
this relation extends to four-point amplitudes for arbitrary external
states~\cite{SWI,BDDKSelfDual}.)  By demanding that the amplitude have
proper factorization as any two external momenta become collinear,
this relation was extended to an arbitrary number of external legs,
valid for the simplest configuration of gluon helicities, the
maximally helicity violating (MHV) one.  Smirnov and two of the
authors (BDS) formulated~\cite{BDS} an all-orders version of this
relation, based on the idea that in the $\NeqFour$ theory the finite
terms should obey the same (exponential) relations as the divergent
terms.  This ansatz gave a prediction to all orders in the coupling
for planar MHV amplitudes.

In a remarkable paper, Alday and Maldacena~\cite{AldayMaldacena}
suggested a way to compute the dimensionally-regulated planar four-point 
amplitude at strong coupling, using dual string theory.
Their calculation reproduced the strong-coupling limit of the BDS ansatz.

The Alday--Maldacena calculation also provided a link between
scattering amplitudes at strong coupling 
and a special kind of Wilson loop, one composed of 
light-like segments in a dual coordinate space.  
Drummond, Korchemsky, and
Sokatchev~\cite{DrummondVanishing} found the surprising result 
that at weak coupling the lowest-order
(one-loop) contribution to a rectangular Wilson loop with 
light-like edges is equal to the four-point one-loop amplitude 
(normalized by the tree amplitude, and up to a constant term).
Brandhuber, Heslop, and Travaglini~\cite{BrandhuberWilson} 
showed that the equality extends to one between Wilson polygons with
$n$ sides and one-loop MHV $n$-point amplitudes for all $n$.

Does this equality extend beyond one loop?  The perturbative planar
amplitudes exhibit a ``dual conformal
invariance''~\cite{MagicIdentities,BCDKS}, distinct from the usual
conformal invariance of the $\NeqFour$ theory.  Roughly speaking, it
corresponds to conformal invariance in momentum space.  Although the
origin of this symmetry is not yet understood, we can use it to guide
the calculations.
Scattering amplitudes with more than four gluons, when normalized
by the tree amplitude, may contain odd powers of the Levi-Civita
tensor contracted with the external momenta.  These terms flip sign
under a parity transformation, which reverses all helicities.
We refer to them as odd terms, and to the remaining terms as even.
One may write the four-point amplitude through five
loops, and the even part of the five-point amplitude at two loops, 
purely in terms of {\it pseudo-conformal\/} integrals.  These
are dimensionally-regulated integrals whose off-shell continuation is
invariant under conformal transformations of dual coordinates, whose
differences are momenta.

As shown by Drummond, Henn, Korchemsky, and Sokatchev (DHKS), the dual
conformal symmetry gives
rise~\cite{DHKSTwoloopBoxWilson,ConformalWard} to an anomalous Ward
identity in dimensional regularization.  This Ward identity fixes, to
all loop orders, the finite parts of the expectation values of the
Wilson loops corresponding to four- and five-point scattering
amplitudes (up to a constant term interpretable as an ultraviolet
scale adjustment).  The same authors also showed that these unique
solutions of the anomalous Ward identity coincide with the four- and
five-point two-loop Wilson loops and 
amplitudes as computed in perturbation theory (and
again normalized by the tree amplitude).  Given as well the one-loop
equivalence between Wilson loops and
amplitudes~\cite{DrummondVanishing,BrandhuberWilson} and the 
strong coupling results of Alday and Maldacena~\cite{AldayMaldacena}, 
we may expect that the same equivalence holds for the four-point 
amplitude
and the even part of the five-point amplitude, to all orders in
perturbation theory;  indeed, into the strong-coupling regime.
The BDS ansatz is in fact a solution to the anomalous Ward 
identity~\cite{DHKSTwoloopBoxWilson}.  Because
of the uniqueness imposed by the Ward identity, the BDS
ansatz too should be expected to hold to all orders for four- and
five-point amplitudes.  It has been checked for the four-point
amplitude through three loops~\cite{ABDK,BDS} and for the five-point
amplitude through two loops~\cite{TwoLoopFiveA,TwoLoopFiveB}.

What about amplitudes with a larger number of legs?  Alday and
Maldacena have shown~\cite{AMTrouble} that in the limit of a large
number of legs, the Wilson loop calculation does not agree with the
BDS ansatz.  This result might imply that the connection between 
Wilson loops and the amplitudes breaks down; or it might mean that the 
BDS ansatz breaks down for more than five external legs.
But at how many legs and how many loops might the breakdown occur?  

It is possible to examine the ansatz for consistency in different
kinematic limits.  Recently, the BDS ansatz has been examined
in various types of Regge, or high-energy, limits of the scattering.
The four- and five-point amplitudes appear to be consistent
in all such limits~\cite{DrummondVanishing,Schnitzer,BNST,Lipatov}.  Indeed,
higher-order coefficients in the Regge slope parameter and
other high-energy quantities can be extracted from such limits.
On the other hand, study of a particular multi-Regge limit of 
$2\to4$ scattering, and also of $3\to3$ scattering, appears to 
indicate a difficulty with the ansatz for the six-gluon 
amplitude starting at two loops~\cite{Lipatov}.

In order to test the BDS ansatz directly, 
we have computed the parity-even part of the
two-loop six-point MHV amplitude in the $\NeqFour$ supersymmetric gauge theory.
With assistance from the work of Drummond, Henn, Korchemsky, and
Sokatchev~\cite{HexagonWilson,WilsonValues}, 
we can also test the correspondence of the parity-even part
with the calculation of a hexagonal Wilson loop.  
Six external legs marks the first
appearance of cross ratios invariant under the dual conformal
transformations; the finite part of the Wilson loop is no longer fixed
by the anomalous Ward identity,  but is determined only up to 
a function of these cross ratios.  Six external legs also marks the first
appearance of non-MHV amplitudes.  The basic Wilson loop
is insensitive to the helicities of the external gluons; hence
it cannot equal these other six-point helicity amplitudes, even at
one loop.  The question of an iterative or exponential
structure for the non-MHV amplitudes is an interesting one, 
but we shall not explore it in the present paper.

Another open question, not addressed in this paper, is the behavior of
the parity-odd part of the two-loop six-gluon MHV amplitude.
In the five-gluon case, the parity-odd part vanishes in the
logarithm of the full amplitude, due to a cancellation between
one- and two-loop terms~\cite{TwoLoopFiveA,TwoLoopFiveB}.
We expect the same cancellation to take place for MHV amplitudes with
six or more gluons, but this expectation remains to be established.
Strictly speaking, the parity-odd cancellation is required to 
establish full correspondence with a Wilson loop, which obviously
does not change with the reversal of all gluon helicities.

We perform the calculation using the unitarity-based method, employing
a variety of four-dimensional and $D$-dimensional cuts (with $D=4-2\ep$)
to express the
amplitude in terms of a selected set of six-point two-loop Feynman
integrals.  The result may be expressed as a sum of pseudo-conformal
integrals~\cite{MagicIdentities}, in close analogy with the four-point
amplitude through five loops~\cite{GSB,BRY,BDS,BCDKS,FiveLoop} and the
parity-even part of the five-point amplitude through two
loops~\cite{FiveGluonOneLoop,BDDKSelfDual,TwoLoopFiveA,TwoLoopFiveB}.
There are some additional integrals in the one- and two-loop six-point
amplitudes, whose pseudo-conformal nature is less clear.  Their
integrands vanish as $D\to4$, yet their integrals can be nonvanishing in
this limit.  However, their one- and two-loop contributions conspire
to cancel in the logarithm of the amplitude (which is what really is
needed to test the BDS ansatz) in the limit $\eps\to0$.  We then
evaluate the integrals using the packages {\tt AMBRE}~\cite{AMBRE} and
{\tt MB}~\cite{MB} and compute the amplitude numerically at a variety
of kinematic points.  The structure of the infrared singularities is
known~\cite{KnownIR,KorchemskyMarchesini}, and agrees with the pole
terms in our expression.  The finite remainders are tested numerically
against the BDS ansatz, and against values for the corresponding
Wilson loop~\cite{WilsonValues}.

The paper is organized as follows.  In \sect{ReviewSection}, we review
the ABDK and BDS ans\"atze, the structure of scattering amplitudes at
strong coupling, the difficulty that appears as the number of external
legs becomes large, and dual conformal invariance.  In
\sect{IntegrandSection}, we give the integrand, and outline its
calculation via the unitarity method.  In \sect{ResultsSection} we
present our results for the amplitude and compare these to the results
of the hexagonal Wilson loop calculation.
\Sect{RemainderPropertiesSection} gives properties that the remainder
function must satisfy.  We give our conclusions and summarize open
problems in \sect{ConclusionSection}.  The appendices contain results
for the integrals.


\section{Review}
\label{ReviewSection}

\subsection{ABDK/BDS Ansatz}

In an $SU(N_c)$ gauge theory the leading-color contributions to the
$L$-loop gauge-theory $n$-point amplitudes can be written as
\begin{eqnarray}
{\cal A}_n^{(L)} &=& g^{n-2} \, a^L
 \sum_{\rho}
\Tr( T^{a_{\rho(1)}} T^{a_{\rho(2)}}
   \ldots T^{a_{\rho(n)}} ) \,
               A_n^{(L)}(\rho(1), \rho(2), \ldots, \rho(n))\,, \hskip .5 cm
\label{LeadingColorDecomposition}
\end{eqnarray}
where
\begin{equation} 
a \equiv (4\pi e^{-\gamma})^\e {\lambda \over 8\pi^2} \,.
\label{adef}
\end{equation}
Here $\lambda = g^2 N_c$ is the 't Hooft parameter, 
$g$ is the Yang-Mills coupling, and $\gamma$ is Euler's constant.  
The sum is over non-cyclic
permutations of the external legs. We have suppressed the momenta
and helicities $k_i$ and $\lambda_i$, leaving only the index $i$ as a
label.  This decomposition holds for any amplitude when all particles
are in the adjoint representation.  We will find it convenient to scale out
the tree amplitude, defining
\begin{equation}
M_n^{(L)}(\e) \equiv A_n^{(L)}/A_n^{(0)}\,.
\label{RescaledLoopAmplitude}
\end{equation}

In planar $\NeqFour$ supersymmetric gauge theory, amplitudes computed to
date satisfy an iteration relation.
At two loops, the iteration conjecture expresses
$n$-point amplitudes entirely in terms of one-loop
amplitudes and a set of constants~\cite{ABDK}.  For two-loop 
MHV amplitudes the ABDK conjecture reads
\be
M_n^{\twoloop}(\e)
= {1 \over 2} \bigl(M_n^{\oneloop}(\e) \bigr)^2
 + f^\twoloop(\e) \, M_n^{\oneloop}(2\e) + C^{(2)} + \Ord(\e)\,,
\label{TwoloopOneloop}
\ee
where
\begin{equation}
f^\twoloop(\e) = - (\zeta_2 + \zeta_3 \e + \zeta_4 \e^2 + \cdots)\,,
\hskip 2 cm
C^{(2)} = - \zeta_2^2/2 \,.
\end{equation}

The form~(\ref{TwoloopOneloop}) was based on explicit computations of
both the four-point two-loop amplitude~\cite{BRY,SmirnovTwoLoop} and
of the splitting amplitudes~\cite{ABDK}, which control the behavior of
the amplitudes as two momenta become collinear.  The collinear 
splitting amplitude has an iterative property, analogous to 
\eqn{TwoloopOneloop}, which guarantees that the
ansatz~(\ref{TwoloopOneloop}) for MHV amplitudes has the correct 
collinear limits for any $n$.  In addition,
MHV amplitudes have no poles as a single multi-particle
kinematic invariant vanishes, 
$(k_i+\cdots +k_j)^2 \to 0$ for $j>(i+1)$ mod $n$;
so factorization in such channels is satisfied trivially 
by~\eqn{TwoloopOneloop}.  For the five-point
amplitude, \eqn{TwoloopOneloop} has also been confirmed by direct
calculation~\cite{TwoLoopFiveA,TwoLoopFiveB}.  Any violation of
\eqn{TwoloopOneloop} beyond five external legs must necessarily be
expressed by a function which vanishes as pairs of color-adjacent
momenta become collinear.  We shall see in
\sect{RemainderPropertiesSection} that such a remainder function is,
however, detectable in the triple-collinear limit in which three 
color-adjacent momenta become collinear.  This limit can first be 
achieved for $n=6$.

The iterative structure in \eqn{TwoloopOneloop}, together with the
exponential nature of infrared
divergences~\cite{KnownIR,KorchemskyMarchesini}, suggest that an
all-orders resummation should be possible.  In ref.~\cite{BDS} the
three-loop generalization for $n=4$ was found by direct calculation,
guiding the all-loop order BDS proposal,
\begin{equation}
\ln{\cal M}_n 
= \sum_{l=1}^\infty a^l 
      \Bigl(f^{(l)}(\e) M_n^{(1)}(l \e) + C^{(l)} + \Ord(\e) \Bigr) \,,
\label{BDSAnsatz}
\end{equation}
where
\begin{equation}
{\cal M}_n = \sum_{L=0}^\infty a^L M_n^{(L)}(\eps)
\end{equation}
is the resummed all-loop amplitude.
The quantity $M_n^{(1)}(l\e)$ is the dimensionally-regulated
one-loop amplitude, with the tree scaled out
according to \eqn{RescaledLoopAmplitude}, and with $\e \to l\e$.  Each
$f^{(l)}(\e)$ is a three-term series in $\e$, beginning at
$\Ord(\e^0)$,
\begin{equation}
 f^{(l)}(\e) = f_0^{(l)} + \e f_1^{(l)} + \e^2 f_2^{(l)} \,.
\label{flexp}
\end{equation}
The  constant $f_0^{(l)}$ is the planar cusp anomalous 
dimension~\cite{KorchemskyRadyushkin},
$f^{(l)}_0 = {1\over4} \, \hat\gamma_K^{(l)} \, $.

In order to test the ABDK/BDS ansatz, 
it is convenient to define an $l$-loop {\it remainder function}
$R_n^{(l)}$ to be the difference between the actual $l$-loop 
rescaled amplitude $M_n^{(l)}$ and the ABDK/BDS prediction for it,
in the limit $\e\to0$.  This function is
finite as $\e\to0$, because the BDS ansatz has all the
correct infrared singularities.  It is only defined in the limit $\e\to0$
because the two-loop ansatz does not hold beyond $\Ord(\ep^0)$, even for
$n=4$.  For example, at two loops the remainder
function is defined by,
\begin{equation}
\Remainder^{(2)}_n \equiv \lim_{\epsilon \to 0} \left[ M_n^{\twoloop}(\e)
- \left( {1 \over 2} \bigl(M_n^{\oneloop}(\e) \bigr)^2
+  f^\twoloop(\e) \, M_n^{\oneloop}(2\e) + C^{(2)}\right)
\right]\,.
\label{def_remainder}
\end{equation}
Notice that the combination $M_n^{(2)} - 1/2 (M_n^{(1)})^2$ appearing
in \eqn{def_remainder} is the order $a^2$
 term in the logarithm of the amplitude (\ref{BDSAnsatz}).

The one-loop MHV amplitudes entering the ABDK/BDS ansatz 
were computed some time ago~\cite{NeqFourOneLoop}, with the result
\begin{equation}
M_n^\oneloop(\ep) = -{1\over 2} {1\over\e^2} \sum_{i=1}^n 
    \biggl({\mu^2 \over -s_{i,i+1}}\biggr)^{\e} + F_n^{(1)}(0) 
+ \Ord(\e) \,,
\label{OneloopMHVAmplitude}
\end{equation}
where we use the normalizations of refs.~\cite{ABDK,BDS,BCDKS}.
The ``0'' argument in the finite part  $F_n^{(1)}(0)$ signifies
that we have taken $\ep \rightarrow 0$.  These terms have the form,
\begin{equation}
F_n^{(1)}(0) = {1 \over 2} \sum_{i=1}^n g_{n,i} \,,
\label{OneLoopFiniteRemainder}
\end{equation}
where
\begin{eqnarray}
g_{n,i} &=&
-\sum_{r=2}^{\lfloor n/2 \rfloor -1}
  \ln \Biggl({ -s_{i\cdots(i+r-1)}\over -s_{i\cdots(i+r)} }\Biggr)
  \ln \Biggl({ -s_{(i+1)\cdots(i+r)}\over -s_{i\cdots(i+r)} }\Biggr) +
D_{n,i} + L_{n,i} + {3\over2} \zeta_2 \,,
\label{UniversalFunci}
\end{eqnarray}
in which $\lfloor x \rfloor$ is the greatest integer less than or equal
to $x$.  Here $s_{i\cdots j} = (k_i + \cdots + k_j)^2$
are the momentum invariants.
(All indices are understood to be $\mod n$.)
The form of $D_{n,i}$ and $L_{n,i}$ depends upon whether $n$ is odd or even.
For the even case ($n=2m$) these quantities are given by
\begin{eqnarray}
D_{2m,i} &=& 
  -\sum_{r=2}^{m-2}
\Li_2 \Biggl( 1- { s_{i\cdots(i+r-1)} s_{(i-1)\cdots(i+r)}\over 
                  s_{i\cdots(i+r)} s_{(i-1)\cdots(i+r-1)} }  \Biggr)
- {1 \over 2} \Li_2 \Biggl( 1- { s_{i\cdots(i+m-2)} s_{(i-1)\cdots(i+m-1)}
                   \over s_{i\cdots(i+m-1)} s_{(i-1)\cdots(i+m-2)}} \Biggr)
 \,,
\nn \\
L_{2m,i} &=& 
   {1\over 4}
  \ln^2 \Biggl({ -s_{i\cdots(i+m-1)}\over -s_{(i+1)\cdots(i+m)} } \Biggr) \,.
\label{DLeven}
\end{eqnarray}
In the odd case ($n=2m+1$), we have,
\begin{eqnarray}
D_{2m+1,i} &=& 
  -\sum_{r=2}^{m-1}
\Li_2 \Biggl( 1- { s_{i\cdots(i+r-1)} s_{(i-1)\cdots(i+r)}\over 
                  s_{i\cdots(i+r)} s_{(i-1)\cdots(i+r-1)} }  \Biggr)
\,,
\nn \\
L_{2m+1,i} &=& 
  - {1\over 2}
  \ln \Biggl({ -s_{i\cdots(i+m-1)}\over -s_{i\cdots(i+m)} } \Biggr)
  \ln \Biggl({ -s_{(i+1)\cdots(i+m)}\over -s_{(i-1)\cdots(i+m-1)} } \Biggr) \,.
\label{DLodd}
\end{eqnarray}

For $n=4$ the above formula does not hold; in that case the finite
part is simply
\begin{equation}
F_4^{(1)}(0) 
= {1\over 2} \ln^2\biggl({-t\over-s}\biggr) + 4 \zeta_2 \,. 
\label{STYF10}
\end{equation}

To make contact with the string theory
literature~\cite{BES,AldayMaldacena, AMTrouble} we define,
\begin{equation}
\fstring(\lambda) = 4 \sum_{l=1}^\infty a^l f^{(l)}_0 \,, \hskip 2 cm 
\gstring(\lambda)
 = 2 \sum_{l=2}^\infty { a^l \over l } f^{(l)}_1 \,, \hskip 2 cm 
\kstring(\lambda)
 = -{1\over2} \sum_{l=2}^\infty {a^l \over l^2} f^{(l)}_2 \,. \hskip .3 cm 
\end{equation}
In terms of these functions, the BDS ansatz~(\ref{BDSAnsatz}) may be written
as,
\def\Divergent{\mathop{\rm Div}\nolimits}
\begin{equation}
\ln{\cal M}_n 
= \Divergent_n + {\fstring(\lambda)\over 4} F_n^{(1)}(0) + n \kstring(\lambda)
  + \Cstring(\lambda) \,.
\label{BDSAnsatzSimp}
\end{equation}
The infrared-divergent part is
\begin{equation}
\Divergent_n =
-\sum_{i=1}^n \Biggl[ {1\over 8 \ep^2} \fstring^{(-2)}\biggl({\lambda
    \mu_{IR}^{2\eps}\over (-s_{i,i+1})^\eps}\biggr) + {1\over 4 \ep}
\gstring^{(-1)}\biggl({\lambda \mu_{IR}^{2\eps}\over
           (-s_{i,i+1})^\eps}\biggr)\Biggr]\,,
\end{equation}
where 
\begin{equation}
\biggl(\lambda {d \over d \lambda} \biggr)^2 \fstring^{(-2)}(\lambda) 
= \fstring(\lambda) \,, \hskip 2 cm 
\biggl(\lambda {d \over d \lambda} \biggr) \gstring^{(-1)}(\lambda) 
= g(\lambda)\,,
\end{equation}
and $\mu_{IR}^2 = 4 \pi e^{-\gamma} \mu^2$.
The first few orders of both the
weak~\cite{CuspWeak,KorchemskyMarchesini,BDS,BCDKS,CSVcusp} and
strong~\cite{Kruczenski,CuspStrongCoupling,KRTT} coupling expansion 
for $f(\lambda)$ have been computed, with the result
\begin{eqnarray}
f(\lambda) &=& \frac{\lambda}{2\pi^2}\left( 1-\frac{\lambda}{48}
+\frac{11\,\lambda^2}{11520}
-\left(\frac{73}{1290240}+
\frac{\zeta_3^2}{512 \pi^6}\right)\lambda^3+ \cdots   
\right) \,, 
\hskip0.6cm \lambda\to0\,,\\
f(\lambda)&=&\frac{\sqrt{\lambda}}{\pi}\left(1-\frac{3\ln
2}{\sqrt{\lambda}}-\frac{\rm K}{\lambda}+\cdots\right) \,,
\hskip5.6cm \lambda\to\infty\,, \hskip .8 cm 
\label{StrongExpansion}
\end{eqnarray}
where ${\rm K}=\sum_{n\ge 0}\frac{(-1)^n}{(2n+1)^2}\simeq0.9159656\ldots$  
is the Catalan constant%
\footnote{The third term in \eqn{StrongExpansion}
was first found in the expansion of an 
integral equation~\cite{BES} in ref.~\cite{BKK}.}.
The function $g(\lambda)$ has also been computed through four 
loops~\cite{BDS,CSVcollinear}, but is less well known at strong 
coupling~\cite{AldayMaldacena}%
\footnote{
At the next order~\cite{KRTT} in the strong-coupling
expansion of $g(\lambda)$, one encounters difficulties with the 
closed-string version of dimensional regularization used 
in ref.~\cite{AldayMaldacena}.},
\begin{eqnarray}
g(\lambda) &=& -\zeta_3 \biggl({\lambda\over8\pi^2}\biggr)^2
+ {2\over3} \bigl( 6 \zeta_5 + 5 \zeta_2\zeta_3 \bigr)
\biggl({\lambda\over8\pi^2}\biggr)^3 
- (77.56 \pm 0.02) \biggl({\lambda\over8\pi^2}\biggr)^4 + \cdots \,, 
\hskip0.1cm \lambda\to0\,,~~~~~~~~~\\
g(\lambda)&=& (1 - \ln 2) { \sqrt{\lambda} \over 2\pi } + \cdots \,,
\hskip8.6cm \lambda\to\infty\,. 
\end{eqnarray}

Beisert, Eden and Staudacher (BES) proposed~\cite{ES,BES} a striking
integral equation giving the cusp anomalous dimension
$\fstring(\lambda)$ for all values of the coupling.  This integral
equation has passed a number of stringent tests at both
weak~\cite{BCDKS,CSVcusp} and strong
coupling~\cite{CuspStrongCoupling,BESStrong,BKK}, and is therefore very
likely the correct expression.

With the cusp anomalous dimension known, the
BDS ansatz~(\ref{BDSAnsatz})
predicts the MHV amplitudes for {\it all} values of the
coupling, up to the undetermined functions $\gstring(\lambda)$, 
$\kstring(\lambda)$ and $\Cstring(\lambda)$, which are independent 
of the kinematics.  The ansatz has been checked through three loops 
at four points~\cite{ABDK,BDS}. Integral
representations of the four-point amplitude have also been given at
four~\cite{BCDKS} and five loops~\cite{FiveLoop}, though these
expressions have not yet been integrated at $\Ord(\eps^0)$
to yield explicit functions of the external momenta.  
If one assumes the dual conformal invariance mentioned in the
Introduction, which we shall discuss at greater length below, 
then the form of the finite parts in the four- and five-point amplitudes
are fixed, which provides another way of arriving at the
ansatz~(\ref{BDSAnsatz}) for $n=4$ or 5.  However, beyond
five external legs, the assumption of dual conformal invariance
does not suffice to fix the functional form of the finite parts of MHV
amplitudes.

\subsection{Scattering Amplitudes at Strong Coupling}

Alday and Maldacena have proposed a very interesting way to compute 
color-ordered planar scattering amplitudes at strong coupling, using the 
AdS/CFT correspondence~\cite{AldayMaldacena}.  They argue that
the leading dependence of any amplitude on the coupling has the form 
$\exp(-\frac{\sqrt{\lambda}}{2 \pi}A)$, where $A$ is
the regularized area of a special surface in AdS, whose definition and 
properties are reviewed below.  Their result reproduces the 
BDS ansatz for the four-point amplitude at strong coupling.

Long ago, Gross and Mende~\cite{GrossMende} showed that in 
string theory in flat space-time, scattering amplitudes can be computed in the
high-energy and fixed-angle regime using a semiclassical
approach. Alday and Maldacena noticed that, thanks to the properties of
anti-de~Sitter space, a semiclassical calculation suffices to compute
the strong-coupling limit of amplitudes in the $\NeqFour$ theory,
for any energy. Glossing over technical details, scattering amplitudes are
given by a saddle-point approximation to the world-sheet
partition function with certain vertex operator insertions.

Two-dimensional duality transformations map a vertex operator
of momentum $k^\mu_i$ to a null (light-like) segment 
pointing along the direction of the momentum of the
corresponding gluon in a dual AdS space.
The endpoint coordinates $y$ of each null segment obey
\begin{eqnarray}
\Delta y_i^\mu = 2\pi k_i^\mu\,.
\label{winding}
\end{eqnarray}
Momentum conservation then implies that these segments form a closed 
polygon in these dual coordinates. 

Alday and Maldacena do not compute the prefactor of
$\exp(-\frac{\sqrt{\lambda}}{2\pi}A)$, 
which is subleading in the strong-coupling
expansion but must contain all the dependence on the polarizations
of the external particles.  MHV scattering amplitudes in the $\NeqFour$
theory are special because supersymmetry Ward 
identities~\cite{SWI,BDDKSelfDual} imply
that they are all identical up to simple spinor product factors.
Equivalently, MHV amplitudes can be written as a product of the tree
amplitude, and an additional factor dependent only on the momentum
invariants, and not on the polarizations.  In the MHV case,
it is natural to identify the tree amplitude with the prefactor.
(A similar proposal has also been made for the non-MHV case~\cite{AFK},
although it is not clear how it can be consistent with the intricate
structure of the one-loop amplitudes~\cite{Fusing,OneloopTwistorB}.)
Then the additional factor is given, to leading-order in the
strong-coupling expansion, by the saddle-point approximation of a suitably
restricted world-sheet partition function.  The world sheets must be
restricted to unpunctured surfaces whose boundary is the closed
polygon of null dual segments.  That is, up to some technical details,
the factor is just the expectation value of a null Wilson loop in the
dual space.

Alday and Maldacena computed the leading term in the strong-coupling
expansion of the four-gluon amplitude by building an explicit solution
out of cusp solutions written down earlier by
Kruczenski~\cite{Kruczenski}.  The solution can also be found by
solving the sigma-model equations~\cite{Mironov} once the Virasoro
constraints are imposed~\cite{Yang}.  The minimal-surface approach has
not yet yielded complete expressions for higher-point amplitudes, because of
difficulties in solving the minimal surface conditions with the proper
boundary conditions, although there has been some 
progress~\cite{Mironov,MinimalSurfaces}.  
However, Buchbinder~\cite{BuchbinderIR} has
shown that the infrared-divergent terms can be obtained using this
approach, and that (as expected) they are consistent with the known
exponentiation in gauge theories.  Komargodski has
argued~\cite{Komargodski} that the collinear-splitting amplitude
obtained from this approach is consistent with the known perturbative
splitting amplitudes.

While the original arguments have been formulated at strong coupling,
the work of Alday and Maldacena also inspired the discovery of a
surprising connection between MHV amplitudes and Wilson loops at weak
coupling at one loop~\cite{DrummondVanishing,BrandhuberWilson}, for
the simplest two-loop
amplitudes~\cite{DHKSTwoloopBoxWilson,ConformalWard}, and perhaps also
at higher orders.

\subsection{Trouble at Large $n$}

In the limit of a large number of external legs, Alday and
Maldacena~\cite{AMTrouble} argued that the situation simplifies
because one may approximate the null Wilson loop by a smooth one.
They considered a rectangular Wilson loop with space-like edges of
length $L$ and width $T$ as the approximation of a null Wilson loop
zig-zagging around the rectangular one.  Each edge corresponds to many
gluons moving in one direction, alternating with many gluons moving in
an opposite direction.  The expectation value of the Wilson loop is
clearly divergent as the area is scaled to infinity. However, one may focus
on the scale-invariant part, which is proportional to
$T/L$. Moreover, one may consider the kinematic configuration
corresponding to $T/L\gg 1$. In this case one may ignore the
contribution from the sides of length $L$ to leading order.  Thus one may
further approximate the Wilson loop by two parallel lines of length $T$ 
at distance $L$ from each other.  In other words, the dominant
$T/L$-dependent part of the world-sheet area is essentially 
$T$ times the heavy-quark potential, computed at strong 
coupling~\cite{StrongCouplingQQ}:
\begin{eqnarray}
\ln \langle W\rangle = 
\sqrt{\lambda}\;\frac{4\pi^2}{\Gamma\left(\frac{1}{4}\right)^4}\;\frac{T}{L}
\,,
\qquad\lambda\gg1\,.
\label{AM_WL}
\end{eqnarray}
Since this expression is finite as $T$ and $L$ 
are scaled to infinity, one should compare
it with the $T/L$-dependent terms in the logarithm of the finite part of
the BDS ansatz. 

Alday and Maldacena~\cite{AMTrouble} worked out the behavior
of the BDS ansatz in this limit by making use of the all-$n$ one-loop relation
between MHV amplitudes and Wilson loops~\cite{BrandhuberWilson},
and the known value of the rectangular Wilson loop at one loop.
They obtained, at strong coupling and for $T\gg L$,
\begin{eqnarray}
\frac{f(\lambda)}{4} \, F_n^{(1)}(0)~
\stackrel{n\rightarrow\infty}{\longrightarrow}~
\frac{\sqrt{\lambda}}{4}\;\frac{T}{L}\,.
\label{AM_expect}
\end{eqnarray}
This result differs from \eqn{AM_WL}, indicating that the BDS ansatz is
incomplete in this limit.
It is interesting to note that the two formul\ae\ differ by less than
10\%, hinting that it may be possible to systematically correct
the BDS ansatz.

While this difference appeared in a particular (and somewhat singular)
kinematic configuration with a very large number of legs, 
the fact that it depends on $T/L$ makes clear that it arises from 
a nontrivial function of momenta. As we discuss below, dual 
conformal symmetry suggests that the first place such a function 
can occur is at six points.  

\subsection{Pseudo-Conformal Integrals and Dual Conformal Invariance}

We now turn to a review of the observed restrictions that dual
conformal invariance places on the integrals appearing in the 
amplitudes~\cite{MagicIdentities, BCDKS,
FiveLoop, DrummondVanishing}.  This mysterious symmetry
plays an important role in our story.
${\cal N}=4$ super-Yang-Mills theory is a conformal field theory at the
quantum level; conformal invariance may be observed in correlation
functions of operators of definite (anomalous) dimension.  However, the
constraints it imposes on on-shell scattering amplitudes are
obscured, both by the need for an infrared regulator and by anomalies
analogous to the holomorphic anomaly of collinear
operators~\cite{HolomorphicAnomaly}.

By inspecting the known results for the one-, two- and three-loop
four-gluon amplitudes~\cite{GSB,BRY,BDS}, Drummond, Henn, Sokatchev
and Smirnov~\cite{MagicIdentities} observed that after continuing the
external momenta of the integrals off shell (in a sense we will
describe below) and then taking $D=4$, they exhibit an $SO(2,4)$ dual
conformal symmetry. This symmetry is distinct from the
four-dimensional position-space conformal group.  Interestingly it
holds individually for each contributing integral through five
loops~\cite{BCDKS,FiveLoop,DrummondVanishing}.  A discussion of the
consequences of dual conformal symmetry for complete amplitudes,
instead of individual integrals, may be found in
refs.~\cite{DHKSTwoloopBoxWilson, ConformalWard}.

The origin of dual conformal symmetry remains obscure, and
its broad validity for amplitudes remains to be proven.%
\footnote{Dual conformal symmetry was introduced in the context of
multi-loop ladder integrals~\cite{Broadhurst}; it has also cropped up
in the two-dimensional theory of Reggeon interactions, again in the
planar limit~\cite{LipatovDuality}.}  
 Indeed, it is not clear that
dual conformal symmetry holds for all contributions to scattering
amplitudes.  For example, at the five-point level, the dual conformal
properties of the parity-odd pieces are not
apparent~\cite{TwoLoopFiveB}.
(It is conceivable that they might be re-expressed as a
linear combination of integrals with manifest properties,
although it seems unlikely, because the single available 
parity-odd contraction of the Levi-Civita tensor at the 
five-point level does not transform homogeneously under the 
inversion of dual variables described below.)
At least in this case, the parity-odd
pieces do not enter the remainder function $R_5^{(2)}$ defined in
\eqn{def_remainder}.  This feature suggests that, as long as we
subtract appropriate lower-loop contributions (to be specified further
below), we may be able to identify simple conformal properties for the
finite remainders $R_n^{(2)}$ of MHV amplitudes to all loop orders.
In this vein, an anomalous conformal Ward identity has been proven for
the finite remainders of Wilson loops~\cite{ConformalWard}, and later found at
strong coupling~\cite{Komargodski}.
The BDS ansatz was shown to obey the same Ward identity, which was
then proposed to hold for MHV amplitudes~\cite{DHKSTwoloopBoxWilson}.

\begin{figure}
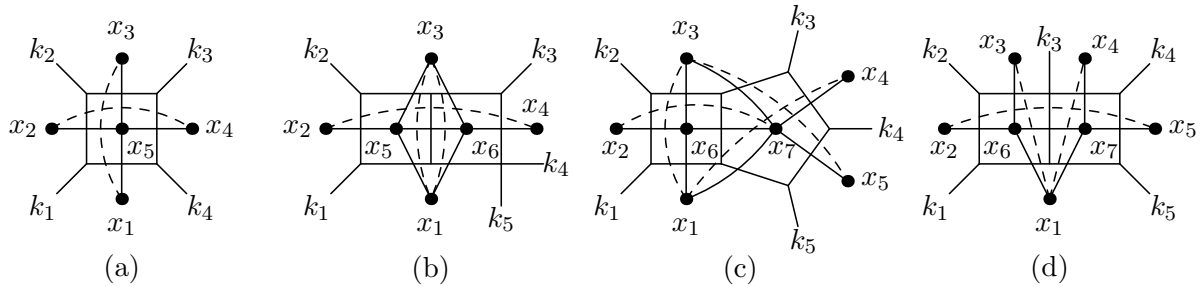

\begin{feynartspicture}(468,117)(4,1)
\FADiagram{}
\FAProp(7.5,10.)(12.5,10.)(0.,){/Straight}{0}
\FAProp(7.5,15.)(12.5,15.)(0.,){/Straight}{0}
\FAProp(7.5,10.)(7.5,15.)(0.,){/Straight}{0}
\FAProp(12.5,10.)(12.5,15.)(0.,){/Straight}{0}
\FAProp(7.5,15.)(5.37878,17.1213)(0.,){/Straight}{0}
\FAProp(12.5,15.)(14.6213,17.1213)(0.,){/Straight}{0}
\FAProp(12.5,10.)(14.6213,7.8787)(0.,){/Straight}{0}
\FAProp(7.5,10)(5.37878,7.8787)(0.,){/Straight}{0}
\FAProp(5.,12.5)(15.,12.5)(0.,){/Straight}{0}
\FAProp(10.,7.5)(10.,17.5)(0.,){/Straight}{0}
\FAProp(5.,12.5)(15.,12.5)(-0.3,){/ScalarDash}{0}
\FAProp(10.,7.5)(10.,17.5)(-0.3,){/ScalarDash}{0}
\FALabel(4.37878,18.1213)[]{$k_2$}
\FALabel(15.6213,18.1213)[]{$k_3$}
\FALabel(15.6213,6.8787)[]{$k_4$}
\FAVert(10.,17.5){0}
\FALabel(10.,19.5)[]{$x_3$}
\FAVert(10.,7.5){0}
\FALabel(10.,5.5)[]{$x_1$}
\FAVert(10.,12.5){0}
\FALabel(11.4,11.)[]{$x_5$}
\FAVert(5.,12.5){0}
\FALabel(3.,12.5)[]{$x_2$}
\FAVert(15.,12.5){0}
\FALabel(17.,12.5)[]{$x_4$}
\FALabel(10.,2.5)[]{(a)}
\FALabel(4.37878,6.8787)[]{$k_1$}
\FADiagram{}
\FAProp(5.,15.)(15.,15.)(0.,){/Straight}{0}
\FAProp(15.,10.)(5.,10.)(0.,){/Straight}{0}
\FAProp(5.,10.)(5.,15.)(0.,){/Straight}{0}
\FAProp(10.,10.)(10.,15.)(0.,){/Straight}{0}
\FAProp(15.,10.)(15.,15.)(0.,){/Straight}{0}
\FAProp(5.,15.)(2.87868,17.1213)(0.,){/Straight}{0}
\FAProp(15.,15.)(17.1213,17.1213)(0.,){/Straight}{0}
\FAProp(15.,10.)(18.,10.)(0.,){/Straight}{0}
\FAProp(15.,10.)(15.,7.)(0.,){/Straight}{0}
\FAProp(5.,10.)(2.87878,7.8787)(0.,){/Straight}{0}
\FAProp(10.,17.5)(10.,7.5)(0.2,){/ScalarDash}{0}
\FAProp(10.,17.5)(7.5,12.5)(0.,){/Straight}{0}
\FAProp(7.5,12.5)(10.,7.5)(0.,){/Straight}{0}
\FAProp(10.,17.5)(10.,7.5)(-0.2,){/ScalarDash}{0}
\FAProp(10.,17.5)(12.5,12.5)(0.,){/Straight}{0}
\FAProp(12.5,12.5)(10.,7.5)(0.,){/Straight}{0}
\FAProp(2.5,12.5)(17.5,12.5)(-0.2,){/ScalarDash}{0}
\FAProp(2.5,12.5)(17.5,12.5)(0.,){/Straight}{0}
\FALabel(1.87868,6.8787)[]{$k_1$}
\FALabel(1.87868,18.1213)[]{$k_2$}
\FALabel(18.1213,18.1213)[]{$k_3$}
\FALabel(18.914,10.)[]{$k_4$}
\FALabel(15.,6.086)[]{$k_5$}
\FAVert(12.5,12.5){0}
\FALabel(13.9,11.)[]{$x_6$}
\FAVert(7.5,12.5){0}
\FALabel(6.3,11.)[]{$x_5$}
\FAVert(10.,17.5){0}
\FALabel(10.,19.5)[]{$x_3$}
\FAVert(10.,7.5){0}
\FALabel(10.,5.5)[]{$x_1$}
\FAVert(2.5,12.5){0}
\FALabel(0.5,12.5)[]{$x_2$}
\FAVert(17.5,12.5){0}
\FALabel(17.5,14.)[]{$x_4$}
\FALabel(10.,2.5)[]{(b)}
\FADiagram{}
\FAProp(12.5,12.5)(6.1529,17.5)(0.15,){/Straight}{0}
\FAProp(12.5,12.5)(6.1529,7.5)(-0.15,){/Straight}{0}
\FAProp(3.6529,15.)(8.6529,15.)(0.,){/Straight}{0}
\FAProp(8.6529,15.)(8.6529,10.)(0.,){/Straight}{0}
\FAProp(8.6529,10.)(3.6529,10.)(0.,){/Straight}{0}
\FAProp(3.6529,10.)(3.6529,15.)(0.,){/Straight}{0}
\FAProp(8.6529,15.)(13.4082,16.5451)(0.,){/Straight}{0}
\FAProp(8.6529,10.)(13.4082,8.45492)(0.,){/Straight}{0}
\FAProp(13.4082,8.45492)(16.3471,12.5)(0.,){/Straight}{0}
\FAProp(13.4082,16.5451)(16.3471,12.5)(0.,){/Straight}{0}
\FAProp(3.6529,15.)(1.5316,17.1213)(0.,){/Straight}{0}
\FAProp(3.6529,10.)(1.5316,7.87868)(0.,){/Straight}{0}
\FAProp(16.3471,12.5)(19.3471,12.5)(0.,){/Straight}{0}
\FAProp(13.4082,8.45492)(14.1085,5.53781)(0.,){/Straight}{0}
\FAProp(13.4082,16.5451)(14.1085,19.4622)(0.,){/Straight}{0}
\FAProp(6.1529,7.5)(6.1529,17.5)(-0.3,){/ScalarDash}{0}
\FAProp(6.1529,7.5)(6.1529,17.5)(0,){/Straight}{0}
\FAProp(1.1529,12.5)(12.5,12.5)(-0.3,){/ScalarDash}{0}
\FAProp(1.1529,12.5)(12.5,12.5)(0.,){/Straight}{0}
\FAProp(6.1529,17.5)(17.6349,8.7693)(-0.2,){/ScalarDash}{0}
\FAProp(6.1529,7.5)(17.6349,16.2307)(-0.15,){/ScalarDash}{0}
\FALabel(0.5316,6.87868)[]{$k_1$}
\FALabel(0.5316,18.1213)[]{$k_2$}
\FALabel(14.5085,4.53781)[]{$k_5$}
\FALabel(14.5085,20.4622)[]{$k_3$}
\FALabel(20.7611,12.5)[]{$k_4$}
\FAVert(6.1529,12.5){0}
\FALabel(7.5259,11.)[]{$x_6$}
\FAVert(12.5,12.5){0}
\FAVert(6.1529,7.5){0}
\FALabel(6.1529,5.5)[]{$x_1$}
\FAVert(6.1529,17.5){0}
\FALabel(6.1529,19.5)[]{$x_3$}
\FAVert(1.1529,12.5){0}
\FALabel(1.1529,11.)[]{$x_2$}
\FAProp(17.6349,8.7693)(12.5,12.5)(0.,){/Straight}{0}
\FAProp(17.6349,16.2307)(12.5,12.5)(0.,){/Straight}{0}
\FALabel(13.,11.)[0]{$x_7$}
\FAVert(17.6349,8.7693){0}
\FALabel(19.6349,8.7693)[]{$x_5$}
\FAVert(17.6349,16.2307){0}
\FALabel(19.6349,16.2307)[]{$x_4$}
\FALabel(10.,2.5)[]{(c)}
\FADiagram{}
\FAProp(5.,15.)(15.,15.)(0.,){/Straight}{0}
\FAProp(15.,10.)(5.,10.)(0.,){/Straight}{0}
\FAProp(5.,10.)(5.,15.)(0.,){/Straight}{0}
\FAProp(10.,10.)(10.,15.)(0.,){/Straight}{0}
\FAProp(15.,10.)(15.,15.)(0.,){/Straight}{0}
\FAProp(5.,15.)(2.87868,17.1213)(0.,){/Straight}{0}
\FAProp(15.,15.)(17.1213,17.1213)(0.,){/Straight}{0}
\FAProp(15.,10.)(17.1213,7.87868)(0.,){/Straight}{0}
\FAProp(5.,10.)(2.87868,7.87868)(0.,){/Straight}{0}
\FAProp(10.,15.)(10.,18.)(0.,){/Straight}{0}
\FAProp(2.5,12.5)(17.5,12.5)(-0.2,){/ScalarDash}{0}
\FAProp(2.5,12.5)(17.5,12.5)(0.,){/Straight}{0}
\FAProp(10.,7.5)(12.5,17.5)(0.,){/ScalarDash}{0}
\FAProp(10.,7.5)(7.5,17.5)(0.,){/ScalarDash}{0}
\FAVert(12.5,12.5){0}
\FAProp(12.5,12.5)(12.5,17.5)(0.,){/Straight}{0}
\FAProp(12.5,12.5)(10.,7.5)(0.,){/Straight}{0}
\FAProp(7.5,12.5)(10.,7.5)(0.,){/Straight}{0}
\FAProp(7.5,12.5)(7.5,17.5)(0.,){/Straight}{0}
\FAVert(7.5,12.5){0}
\FAVert(7.5,17.5){0}
\FALabel(6.3,11.)[]{$x_6$}
\FALabel(13.9,11.)[]{$x_7$}
\FALabel(6.,18.5)[]{$x_3$}
\FAVert(12.5,17.5){0}
\FALabel(14.,18.5)[]{$x_4$}
\FAVert(10.,7.5){0}
\FALabel(10.,5.5)[]{$x_1$}
\FAVert(2.5,12.5){0}
\FALabel(2.5,11.)[]{$x_2$}
\FAVert(17.5,12.5){0}
\FALabel(19.5,12.5)[]{$x_5$}
\FALabel(1.87868,6.8787)[]{$k_1$}
\FALabel(1.87868,18.1213)[]{$k_2$}
\FALabel(18.1213,18.1213)[]{$k_4$}
\FALabel(18.1213,6.8787)[]{$k_5$}
\FALabel(10.,18.914)[]{$k_3$}
\FALabel(10.,2.5)[]{(d)}
\end{feynartspicture}
\vskip -0.6125 cm
\caption[a]{\small Examples of pseudo-conformal integrals. Points $x_i$
label the vertices of the dual graph, a solid line connecting two points
$x_i$ and $x_j$ corresponds to a factor of $1/x_{ij}^2$, while a dashed
line corresponds to a factor of $x_{ij}^2$. An integral is
pseudo-conformal if the difference between the number of solid lines and
dashed lines at a vertex equals $4$ at the internal vertices and zero at
the external vertices.  Graphs $(b)$, $(c)$ and $(d)$ show that the
integrals appearing in the even part of the five-point two-loop amplitude
are pseudo-conformal.  (In fact, $(d)$ only appears in the odd part.)}
\label{ConformalIntegralsFigure}
\end{figure}

Dual conformal symmetry is most transparent in terms of dual variables $x_i$
which are related to the gluon momenta in a way analogous
to \eqn{winding},
\begin{eqnarray}
k_i=x_{i+1}-x_{i}\,, 
\label{DualMomenta}
\end{eqnarray}
and similarly for the loop momenta. Formally, in an integral
we may identify the variables $x_i$
as the positions of the vertices of a dual graph. In this construction
the momentum conservation constraint is replaced by an invariance 
under uniform shifts of the dual coordinates $x_i$. 
Since the dual variables are unconstrained and the parametrization
(\ref{DualMomenta}) automatically satisfies momentum conservation,
one may define an inversion 
operator
\begin{eqnarray}
I=\sum_i I_i\,, \qquad I_i:\ x_i^\mu\mapsto \frac{x_i^\mu}{x_i^2}\,.
\end{eqnarray}
One may also define conformal boost transformations of the dual
variables $x$; they are generated by
\begin{eqnarray}
K^\mu=\sum_i K^\mu_i\,, \qquad
K^\mu_i=2 x_i^\mu x_i\cdot \partial_i-x_i^2\partial_i^\mu\,.
\end{eqnarray}
Invariance under $I$ implies invariance under $K$,
because the conformal boosts are generated by two inversions, with
an infinitesimal translation of $x$ in between.

The principal conformal-invariance constraints on integrals
constructed from the invariants $x_{ij}^2$ are exposed by performing
the inversion $I$.
Since dimensional regularization breaks the dual conformal invariance,
for the purposes of exposing the symmetry we adopt a different
infrared regularization of the integrals.  We take the external legs
off shell, letting $k_i^2\neq0$, instead of using a dimensional
regulator as in the rest of the paper.  Under the inversion, the
Mandelstam invariants and the integration measure transform as
\begin{equation}
x_{ij}^2 \rightarrow {x_{ij}^2 \over x_i^2 x_j^2} \,, \hskip 1 cm
 {d^4 x_5} \rightarrow {d^4 x_5 \over (x_5^2)^{4}} \,, \hskip 1 cm
{d^4 x_6} \rightarrow {d^4 x_6 \over (x_6^2)^{4}} \,.
\end{equation}
Simple dual diagrams may be used to identify all the integrals that
are invariant under $I$; see refs.~\cite{MagicIdentities,BCDKS,FiveLoop}
for a detailed discussion.
We call such integrals ``pseudo-conformal'' because in the amplitudes
we actually use the dimensionally-regulated versions of
the integrals, which breaks the conformal invariance but guarantees
the gauge-invariance of the amplitudes.

While pseudo-conformal integrals were initially identified in
four-point scattering amplitudes, they apparently play a more general
role in the planar ${\cal N}=4$ theory; they also appear in $n$-point
one-loop MHV amplitudes~\cite{NeqFourOneLoop}, as well as the even
part of the two-loop five-point
amplitude~\cite{TwoLoopFiveA,TwoLoopFiveB}
(see~\fig{ConformalIntegralsFigure}).  Their relevance for non-MHV
amplitudes, for example at one loop~\cite{Fusing}, remains to be
clarified.  In this paper we show that for the two-loop six-point MHV
amplitude, they again appear in the even part though some 
additional integrals also appear.  (We shall not compute
the odd part in this paper.)

More generally, conformal invariance 
implies that, as for two-dimensional conformal field theories,
conformally invariant quantities can depend only on the cross
ratios,
\begin{eqnarray}
u_{ijkl} = \frac{x_{ij}^2x_{kl}^2}{x_{ik}^2x_{jl}^2} \,,
\label{CrossRatios}
\end{eqnarray}
where $i, j, k$ and $l$ are any four vertices of the dual graph.  In
scattering amplitudes we must exclude cross ratios
that vanish or diverge on account of an $x_{ij}^2$
vanishing due to the on-shell conditions, $x_{i,i+1}^2 = k_i^2 \to 0$.
Consequently, the only cross ratios that may appear in a conformally
invariant part of an on-shell amplitude are
those with $|i-j|\ge 2\; \mod n$ for all 
$i,j=1,\dots,n$; otherwise $x_{i j}$ would correspond to a massless
momentum.  We remark that there are no conformal
cross ratios for $n=4, 5$, as it is not possible to form a cross
ratio without encountering a vanishing $x_{ij}^2$~\cite{DHKSTwoloopBoxWilson}.

\section{Six-point Integrand}
\label{IntegrandSection}

\begin{figure}
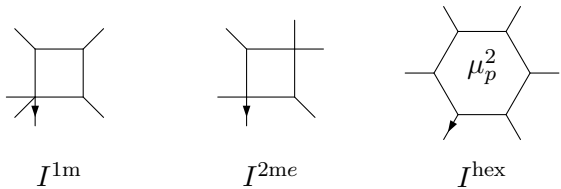

\begin{feynartspicture}(240,80)(3,1)
\FADiagram{}
\FAProp(7.5,10.5)(12.5,10.5)(0.,){/Straight}{0}
\FAProp(7.5,15.5)(12.5,15.5)(0.,){/Straight}{0}
\FAProp(7.5,10.5)(7.5,15.5)(0.,){/Straight}{0}
\FAProp(12.5,10.5)(12.5,15.5)(0.,){/Straight}{0}
\FAProp(7.5,15.5)(5.37878,17.6213)(0.,){/Straight}{0}
\FAProp(12.5,15.5)(14.6213,17.6213)(0.,){/Straight}{0}
\FAProp(12.5,10.5)(14.6213,8.3787)(0.,){/Straight}{0}
\FAProp(7.5,10.5)(4.5,10.5)(0.,){/Straight}{0}
\FAProp(7.5,10.5)(7.5,7.5)(0.,){/Straight}{1}
\FAProp(7.5,10.5)(5.3787,8.3787)(0.,){/Straight}{0}
\FALabel(10.,2.5)[]{$I^{1\rm m}$}
\FADiagram{}
\FAProp(7.5,10.5)(12.5,10.5)(0.,){/Straight}{0}
\FAProp(7.5,15.5)(12.5,15.5)(0.,){/Straight}{0}
\FAProp(7.5,10.5)(7.5,15.5)(0.,){/Straight}{0}
\FAProp(12.5,10.5)(12.5,15.5)(0.,){/Straight}{0}
\FAProp(7.5,15.5)(5.37878,17.6213)(0.,){/Straight}{0}
\FAProp(12.5,15.5)(15.5,15.5)(0.,){/Straight}{0}
\FAProp(12.5,15.5)(12.5,18.5)(0.,){/Straight}{0}
\FAProp(12.5,10.5)(14.6213,8.3787)(0.,){/Straight}{0}
\FAProp(7.5,10.5)(4.5,10.5)(0.,){/Straight}{0}
\FAProp(7.5,10.5)(7.5,7.5)(0.,){/Straight}{1}
\FALabel(10.,2.5)[]{$I^{2{\rm m}e}$}
\FADiagram{}
\FAProp(7.5,8.6699)(12.5,8.6699)(0.,){/Straight}{0}
\FAProp(12.5,8.6699)(15.,13.)(0.,){/Straight}{0}
\FAProp(15.,13.)(12.5,17.3301)(0.,){/Straight}{0}
\FAProp(12.5,17.3301)(7.5,17.3301)(0.,){/Straight}{0}
\FAProp(7.5,17.3301)(5.,13.)(0.,){/Straight}{0}
\FAProp(5.,13.)(7.5,8.6699)(0.,){/Straight}{0}
\FAProp(7.5,8.6699)(6.,6.0718)(0.,){/Straight}{1}
\FAProp(12.5,8.6699)(14.,6.0718)(0.,){/Straight}{0}
\FAProp(15.,13.)(18.,13.)(0.,){/Straight}{0}
\FAProp(12.5,17.3301)(14.,19.9282)(0.,){/Straight}{0}
\FAProp(7.5,17.3301)(6.,19.9282)(0.,){/Straight}{0}
\FAProp(5.,13.)(2.,13.)(0.,){/Straight}{0}
\FALabel(10.,13.6699)[]{$\mu_p^2$}
\FALabel(10.,2.5)[]{$I^{\rm hex}$}
\end{feynartspicture}
\vskip -0.6125 cm
\caption{The three independent integrals which contribute to the
six-particle amplitude at one loop.
The external momenta are labeled clockwise, 
beginning with $k_1$ which is denoted by an arrow.
The $\mu_p^2$ in the hexagon integral indicates that
a factor of the square of the $(-2\eps)$-dimensional components
of the loop momentum is to be inserted in the numerator of the integrand.
}
\label{OneLoopIntegralsFigure}
\end{figure}

Before turning to our calculation of the two-loop amplitude,
we present the one-loop six-point 
amplitude~\cite{BDDKSelfDual}.  
We drop the parity-odd terms, which are proportional to the Levi-Civita 
tensor and are also $\Ord(\eps)$.
We split the even part into two pieces,
\begin{equation}
M_6^{(1), D=4-2\ep}(\epsilon) = M_6^{(1), D=4} (\epsilon) +
                               M_6^{(1), \mu} (\epsilon)\,,
\label{OneLoopAssembly}
\end{equation}
where
\begin{equation}
M_6^{(1), D=4} (\epsilon) = - \frac{1}{4} \sum_{6~{\rm perms.}}
\left[
s_{45} s_{56} I^{1\rm m}(\epsilon)
+ \frac{1}{2} (s_{123} s_{345} - s_{12} s_{45}) I^{2{\rm m}e}(\epsilon)
\right] \,,
\label{OneLoopAssemblyA}
\end{equation}
and
\begin{equation}
M_6^{(1),\mu} (\epsilon) = 
- \frac{1}{4} \tr[123456] I^{\rm hex}(\epsilon)\,.
\label{OneLoopAssemblyB}
\end{equation}
The first piece contains scalar box integrals, which are constructible solely 
from cuts in $D=4$.  The hexagon integral in the second piece
contains a numerator factor of $\mu_p^2$, as indicated in
\fig{OneLoopIntegralsFigure}, which can only be detected by cuts
in which the cut loop momentum $p$ has a nonvanishing 
$(-2\ep)$-dimensional component $\mu_p$:  $p^2 \equiv p_{[4]}^2 - \mu_p^2$,
so that $\mu_p^2$ is positive.  (The overall scale $\mu$ arising from 
dimensional regularization, {\it e.g.} in \eqn{OneloopMHVAmplitude}, should
not be confused with $\mu_p$.)

The three integrals appearing in the one-loop amplitude
are defined in \fig{OneLoopIntegralsFigure}, using the convention
that each loop momentum integral is normalized according to
\begin{equation}
-i \pi^{-D/2} e^{\ep \gamma} \int d^D p\,.
\end{equation}
The coefficient of the hexagon integral involves the quantity
\begin{equation}
{\rm tr}[123456] \equiv {\rm tr}[ \s k_1 \s k_2 \s k_3 \s k_4
\s k_5 \s k_6] = s_{123} s_{234} s_{345}
- s_{61} s_{34} s_{123} - s_{12} s_{45} s_{234}
- s_{23} s_{56} s_{345}\,.
\end{equation}
The sum in \eqn{OneLoopAssemblyA}
runs over the six cyclic permutations of the external momenta $k_i$.
There is a symmetry factor of $1/2$ in front of $I^{2{\rm m}e}$
to correct for a double count in this case --- there are only three
separate $I^{2{\rm m}e}$ integrals.

\subsection{Construction of the Integrand from Unitarity Cuts}

We computed the two-loop six-point MHV amplitude using the unitarity
method~\cite{NeqFourOneLoop,Fusing,GeneralizedUnitarity}.  We know on
general grounds that the amplitude can be written in terms of Feynman
integrals multiplied by rational coefficients. 

In principle, we should consider all two-loop six-point integrals.
We can reduce this rather large set of integrals using the no-triangle
constraint, which states that one can find a representation for 
the amplitude in which no integral with a triangle (or bubble) subintegral
appears.  Using generalized unitarity, we may establish this constraint
for two classes of potential triangle contributions:  those which may be excised from
a multi-loop topology by cutting only gluon lines, and 
(using supersymmetry 
Ward identities) those which may be excised by cutting at most 
five legs~\cite{BCDKS}.
We shall assume that the no-triangle constraint holds as well for the other 
contributions.

\begin{figure}
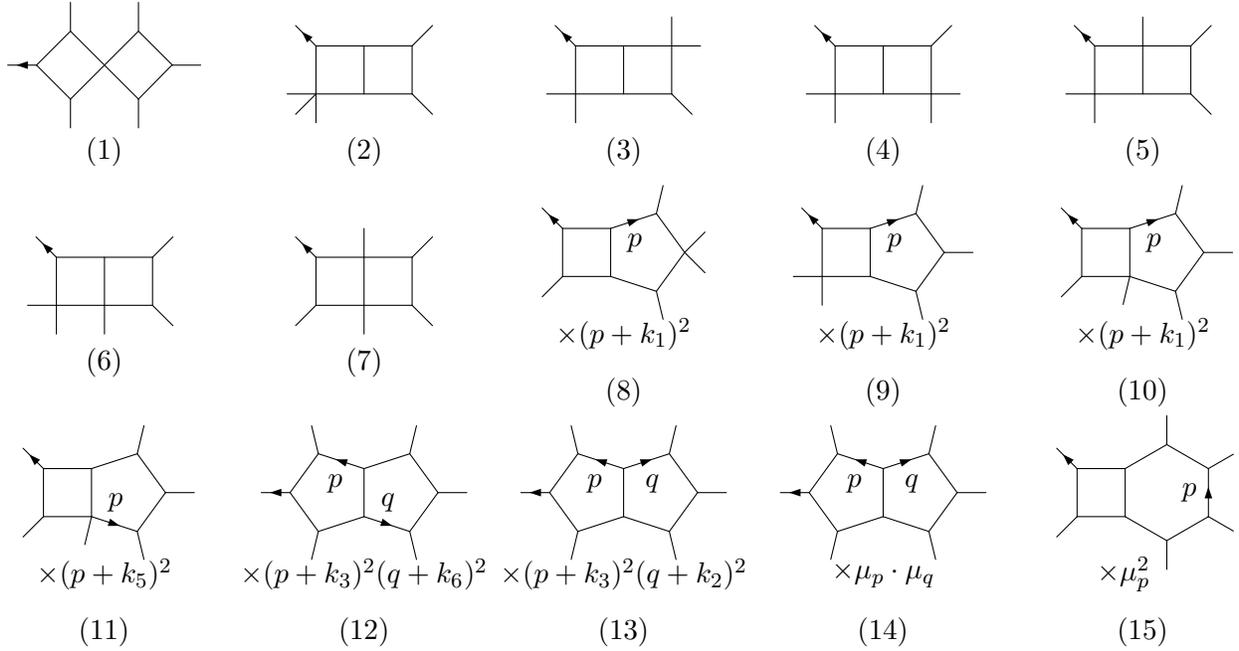

\begin{feynartspicture}(400,240)(5,3)
\FADiagram{}
\FAProp(0.,13.)(3.5355,9.46447)(0.,){/Straight}{0}
\FAProp(0.,13.)(-3.5355,16.5355)(0.,){/Straight}{0}
\FAProp(-3.5355,16.5355)(-7.0711,13.)(0.,){/Straight}{0}
\FAProp(-7.0711,13.)(-3.5355,9.46447)(0.,){/Straight}{0}
\FAProp(-3.5355,9.46447)(0.,13.)(0.,){/Straight}{0}
\FAProp(0.,13.)(3.5355,16.5355)(0.,){/Straight}{0}
\FAProp(3.5355,16.5355)(7.0711,13.)(0.,){/Straight}{0}
\FAProp(7.011,13.)(3.5355,9.46447)(0.,){/Straight}{0}
\FAProp(-3.5355,16.5355)(-3.5355,19.5355)(0.,){/Straight}{0}
\FAProp(3.5355,16.5355)(3.5355,19.5355)(0.,){/Straight}{0}
\FAProp(7.011,13.)(10.011,13.)(0.,){/Straight}{0}
\FAProp(3.5355,9.46447)(3.5355,6.46447)(0.,){/Straight}{0}
\FAProp(-3.5355,9.46447)(-3.5355,6.46447)(0.,){/Straight}{0}
\FAProp(-7.0711,13.)(-10.07107,13.)(0.,){/Straight}{1}
\FALabel(0.,4.)[]{(1)}
\FADiagram{}
\FAProp(0.,15.)(10.,15.)(0.,){/Straight}{0}
\FAProp(10.,10.)(0.,10.)(0.,){/Straight}{0}
\FAProp(0.,10.)(0.,15.)(0.,){/Straight}{0}
\FAProp(5.,10.)(5.,15.)(0.,){/Straight}{0}
\FAProp(10.,10.)(10.,15.)(0.,){/Straight}{0}
\FAProp(0.,15.)(-2.1213,17.1213)(0.,){/Straight}{1}
\FAProp(10.,15.)(12.1213,17.1213)(0.,){/Straight}{0}
\FAProp(10.,10.)(12.1213,7.87868)(0.,){/Straight}{0}
\FAProp(0.,10.)(0.,7.)(0.,){/Straight}{0}
\FAProp(0.,10.)(-3.,10.)(0.,){/Straight}{0}
\FAProp(0.,10.)(-2.1213,7.87868)(0.,){/Straight}{0}
\FALabel(5.,4.)[]{(2)}
\FADiagram{}
\FAProp(5.,15.)(15.,15.)(0.,){/Straight}{0}
\FAProp(15.,10.)(5.,10.)(0.,){/Straight}{0}
\FAProp(5.,10.)(5.,15.)(0.,){/Straight}{0}
\FAProp(10.,10.)(10.,15.)(0.,){/Straight}{0}
\FAProp(15.,10.)(15.,15.)(0.,){/Straight}{0}
\FAProp(5.,15.)(2.87868,17.1213)(0.,){/Straight}{1}
\FAProp(15.,15.)(18.,15.)(0.,){/Straight}{0}
\FAProp(15.,15.)(15.,18.)(0.,){/Straight}{0}
\FAProp(15.,10.)(17.1213,7.87868)(0.,){/Straight}{0}
\FAProp(5.,10.)(5.,7.)(0.,){/Straight}{0}
\FAProp(5.,10.)(2.,10.)(0.,){/Straight}{0}
\FALabel(10.,4.)[]{(3)}
\FADiagram{}
\FAProp(10.,15.)(20.,15.)(0.,){/Straight}{0}
\FAProp(20.,10.)(10.,10.)(0.,){/Straight}{0}
\FAProp(10.,10.)(10.,15.)(0.,){/Straight}{0}
\FAProp(15.,10.)(15.,15.)(0.,){/Straight}{0}
\FAProp(20.,10.)(20.,15.)(0.,){/Straight}{0}
\FAProp(10.,15.)(7.87868,17.1213)(0.,){/Straight}{1}
\FAProp(20.,15.)(22.1213,17.1213)(0.,){/Straight}{0}
\FAProp(20.,10.)(23.,10.)(0.,){/Straight}{0}
\FAProp(20.,10.)(20.,7.)(0.,){/Straight}{0}
\FAProp(10.,10.)(10.,7.)(0.,){/Straight}{0}
\FAProp(10.,10.)(7.,10.)(0.,){/Straight}{0}
\FALabel(15.,4.)[]{(4)}
\FADiagram{}
\FAProp(15.,15.)(25.,15.)(0.,){/Straight}{0}
\FAProp(25.,10.)(15.,10.)(0.,){/Straight}{0}
\FAProp(15.,10.)(15.,15.)(0.,){/Straight}{0}
\FAProp(20.,10.)(20.,15.)(0.,){/Straight}{0}
\FAProp(25.,10.)(25.,15.)(0.,){/Straight}{0}
\FAProp(15.,15.)(12.87868,17.1213)(0.,){/Straight}{1}
\FAProp(25.,15.)(27.1213,17.1213)(0.,){/Straight}{0}
\FAProp(25.,10.)(27.1213,7.87868)(0.,){/Straight}{0}
\FAProp(15.,10.)(15.,7.)(0.,){/Straight}{0}
\FAProp(15.,10.)(12.,10.)(0.,){/Straight}{0}
\FAProp(20.,15.)(20.,18.)(0.,){/Straight}{0}
\FALabel(20.,4.)[]{(5)}
\FADiagram{}
\FAProp(-5.,15.)(5.,15.)(0.,){/Straight}{0}
\FAProp(5.,10.)(-5.,10.)(0.,){/Straight}{0}
\FAProp(-5.,10.)(-5.,15.)(0.,){/Straight}{0}
\FAProp(0.,10.)(0.,15.)(0.,){/Straight}{0}
\FAProp(5.,10.)(5.,15.)(0.,){/Straight}{0}
\FAProp(-5.,15.)(-7.1213,17.1213)(0.,){/Straight}{1}
\FAProp(5.,15.)(7.1213,17.1213)(0.,){/Straight}{0}
\FAProp(5.,10.)(7.1213,7.87868)(0.,){/Straight}{0}
\FAProp(-5.,10.)(-5.,7.)(0.,){/Straight}{0}
\FAProp(-5.,10.)(-8.,10.)(0.,){/Straight}{0}
\FAProp(0.,10.)(0.,7.)(0.,){/Straight}{0}
\FALabel(0.,4.)[]{(6)}
\FADiagram{}
\FAProp(0.,15.)(10.,15.)(0.,){/Straight}{0}
\FAProp(10.,10.)(0.,10.)(0.,){/Straight}{0}
\FAProp(0.,10.)(0.,15.)(0.,){/Straight}{0}
\FAProp(5.,10.)(5.,15.)(0.,){/Straight}{0}
\FAProp(10.,10.)(10.,15.)(0.,){/Straight}{0}
\FAProp(0.,15.)(-2.1213,17.1213)(0.,){/Straight}{1}
\FAProp(10.,15.)(12.1213,17.1213)(0.,){/Straight}{0}
\FAProp(10.,10.)(12.1213,7.87868)(0.,){/Straight}{0}
\FAProp(0.,10.)(-2.1213,7.87868)(0.,){/Straight}{0}
\FAProp(5.,10.)(5.,7.)(0.,){/Straight}{0}
\FAProp(5.,15.)(5.,18.)(0.,){/Straight}{0}
\FALabel(5.,4.)[]{(7)}
\FADiagram{}
\FAProp(3.6529,18.)(8.6529,18.)(0.,){/Straight}{0}
\FAProp(8.6529,18.)(8.6529,13.)(0.,){/Straight}{0}
\FAProp(8.6529,13.)(3.6529,13.)(0.,){/Straight}{0}
\FAProp(3.6529,13.)(3.6529,18.)(0.,){/Straight}{0}
\FAProp(8.6529,18.)(13.4082,19.5451)(0.,){/Straight}{1}
\FAProp(8.6529,13.)(13.4082,11.45492)(0.,){/Straight}{0}
\FAProp(13.4082,11.45492)(16.3471,15.5)(0.,){/Straight}{0}
\FAProp(13.4082,19.5451)(16.3471,15.5)(0.,){/Straight}{0}
\FAProp(3.6529,18.)(1.5316,20.1213)(0.,){/Straight}{1}
\FAProp(3.6529,13.)(1.5316,10.87868)(0.,){/Straight}{0}
\FAProp(16.3471,15.5)(18.4684,17.6213)(0.,){/Straight}{0}
\FAProp(16.3471,15.5)(18.4684,13.3787)(0.,){/Straight}{0}
\FAProp(13.4082,11.45492)(14.1085,8.53781)(0.,){/Straight}{0}
\FAProp(13.4082,19.5451)(14.1085,22.4622)(0.,){/Straight}{0}
\FALabel(11.1529,16.5)[]{$p$}
\FALabel(10.,7.)[]{$\times(p + k_1)^2$}
\FALabel(10.,1.)[]{(8)}
\FADiagram{}
\FAProp(8.6529,18.)(13.6529,18.)(0.,){/Straight}{0}
\FAProp(13.6529,18.)(13.6529,13.)(0.,){/Straight}{0}
\FAProp(13.6529,13.)(8.6529,13.)(0.,){/Straight}{0}
\FAProp(8.6529,13.)(8.6529,18.)(0.,){/Straight}{0}
\FAProp(13.6529,18.)(18.4082,19.5451)(0.,){/Straight}{1}
\FAProp(13.6529,13.)(18.4082,11.45492)(0.,){/Straight}{0}
\FAProp(18.4082,11.45492)(21.3471,15.5)(0.,){/Straight}{0}
\FAProp(18.4082,19.5451)(21.3471,15.5)(0.,){/Straight}{0}
\FAProp(8.6529,18.)(6.5316,20.1213)(0.,){/Straight}{1}
\FAProp(8.6529,13.)(8.6529,10.)(0.,){/Straight}{0}
\FAProp(8.6529,13.)(5.6529,13.)(0.,){/Straight}{0}
\FAProp(21.3471,15.5)(24.3471,15.5)(0.,){/Straight}{0}
\FAProp(18.4082,11.45492)(19.1085,8.53781)(0.,){/Straight}{0}
\FAProp(18.4082,19.5451)(19.1085,22.4622)(0.,){/Straight}{0}
\FALabel(16.1529,16.5)[]{$p$}
\FALabel(15.,7.)[]{$\times(p + k_1)^2$}
\FALabel(15.,1.)[]{(9)}
\FADiagram{}
\FAProp(13.6529,18.)(18.6529,18.)(0.,){/Straight}{0}
\FAProp(18.6529,18.)(18.6529,13.)(0.,){/Straight}{0}
\FAProp(18.6529,13.)(13.6529,13.)(0.,){/Straight}{0}
\FAProp(13.6529,13.)(13.6529,18.)(0.,){/Straight}{0}
\FAProp(18.6529,18.)(23.4082,19.5451)(0.,){/Straight}{1}
\FAProp(18.6529,13.)(23.4082,11.45492)(0.,){/Straight}{0}
\FAProp(23.4082,11.45492)(26.3471,15.5)(0.,){/Straight}{0}
\FAProp(23.4082,19.5451)(26.3471,15.5)(0.,){/Straight}{0}
\FAProp(13.6529,18.)(11.5316,20.1213)(0.,){/Straight}{1}
\FAProp(13.6529,13.)(11.5316,10.87868)(0.,){/Straight}{0}
\FAProp(26.3471,15.5)(29.3471,15.5)(0.,){/Straight}{0}
\FAProp(23.4082,11.45492)(24.1085,8.53781)(0.,){/Straight}{0}
\FAProp(23.4082,19.5451)(24.1085,22.4622)(0.,){/Straight}{0}
\FAProp(18.6529,13.)(17.9526,10.0829)(0.,){/Straight}{0}
\FALabel(21.1529,16.5)[]{$p$}
\FALabel(20.,7.)[]{$\times(p + k_1)^2$}
\FALabel(20.,1.)[]{(10)}
\FADiagram{}
\FAProp(-6.3471,15.)(-1.3471,15.)(0.,){/Straight}{0}
\FAProp(-1.3471,15.)(-1.3471,10.)(0.,){/Straight}{0}
\FAProp(-1.3471,10.)(-6.3471,10.)(0.,){/Straight}{0}
\FAProp(-6.3471,10.)(-6.3471,15.)(0.,){/Straight}{0}
\FAProp(-1.3471,15.)(3.4082,16.5451)(0.,){/Straight}{0}
\FAProp(-1.3471,10.)(3.4082,8.45492)(0.,){/Straight}{1}
\FAProp(3.4082,8.45492)(6.3471,12.5)(0.,){/Straight}{0}
\FAProp(3.4082,16.5451)(6.3471,12.5)(0.,){/Straight}{0}
\FAProp(-6.3471,15.)(-8.4684,17.1213)(0.,){/Straight}{1}
\FAProp(-6.3471,10.)(-8.4684,7.87868)(0.,){/Straight}{0}
\FAProp(6.3471,12.5)(9.3471,12.5)(0.,){/Straight}{0}
\FAProp(3.4082,8.45492)(4.1085,5.53781)(0.,){/Straight}{0}
\FAProp(3.4082,16.5451)(4.1085,19.4622)(0.,){/Straight}{0}
\FAProp(-1.3471,10.)(-2.0474,7.08289)(0.,){/Straight}{0}
\FALabel(1.1529,11.5)[]{$p$}
\FALabel(0.,4.)[]{$\times(p + k_5)^2$}
\FALabel(0.,-2.)[]{(11)}
\FADiagram{}
\FAProp(5.,10.)(5.,15.)(0.,){/Straight}{0}
\FAProp(5.,15.)(9.7553,16.5451)(0.,){/Straight}{0}
\FAProp(5.,10.)(9.7553,8.45492)(0.,){/Straight}{1}
\FAProp(9.7553,8.45492)(12.6942,12.5)(0.,){/Straight}{0}
\FAProp(9.7553,16.5451)(12.6942,12.5)(0.,){/Straight}{0}
\FAProp(12.6942,12.5)(15.6942,12.5)(0.,){/Straight}{0}
\FAProp(9.7553,8.45492)(10.4556,5.53781)(0.,){/Straight}{0}
\FAProp(9.7553,16.5451)(10.4556,19.4622)(0.,){/Straight}{0}
\FAProp(5.,15.)(0.2447,16.5451)(0.,){/Straight}{1}
\FAProp(5.,10.)(0.2447,8.45492)(0.,){/Straight}{0}
\FAProp(0.2447,8.45492)(-2.6942,12.5)(0.,){/Straight}{0}
\FAProp(0.2447,16.5451)(-2.6942,12.5)(0.,){/Straight}{0}
\FAProp(0.2447,8.45492)(-0.4556,5.53781)(0.,){/Straight}{0}
\FAProp(0.2447,16.5451)(-0.4556,19.4622)(0.,){/Straight}{0}
\FAProp(-2.6942,12.5)(-5.6942,12.5)(0.,){/Straight}{1}
\FALabel(2.,13.5)[]{$p$}
\FALabel(7.5,11.5)[]{$q$}
\FALabel(5.,4.)[]{$\times(p + k_3)^2 (q+k_6)^2$}
\FALabel(5.,-2.)[]{(12)}
\FADiagram{}
\FAProp(10.,10.)(10.,15.)(0.,){/Straight}{0}
\FAProp(10.,15.)(14.7553,16.5451)(0.,){/Straight}{1}
\FAProp(10.,10.)(14.7553,8.45492)(0.,){/Straight}{0}
\FAProp(14.7553,8.45492)(17.6942,12.5)(0.,){/Straight}{0}
\FAProp(14.7553,16.5451)(17.6942,12.5)(0.,){/Straight}{0}
\FAProp(17.6942,12.5)(20.6942,12.5)(0.,){/Straight}{0}
\FAProp(14.7553,8.45492)(15.4556,5.53781)(0.,){/Straight}{0}
\FAProp(14.7553,16.5451)(15.4556,19.4622)(0.,){/Straight}{0}
\FAProp(10.,15.)(5.2447,16.5451)(0.,){/Straight}{1}
\FAProp(10.,10.)(5.2447,8.45492)(0.,){/Straight}{0}
\FAProp(5.2447,8.45492)(2.3058,12.5)(0.,){/Straight}{0}
\FAProp(5.2447,16.5451)(2.3058,12.5)(0.,){/Straight}{0}
\FAProp(5.2447,8.45492)(4.5444,5.53781)(0.,){/Straight}{0}
\FAProp(5.2447,16.5451)(4.5444,19.4622)(0.,){/Straight}{0}
\FAProp(2.3058,12.5)(-0.6942,12.5)(0.,){/Straight}{1}
\FALabel(7.,13.5)[]{$p$}
\FALabel(13.,13.5)[]{$q$}
\FALabel(10.,4.)[]{$\times(p + k_3)^2 (q+k_2)^2$}
\FALabel(10.,-2.)[]{(13)}
\FADiagram{}
\FAProp(15.,10.)(15.,15.)(0.,){/Straight}{0}
\FAProp(15.,15.)(19.7553,16.5451)(0.,){/Straight}{1}
\FAProp(15.,10.)(19.7553,8.45492)(0.,){/Straight}{0}
\FAProp(19.7553,8.45492)(22.6942,12.5)(0.,){/Straight}{0}
\FAProp(19.7553,16.5451)(22.6942,12.5)(0.,){/Straight}{0}
\FAProp(22.6942,12.5)(25.6942,12.5)(0.,){/Straight}{0}
\FAProp(19.7553,8.45492)(20.4556,5.53781)(0.,){/Straight}{0}
\FAProp(19.7553,16.5451)(20.4556,19.4622)(0.,){/Straight}{0}
\FAProp(15.,15.)(10.2447,16.5451)(0.,){/Straight}{1}
\FAProp(15.,10.)(10.2447,8.45492)(0.,){/Straight}{0}
\FAProp(10.2447,8.45492)(7.3058,12.5)(0.,){/Straight}{0}
\FAProp(10.2447,16.5451)(7.3058,12.5)(0.,){/Straight}{0}
\FAProp(10.2447,8.45492)(9.5444,5.53781)(0.,){/Straight}{0}
\FAProp(10.2447,16.5451)(9.5444,19.4622)(0.,){/Straight}{0}
\FAProp(7.3058,12.5)(4.3058,12.5)(0.,){/Straight}{1}
\FALabel(12.,13.5)[]{$p$}
\FALabel(18.,13.5)[]{$q$}
\FALabel(15.,4.)[]{$\times \mu_p \cdot \mu_q$}
\FALabel(15.,-2.)[]{(14)}
\FADiagram{}
\FAProp(13.1699,15.)(18.1699,15.)(0.,){/Straight}{0}
\FAProp(18.1699,15.)(18.1699,10.)(0.,){/Straight}{0}
\FAProp(18.1699,10.)(13.1699,10.)(0.,){/Straight}{0}
\FAProp(13.1699,10.)(13.1699,15.)(0.,){/Straight}{0}
\FAProp(18.1699,15.)(22.5,17.5)(0.,){/Straight}{0}
\FAProp(18.1699,10.)(22.5,7.5)(0.,){/Straight}{0}
\FAProp(22.5,17.5)(26.8302,15.)(0.,){/Straight}{0}
\FAProp(22.5,7.5)(26.8302,10.)(0.,){/Straight}{0}
\FAProp(26.8302,15.)(26.8302,10.)(0.,){/Straight}{-1}
\FAProp(13.1699,15.)(11.0486,17.1213)(0.,){/Straight}{1}
\FAProp(13.1699,10.)(11.0486,7.87868)(0.,){/Straight}{0}
\FAProp(22.5,17.5)(22.5,20.5)(0.,){/Straight}{0}
\FAProp(22.5,7.5)(22.5,4.5)(0.,){/Straight}{0}
\FAProp(26.8302,15.)(29.4283,16.5)(0.,){/Straight}{0}
\FAProp(26.8302,10.)(29.4283,8.5)(0.,){/Straight}{0}
\FALabel(24.8302,12.5)[]{$p$}
\FALabel(18.,4.)[]{$\times \mu_p^2$}
\FALabel(20.,-2.)[]{(15)}
\end{feynartspicture}
\caption{The 15 independent integrals which contribute to the
even part of the six-particle amplitude at two loops.
The external momenta are labeled clockwise with $k_1$ 
denoted by an arrow.  Integrals (8)--(15) are defined to include
the indicated numerator factors involving the loop momenta.
In the last two integrals, $\mu_p$ denotes the
$(-2 \e)$-dimensional component of the loop momentum $p$.
}
\label{ContributingIntegralsFigure}
\end{figure}

\begin{figure}
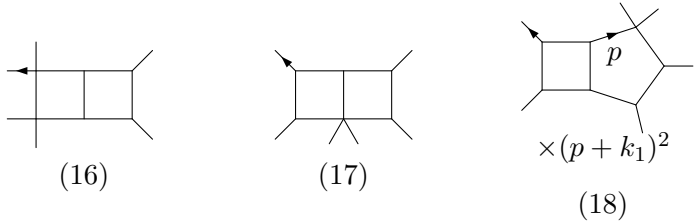

\begin{feynartspicture}(240,80)(3,1)
\FADiagram{}
\FAProp(0.,15.)(10.,15.)(0.,){/Straight}{0}
\FAProp(10.,10.)(0.,10.)(0.,){/Straight}{0}
\FAProp(0.,10.)(0.,15.)(0.,){/Straight}{0}
\FAProp(5.,10.)(5.,15.)(0.,){/Straight}{0}
\FAProp(10.,10.)(10.,15.)(0.,){/Straight}{0}
\FAProp(0.,15.)(-3.,15.)(0.,){/Straight}{1}
\FAProp(0.,15.)(0.,18.)(0.,){/Straight}{0}
\FAProp(10.,15.)(12.1213,17.1213)(0.,){/Straight}{0}
\FAProp(10.,10.)(12.1213,7.87868)(0.,){/Straight}{0}
\FAProp(0.,10.)(0.,7.)(0.,){/Straight}{0}
\FAProp(0.,10.)(-3.,10.)(0.,){/Straight}{0}
\FALabel(5.,4.)[]{(16)}
\FADiagram{}
\FAProp(5.,15.)(15.,15.)(0.,){/Straight}{0}
\FAProp(15.,10.)(5.,10.)(0.,){/Straight}{0}
\FAProp(5.,10.)(5.,15.)(0.,){/Straight}{0}
\FAProp(10.,10.)(10.,15.)(0.,){/Straight}{0}
\FAProp(15.,10.)(15.,15.)(0.,){/Straight}{0}
\FAProp(5.,15.)(2.87868,17.1213)(0.,){/Straight}{1}
\FAProp(15.,15.)(17.1213,17.1213)(0.,){/Straight}{0}
\FAProp(15.,10.)(17.1213,7.87868)(0.,){/Straight}{0}
\FAProp(5.,10.)(2.87868,7.87868)(0.,){/Straight}{0}
\FAProp(10.,10.)(8.5,7.4019)(0.,){/Straight}{0}
\FAProp(10.,10.)(11.5,7.4019)(0.,){/Straight}{0}
\FALabel(10.,4.)[]{(17)}
\FADiagram{}
\FAProp(8.6529,18.)(13.6529,18.)(0.,){/Straight}{0}
\FAProp(13.6529,18.)(13.6529,13.)(0.,){/Straight}{0}
\FAProp(13.6529,13.)(8.6529,13.)(0.,){/Straight}{0}
\FAProp(8.6529,13.)(8.6529,18.)(0.,){/Straight}{0}
\FAProp(13.6529,18.)(18.4082,19.5451)(0.,){/Straight}{1}
\FAProp(13.6529,13.)(18.4082,11.45492)(0.,){/Straight}{0}
\FAProp(18.4082,11.45492)(21.3471,15.5)(0.,){/Straight}{0}
\FAProp(18.4082,19.5451)(21.3471,15.5)(0.,){/Straight}{0}
\FAProp(8.6529,18.)(6.5316,20.1213)(0.,){/Straight}{1}
\FAProp(8.6529,13.)(6.5316,10.87868)(0.,){/Straight}{0}
\FAProp(21.3471,15.5)(24.3471,15.5)(0.,){/Straight}{0}
\FAProp(18.4082,11.45492)(19.1085,8.53781)(0.,){/Straight}{0}
\FAProp(18.4082,19.5451)(20.7396,21.4331)(0.,){/Straight}{0}
\FAProp(18.4082,19.5451)(16.7743,22.0611)(0.,){/Straight}{0}
\FALabel(16.1529,16.5)[]{$p$}
\FALabel(15.,7.)[]{$\times (p + k_1)^2$}
\FALabel(15.,1.)[]{(18)}
\end{feynartspicture}
\vskip -0.6125 cm
\caption{The three independent two-loop
diagrams which can be made pseudo-conformal by including appropriate
numerators but which do not contribute to the amplitude ({\it i.e.}, 
they enter with zero coefficient).}
\label{NonContributingIntegralsFigure}
\end{figure}

We expect, based on previous calculations of four-point amplitudes at
two, three, and four loops (as well as the consistency of the
five-loop construction) and of the five-point amplitude at two loops,
that essentially only pseudo-conformal integrals will be required for
the six-point amplitude.  The basic topologies for the integrals that
appear in the planar six-point MHV amplitude are shown in
\figs{ContributingIntegralsFigure}{NonContributingIntegralsFigure}.
Of these, all are pseudo-conformal, except for (14) and (15), whose
integrands vanish as $D\to4$, yet integral (15) is nonvanishing in
this limit.  The complete set of inequivalent pseudo-conformal
integrals (including numerator factors) is presented in
\fig{MotherFigure}.  The no-triangle constraint would also follow from
the stronger assumption of pseudo-conformality, were we to make it.
We will not; our calculation can be seen as a test of it instead. The
appearance of integrals (14) and (15) shows that additional pieces,
unobvious from naive considerations of dual conformal invariance,
enter.

\begin{figure}[t]
\centerline{\epsfxsize 5. truein \epsfbox{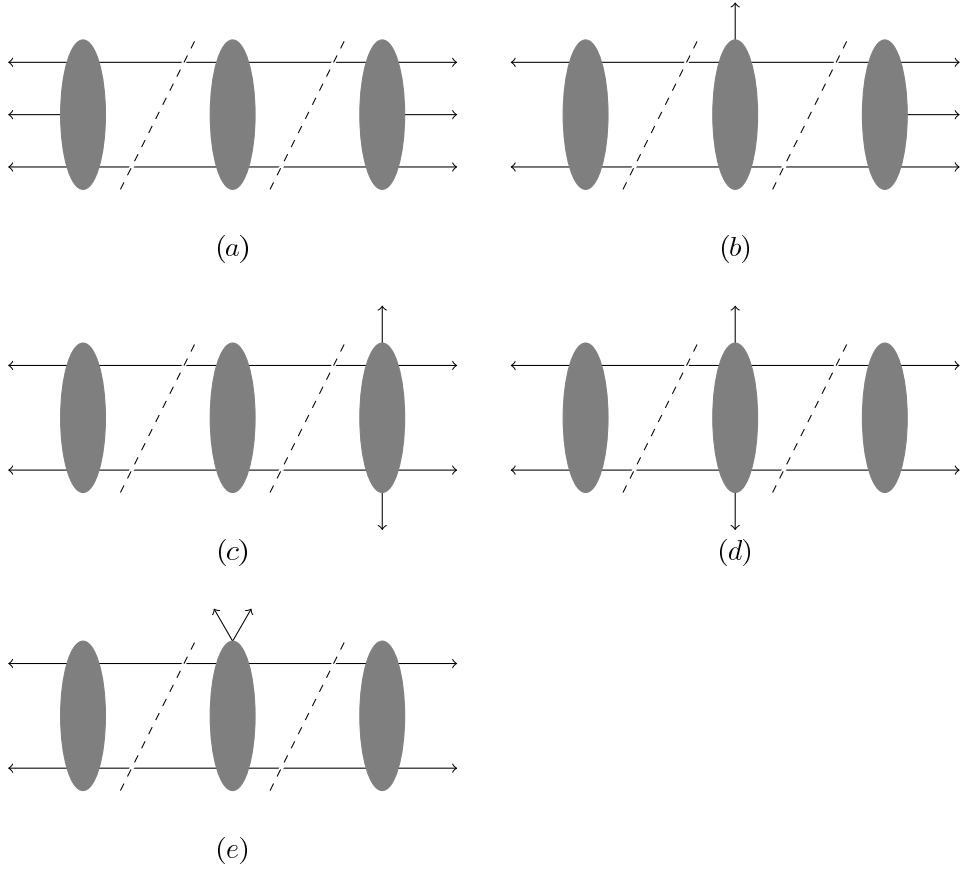}}
\caption[a]{\small The double two-particle cuts used to determine
the integrand.}
\label{DoubleTwoParticleCutsFigure}
\end{figure}

Given the no-triangle constraint, iterated two-particle cuts suffice
to determine the full integrand.  While the four-dimensional cuts
do not suffice to determine the integrand completely, they can
determine most terms.  The four-dimensional double two-particle cuts,
shown in \fig{DoubleTwoParticleCutsFigure}, are particularly convenient
to compute because they can be built out of MHV tree amplitudes.

\begin{figure}[t]
\centerline{\epsfxsize 4.3 truein \epsfbox{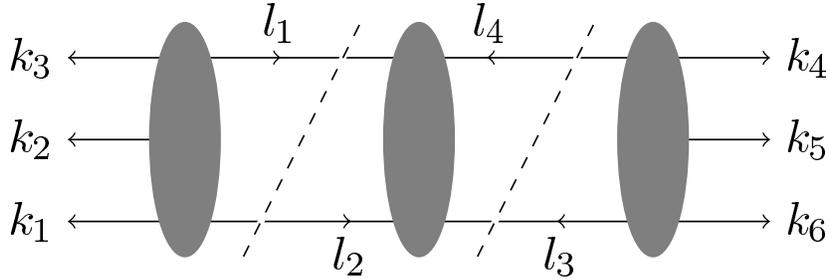}}
\caption[a]{\small Labeling of momenta for cut {\it a}.}
\label{ExampleLabelingFigure}
\end{figure}

As example, consider the helicity assignment $(1^-,2^-,3^+,4^+,5^+,6^+)$,
and compute cut {\it a\/}.
Because of a Ward identity for $\NeqFour$ 
supersymmetry~\cite{SWI,BDDKSelfDual}, once we have divided by the tree 
amplitude, the expression is in fact independent of the placement of 
the negative helicities.  
The labeling of the external legs is shown in \fig{ExampleLabelingFigure}.

The product of the three tree amplitudes corresponding to cut {\it a\/} is
\begin{equation}
i\frac{\spa1.2^3}{
  \spa2.3\spa3.{l_1}\spa{l_1}.{l_2}\spa{l_2}.1} 
   \times i\frac{\spa{(-l_2)}.{(-l_1)}^3}{
    \spa{(-l_1)}.{(-l_4)}\spa{(-l_4)}.{(-l_3)}\spa{(-l_3)}.{(-l_2)}} 
   \times i\frac{\spa{l_3}.{l_4}^3}{\spa{l_4}.4\spa4.5\spa5.6\spa6.{l_3}}\,.
\label{Cut_a}
\end{equation} 
After spinor simplifications, dividing by the tree amplitude, and
rationalizing denominators to Lorentz products, we find that the numerator
can be written as follows,
\begin{equation}
  \tr_+\left[\s l_2 \, \s k_1 \, \s k_6 \, \s l_3 \,
             \s l_4 \, \s k_4 \, \s k_3 \, \s l_1\right] 
  \tr_+\left[\s l_2 \, \s l_3 \, \s l_4 \, \s l_1\right]\,,
\end{equation} 
where $\tr_+[\cdots] = \frac 1 2 \tr[(1 + \gamma_5) \cdots]$.

Upon expanding the traces, we find both even and odd terms.
The odd terms contain a factor $\epsilon(a, b, c, d)
\equiv \epsilon_{\mu \nu \rho \sigma} a^\mu b^\nu c^\rho d^\sigma$,
whose origin lies in the presence of the $\gamma_5$ matrix inside the
traces.  (The product of two epsilon tensors would yield an even term,
but only the longer trace here can actually produce an epsilon tensor,
as $\epsilon(l_1, l_2, l_3, l_4)$ vanishes because of momentum conservation.)
We ignore the odd terms in our calculation.

In order to identify the coefficients of the integrals in 
\fig{ContributingIntegralsFigure}, we should use momentum conservation
to re-express all Lorentz invariants in terms of independent invariants.
The required simplifications can be done analytically, but in some cases
(for example cut {\it d\/})
it is easier to do them numerically, by matching to a target expression.

Doing so, we obtain for the final result of cut {\it a}, in the
$(3,4,5,6,1,2)$
permutation with respect to fig.~\ref{ContributingIntegralsFigure},
\def\indentA{\hphantom{\frac14 \Biggl[]}}
\begin{eqnarray}
  &&\frac 1 4 \Biggl[
    \frac {s_{123}^2 s_{34} s_{61} - s_{123}^2 s_{234} s_{345} 
           + s_{123} s_{234} s_{12} s_{45} + s_{123} s_{345} s_{23} s_{56}}
          {(k_1 + l_2)^2 (k_3 + l_1)^2 (k_4 + l_4)^2 (k_6 + l_3)^2} \nn\\ 
  &&\indentA
    + \frac {s_{123}^2 s_{345} - s_{123} s_{12} s_{45}}
          {(k_3 + l_1)^2 (l_2 + l_3)^2 (k_6 + l_3)^2}  
    + \frac {s_{123}^2 s_{234} - s_{123} s_{23} s_{56}}
          {(k_1 + l_2)^2 (l_2 + l_3)^2 (k_4 + l_4)^2} \nn\\ 
  &&\indentA
	+ \frac {s_{123}^2 s_{34}}
		{(k_3 + l_1)^2 (l_2 + l_3)^2 (k_4 +l_4)^2}  
    + \frac {s_{123}^2 s_{61}}
		{(k_1 + l_2)^2 (l_2 + l_3)^2 (k_6 +l_3)^2} \nn\\
   &&\indentA
	+ \frac {s_{123} s_{12} s_{23} (k_6 - l_2)^2}
          {(k_1 + l_2)^2 (k_3 + l_1)^2 (l_2 + l_3)^2 (k_6 + l_3)^2} 
    +\frac {s_{123} s_{12} s_{23} (k_4 - l_1)^2}
          {(k_1 + l_2)^2 (k_3 + l_1)^2 (l_2 + l_3)^2 (k_4 + l_4)^2} \nn\\ 
   &&\indentA
    +\frac {s_{123} s_{45} s_{56} (k_3 - l_4)^2}
          {(k_3 + l_1)^2 (l_2 + l_3)^2 (k_4 + l_4)^2 (k_6 + l_3)^2} 
    +\frac {s_{123} s_{45} s_{56} (k_1 - l_3)^2}
          {(k_1 + l_2)^2 (l_2 + l_3)^2 (k_4 + l_4)^2 (k_6 + l_3)^2}\nn\\ 
  &&\indentA
    +\frac {1}
          {(k_1 + l_2)^2 (k_3 + l_1)^2 (l_2 + l_3)^2 (k_4 + l_4)^2 
                (k_6 + l_3)^2} \nn\\ 
  &&\indentA\hskip 20mm \times\Bigl( 
         - s_{123}^2 s_{61} (k_3 - l_4)^2 (k_4 - l_1)^2 
           - s_{123}^2 s_{34} (k_1 - l_3)^2 (k_6 - l_2)^2 \nn\\
  &&\indentA \hskip 20mm\hphantom{\times \Bigl()}
           + s_{123} (s_{123} s_{234} - s_{23} s_{56}) 
                (k_3 - l_4)^2 (k_6 - l_2)^2 \nn\\
  &&\indentA \hskip 20mm\hphantom{\times \Bigl()}
           + s_{123} (s_{123} s_{345} - s_{12} s_{45}) 
                (k_1 - l_3)^2 (k_4 - l_1)^2 \Bigr)
  \Biggr]\,.
\label{cutabig}
\end{eqnarray}
{}From this expression we can read off the coefficients of every integral
in \figs{ContributingIntegralsFigure}{NonContributingIntegralsFigure} 
that is nonvanishing on cut {\it a\/} for four-dimensional
values of the cut momenta.  These integrals are, in order of their
appearance in \eqn{cutabig}, $I^{(1)}$, $I^{(3)}$ (twice), 
$I^{(4)}$ (twice), $I^{(9)}$ (four times), $I^{(12)}$ and $I^{(13)}$
(each twice). 
For example, from the second term,
up to a symmetry factor,
we can simply read off the coefficient of $I^{(3)}$ to be 
$s_{123} (s_{123} s_{345}-s_{12} s_{45})$, or equivalently
$s_{234} (s_{123} s_{234}-s_{23} s_{56})$ in the
labeling of \fig{ContributingIntegralsFigure}.  The numerator in the 
second term on the fourth line produces the required loop-momentum dependent
factor for a reflected version of $I^{(9)}$, corresponding to the
$(4,3,2,1,6,5)$ permutation.  The coefficient is $s_{123} s_{12} s_{23}$,
or $s_{234} s_{34} s_{23}$ in the figure's labeling.
The coefficients of the remaining integrals
can be determined by the other cuts in \fig{DoubleTwoParticleCutsFigure}.

At one loop in any supersymmetric theory, the improved ultraviolet
power-counting ensures that any rational terms in an amplitude are
linked to terms with branch cuts.  That is, all terms in an amplitude
can be determined solely from the standard integral basis, which can
be detected in four-dimensional cuts.  Beyond one loop, four-dimensional
cuts no longer suffice for $\NeqOne$ supersymmetric 
theories~\cite{NeqOneTwoLoop}.  In the $\NeqFour$ theory, four-point
amplitudes through five loops
are determined solely by their four-dimensional cuts.
The same is true for the even terms in the five-point amplitude.  (It
is no longer true for the odd terms, but in any case we are ignoring 
the corresponding terms in the six-point amplitude.)  
However, there is no proof other
than explicit computation of this observation.  Accordingly, we
cannot be certain that four-dimensional cuts will suffice for our
calculation.

Indeed, the hexagon integral in fig.~\ref{OneLoopIntegralsFigure}
will contribute terms of $\Ord(\e)$ to the one-loop amplitude.  
In the term in the iteration relation~(\ref{TwoloopOneloop})
in which the one-loop amplitude appears squared, 
the product of such terms with singular terms in $M_6^{\oneloop}(\e)$
survives to give $\Ord(\e^{-1})$ and finite contributions.
We will see in \sect{ResultsSectionMu} that
such contributions are offset by those coming from 
the last integral~(15) from fig.~\ref{ContributingIntegralsFigure},
induced by the $(-2\e)$-dimensional components of the loop momentum.
These contributions must be computed using $D$-dimensional cuts,
either making use of prior computations~\cite{BDDKSelfDual}, or
by direct computation.  We have computed two cuts, corresponding
to fig.~\ref{DoubleTwoParticleCutsFigure}({\it a})
and~({\it c}), using $D$-dimensional cuts.
These cuts
determine the coefficients of integrals~(14) and~(15), respectively,
in fig.~\ref{ContributingIntegralsFigure}.  The calculations were
done using the same approach used in refs.~\cite{ABDK,TwoLoopFiveB}.
While we can no longer use standard helicity states for the computation,
we can take advantage of the equivalence between the $\NeqFour$
theory and ten-dimensional $\NeqOne$ super-Yang-Mills theory
compactified on a torus.  We
compute the cuts with the spin algebra performed in the ten-dimensional
theory, keeping loop momenta in $D$ dimensions.  (External momenta can
be taken to be four-dimensional.)  The ten-dimensional gluon corresponds
to a four-dimensional gluon and six real scalar degrees of freedom, 
while the ten-dimensional Majorana-Weyl fermions correspond to 
four flavors of gluinos.

\subsection{Presentation of the Integrand}

By analyzing the cuts outlined in the previous section, we find the
complete expression for the parity-even part of the two-loop
six-particle amplitude to be
\begin{equation}
M_6^{(2), D=4-2\ep}(\epsilon) = M_6^{(2), D=4} (\epsilon) +
                               M_6^{(2), \mu} (\epsilon)\,,
\label{TwoLoopAssembly}
\end{equation}
where 
\begin{eqnarray}
M_6^{(2),D=4}(\epsilon) &=& \frac{1}{16} \sum_{12~{\rm perms.}}
\Bigg[
\frac{1}{4} c_1 I^{(1)}(\epsilon)
+ c_2 I^{(2)}(\epsilon)
+ \frac{1}{2} c_3 I^{(3)}(\epsilon)
+ \frac{1}{2} c_4 I^{(4)}(\epsilon)
+ c_5 I^{(5)}(\epsilon)
\cr
&&\qquad\qquad\quad\null
+ c_6 I^{(6)}(\epsilon)
+ \frac{1}{4} c_7 I^{(7)}(\epsilon)
+ \frac{1}{2} c_8 I^{(8)}(\epsilon)
+ c_9 I^{(9)}(\epsilon)
+ c_{10} I^{(10)}(\epsilon)
\cr
&&\qquad\qquad\quad\null
+ c_{11} I^{(11)}(\epsilon)
+ \frac{1}{2} c_{12} I^{(12)}(\epsilon)
+ \frac{1}{2} c_{13} I^{(13)}(\epsilon)
\Bigg] \,, 
\label{TwoLoopAssemblyA}
\end{eqnarray}
and 
\begin{eqnarray}
M_6^{(2),\mu}(\epsilon) &=& \frac{1}{16} \sum_{12~{\rm perms.}}
\Bigg[
  \frac{1}{4} c_{14} I^{(14)}(\epsilon)
+ \frac{1}{2} c_{15} I^{(15)}(\epsilon)
\Bigg] \,, 
\label{TwoLoopAssemblyB}
\end{eqnarray}
involving the 15 independent integrals shown in
\fig{ContributingIntegralsFigure}.  As with the one-loop
amplitude~(\ref{OneLoopAssembly}), we have separated the integrals
(1)--(13), which are constructible solely from four-dimensional cuts,
from integrals (14) and (15), in which the $(-2\ep)$-dimensional
components of loop momenta $\mu_p$ and $\mu_q$ are explicitly present.
(The $D=4$ superscript in $M_6^{(2),D=4}(\epsilon)$ refers to the lack of
$\mu_p$ and $\mu_q$ terms in the numerators of the integrands; the
argument $\ep$ indicates the dependence of the integrals on the
dimensional regularization parameter.)  Of the $1/16$ overall normalization 
in \eqns{TwoLoopAssemblyA}{TwoLoopAssemblyB}, $1/4$ is due to our choice
of normalization in \eqn{LeadingColorDecomposition} and the other factor 
of $1/4$ emerges from the calculation of the unitarity cuts.

The sum runs over the 12 cyclic and reflection permutations of
external legs,
\begin{eqnarray}
&& (1,2,3,4,5,6), \quad
   (2,3,4,5,6,1),  \quad
   (3,4,5,6,1,2),  \quad
(   4,5,6,1,2,3), \nn \\
&& (5,6,1,2,3,4),  \quad
   (6,1,2,3,4,5),  \quad
   (6,5,4,3,2,1),  \quad
   (1,6,5,4,3,2),  \nn \\
&& (2,1,6,5,4,3),   \quad
   (3,2,1,6,5,4),  \quad
   (4,3,2,1,6,5),  \quad
   (5,4,3,2,1,6).  \quad
\end{eqnarray}
The numerical coefficients in each term of
\eqns{TwoLoopAssemblyA}{TwoLoopAssemblyB} are symmetry factors to
remove double counts in the permutation sum.  The coefficients for the
$(1,2,3,4,5,6)$ permutation are,
\begin{eqnarray}
c_1 &=& 
s_{61} s_{34} s_{123} s_{345} + s_{12} s_{45} s_{234} s_{345} +
s_{345}^2 (s_{23} s_{56} - s_{123} s_{234})\,,\cr
c_2 &=& 
2 s_{12} s_{23}^2\,,\cr
c_3 &=& 
s_{234} (s_{123} s_{234} - s_{23} s_{56})\,,\cr
c_4 &=& 
s_{12} s_{234}^2\,,\cr
c_5 &=& 
s_{34} (s_{123} s_{234} - 2 s_{23} s_{56})\,,\cr
c_6 &=& 
- s_{12} s_{23} s_{234}\,,\cr
c_7 &=& 
2 s_{123} s_{234} s_{345} - 4 s_{61} s_{34} s_{123} - s_{12} s_{45}
s_{234} - s_{23} s_{56} s_{345}\,,\cr
c_8 &=& 
2 s_{61} (s_{234} s_{345} - s_{61} s_{34})\,,\cr
c_9 &=& 
s_{23} s_{34} s_{234}\,,\cr
c_{10} &=& 
s_{23} (2 s_{61} s_{34} - s_{234} s_{345})\,,\cr
c_{11} &=& 
s_{12} s_{23} s_{234}\,,\cr
c_{12} &=& 
s_{345} (s_{234} s_{345} - s_{61} s_{34})\,,\cr
c_{13} &=& 
- s_{345}^2 s_{56}\,,\cr
c_{14} &=& 
-2 s_{345} (s_{123} s_{234} s_{345} - s_{61} s_{34} s_{123} - s_{12}
  s_{45} s_{234} - s_{23} s_{56} s_{345})\,,\cr
c_{15} &=& 
2 s_{61} ( s_{123} s_{234} s_{345} - s_{61} s_{34} s_{123} - s_{12}
   s_{45} s_{234} - s_{23} s_{56} s_{345})\,.
\label{IntegralCoefficients}
\end{eqnarray}

\Eqn{TwoLoopAssembly} omits all odd terms proportional to the 
Levi-Civita tensor.  Because neither the strong-coupling string 
theory calculations nor the Wilson loop calculations are sensitive
to such terms, we may drop them without affecting comparisons to either.

Finally we should note that while the set of integrals in
\figs{ContributingIntegralsFigure}{NonContributingIntegralsFigure} is
convenient for calculation,
non-trivial relations between them and other integrals may allow for other
equivalent representations of the amplitude~(\ref{TwoLoopAssembly}) in
which the form of the coefficients~(\ref{IntegralCoefficients}) is
substantially altered.

\subsection{Dual Conformal Structure of the Integrand}

As mentioned above, we expect that planar amplitudes should manifest
dual conformal symmetry.  An even stronger statement, that
an amplitude can be expressed as a linear combination of
integrals, each exhibiting manifest pseudo-conformal invariance,
has been observed to hold for the four-particle amplitude
through five loops, and for the even part of the five-particle
amplitude at two loops.

\begin{figure}
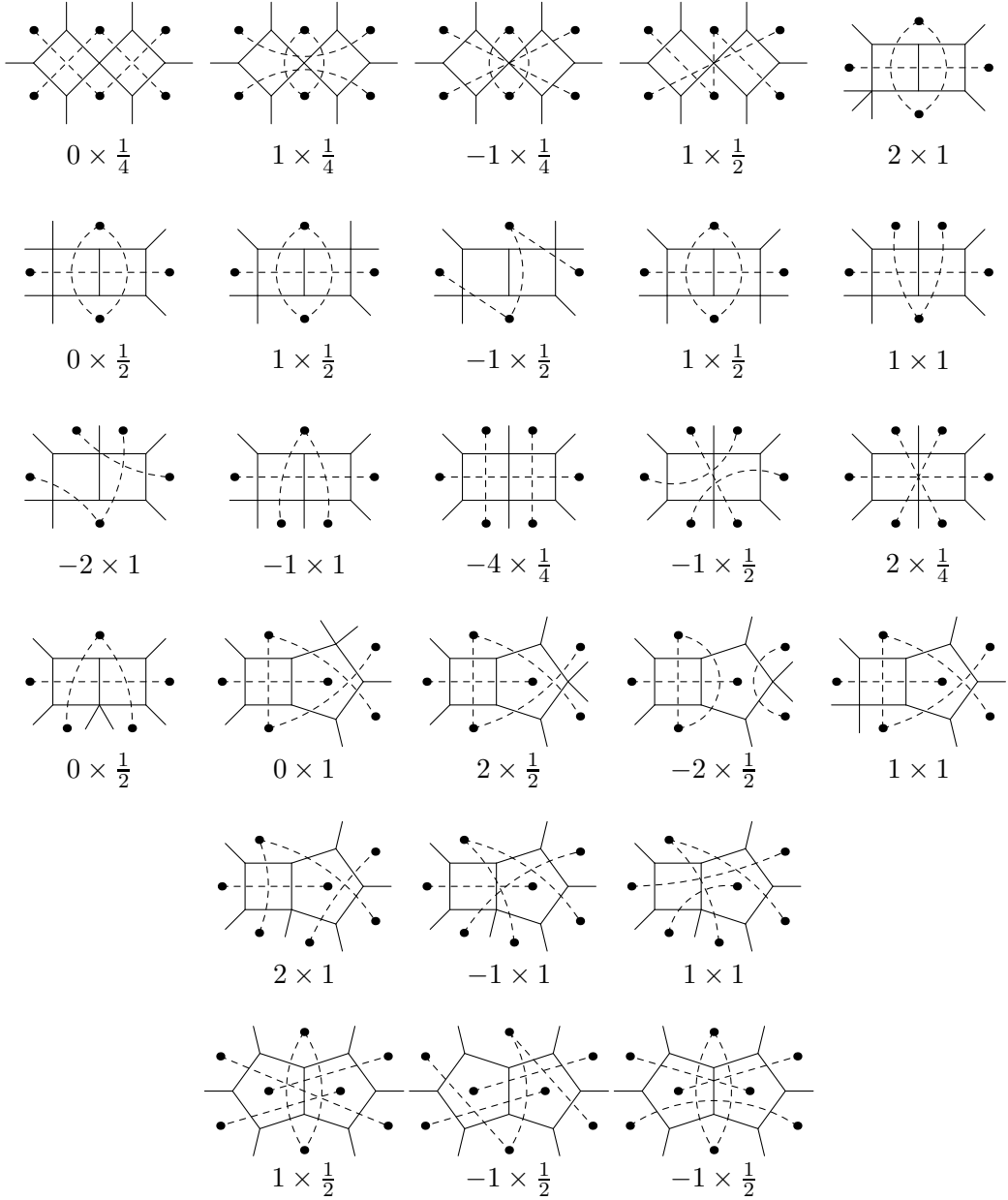

\begin{feynartspicture}(400,480)(5,6)
\FADiagram{}
\FAProp(10.,13.)(13.5355,9.46447)(0.,){/Straight}{0}
\FAProp(10.,13.)(6.46447,16.5355)(0.,){/Straight}{0}
\FAProp(6.46447,16.5355)(2.92893,13.)(0.,){/Straight}{0}
\FAProp(2.92893,13.)(6.46447,9.46447)(0.,){/Straight}{0}
\FAProp(6.46447,9.46447)(10.,13.)(0.,){/Straight}{0}
\FAProp(10.,13.)(13.5355,16.5355)(0.,){/Straight}{0}
\FAProp(13.5355,16.5355)(17.0711,13.)(0.,){/Straight}{0}
\FAProp(17.011,13.)(13.5355,9.46447)(0.,){/Straight}{0}
\FAProp(6.46447,16.5355)(6.46447,19.5355)(0.,){/Straight}{0}
\FAProp(13.5355,16.5355)(13.5355,19.5355)(0.,){/Straight}{0}
\FAProp(17.011,13.)(20.011,13.)(0.,){/Straight}{0}
\FAProp(13.5355,9.46447)(13.5355,6.46447)(0.,){/Straight}{0}
\FAProp(6.46447,9.46447)(6.46447,6.46447)(0.,){/Straight}{0}
\FAProp(2.92893,13.)(-0.07107,13.)(0.,){/Straight}{0}
\FAProp(10.,16.5355)(2.9289,9.4645)(0.,){/ScalarDash}{0}
\FAProp(10.,16.5355)(17.0711,9.4645)(0.,){/ScalarDash}{0}
\FAProp(10.,9.46447)(2.9289,16.5355)(0.,){/ScalarDash}{0}
\FAProp(10.,9.46447)(17.0711,16.5355)(0.,){/ScalarDash}{0}
\FAVert(10.,16.5355){0}
\FAVert(2.9289,9.4645){0}
\FAVert(17.0711,9.4645){0}
\FAVert(10.,9.46447){0}
\FAVert(17.0711,16.5355){0}
\FAVert(2.9289,16.5355){0}
\FALabel(10.,3.)[]{$0 \times {\textstyle{\frac{1}{4}}}$}
\FADiagram{}
\FAProp(10.,13.)(13.5355,9.46447)(0.,){/Straight}{0}
\FAProp(10.,13.)(6.46447,16.5355)(0.,){/Straight}{0}
\FAProp(6.46447,16.5355)(2.92893,13.)(0.,){/Straight}{0}
\FAProp(2.92893,13.)(6.46447,9.46447)(0.,){/Straight}{0}
\FAProp(6.46447,9.46447)(10.,13.)(0.,){/Straight}{0}
\FAProp(10.,13.)(13.5355,16.5355)(0.,){/Straight}{0}
\FAProp(13.5355,16.5355)(17.0711,13.)(0.,){/Straight}{0}
\FAProp(17.011,13.)(13.5355,9.46447)(0.,){/Straight}{0}
\FAProp(6.46447,16.5355)(6.46447,19.5355)(0.,){/Straight}{0}
\FAProp(13.5355,16.5355)(13.5355,19.5355)(0.,){/Straight}{0}
\FAProp(17.011,13.)(20.011,13.)(0.,){/Straight}{0}
\FAProp(13.5355,9.46447)(13.5355,6.46447)(0.,){/Straight}{0}
\FAProp(6.46447,9.46447)(6.46447,6.46447)(0.,){/Straight}{0}
\FAProp(2.92893,13.)(-0.07107,13.)(0.,){/Straight}{0}
\FAProp(2.9289,9.4645)(17.0711,9.4645)(-0.4,){/ScalarDash}{0}
\FAProp(2.9289,16.5355)(17.0711,16.5355)(0.4,){/ScalarDash}{0}
\FAProp(10.,16.5355)(10.,9.46447)(0.6,){/ScalarDash}{0}
\FAProp(10.,16.5355)(10.,9.46447)(-0.6,){/ScalarDash}{0}
\FAVert(10.,16.5355){0}
\FAVert(2.9289,9.4645){0}
\FAVert(17.0711,9.4645){0}
\FAVert(10.,9.46447){0}
\FAVert(17.0711,16.5355){0}
\FAVert(2.9289,16.5355){0}
\FALabel(10.,3.)[]{$1 \times {\textstyle{\frac{1}{4}}}$}
\FADiagram{}
\FAProp(10.,13.)(13.5355,9.46447)(0.,){/Straight}{0}
\FAProp(10.,13.)(6.46447,16.5355)(0.,){/Straight}{0}
\FAProp(6.46447,16.5355)(2.92893,13.)(0.,){/Straight}{0}
\FAProp(2.92893,13.)(6.46447,9.46447)(0.,){/Straight}{0}
\FAProp(6.46447,9.46447)(10.,13.)(0.,){/Straight}{0}
\FAProp(10.,13.)(13.5355,16.5355)(0.,){/Straight}{0}
\FAProp(13.5355,16.5355)(17.0711,13.)(0.,){/Straight}{0}
\FAProp(17.011,13.)(13.5355,9.46447)(0.,){/Straight}{0}
\FAProp(6.46447,16.5355)(6.46447,19.5355)(0.,){/Straight}{0}
\FAProp(13.5355,16.5355)(13.5355,19.5355)(0.,){/Straight}{0}
\FAProp(17.011,13.)(20.011,13.)(0.,){/Straight}{0}
\FAProp(13.5355,9.46447)(13.5355,6.46447)(0.,){/Straight}{0}
\FAProp(6.46447,9.46447)(6.46447,6.46447)(0.,){/Straight}{0}
\FAProp(2.92893,13.)(-0.07107,13.)(0.,){/Straight}{0}
\FAProp(10.,16.5355)(10.,9.46447)(0.6,){/ScalarDash}{0}
\FAProp(10.,16.5355)(10.,9.46447)(-0.6,){/ScalarDash}{0}
\FAProp(2.9289,16.5355)(17.0711,9.4645)(0.,){/ScalarDash}{0}
\FAProp(2.9289,9.4645)(17.0711,16.5355)(0.,){/ScalarDash}{0}
\FAVert(10.,16.5355){0}
\FAVert(2.9289,9.4645){0}
\FAVert(17.0711,9.4645){0}
\FAVert(10.,9.46447){0}
\FAVert(17.0711,16.5355){0}
\FAVert(2.9289,16.5355){0}
\FALabel(10.,3.)[]{$-1 \times {\textstyle{\frac{1}{4}}}$}
\FADiagram{}
\FAProp(10.,13.)(13.5355,9.46447)(0.,){/Straight}{0}
\FAProp(10.,13.)(6.46447,16.5355)(0.,){/Straight}{0}
\FAProp(6.46447,16.5355)(2.92893,13.)(0.,){/Straight}{0}
\FAProp(2.92893,13.)(6.46447,9.46447)(0.,){/Straight}{0}
\FAProp(6.46447,9.46447)(10.,13.)(0.,){/Straight}{0}
\FAProp(10.,13.)(13.5355,16.5355)(0.,){/Straight}{0}
\FAProp(13.5355,16.5355)(17.0711,13.)(0.,){/Straight}{0}
\FAProp(17.011,13.)(13.5355,9.46447)(0.,){/Straight}{0}
\FAProp(6.46447,16.5355)(6.46447,19.5355)(0.,){/Straight}{0}
\FAProp(13.5355,16.5355)(13.5355,19.5355)(0.,){/Straight}{0}
\FAProp(17.011,13.)(20.011,13.)(0.,){/Straight}{0}
\FAProp(13.5355,9.46447)(13.5355,6.46447)(0.,){/Straight}{0}
\FAProp(6.46447,9.46447)(6.46447,6.46447)(0.,){/Straight}{0}
\FAProp(2.92893,13.)(-0.07107,13.)(0.,){/Straight}{0}
\FAProp(10.,16.5355)(10.,9.46447)(0.,){/ScalarDash}{0}
\FAProp(10.,9.46447)(2.9289,16.5355)(0.,){/ScalarDash}{0}
\FAProp(10.,16.5355)(17.0711,9.4645)(0.,){/ScalarDash}{0}
\FAProp(2.9289,9.4645)(17.0711,16.5355)(0.,){/ScalarDash}{0}
\FAVert(10.,16.5355){0}
\FAVert(2.9289,9.4645){0}
\FAVert(17.0711,9.4645){0}
\FAVert(10.,9.46447){0}
\FAVert(17.0711,16.5355){0}
\FAVert(2.9289,16.5355){0}
\FALabel(10.,3.)[]{$1 \times {\textstyle{\frac{1}{2}}}$}
\FADiagram{}
\FAProp(5.,15.)(15.,15.)(0.,){/Straight}{0}
\FAProp(15.,10.)(5.,10.)(0.,){/Straight}{0}
\FAProp(5.,10.)(5.,15.)(0.,){/Straight}{0}
\FAProp(10.,10.)(10.,15.)(0.,){/Straight}{0}
\FAProp(15.,10.)(15.,15.)(0.,){/Straight}{0}
\FAProp(5.,15.)(2.87868,17.1213)(0.,){/Straight}{0}
\FAProp(15.,15.)(17.1213,17.1213)(0.,){/Straight}{0}
\FAProp(15.,10.)(17.1213,7.87868)(0.,){/Straight}{0}
\FAProp(5.,10.)(5.,7.)(0.,){/Straight}{0}
\FAProp(5.,10.)(2.,10.)(0.,){/Straight}{0}
\FAProp(5.,10.)(2.87868,7.87868)(0.,){/Straight}{0}
\FAProp(10.,17.5)(10.,7.5)(0.6,){/ScalarDash}{0}
\FAProp(10.,17.5)(10.,7.5)(-0.6,){/ScalarDash}{0}
\FAProp(2.5,12.5)(17.5,12.5)(0.,){/ScalarDash}{0}
\FAVert(10.,17.5){0}
\FAVert(10.,7.5){0}
\FAVert(2.5,12.5){0}
\FAVert(17.5,12.5){0}
\FALabel(10.,3.)[]{$2 \times 1$}
\FADiagram{}
\FAProp(5.,15.)(15.,15.)(0.,){/Straight}{0}
\FAProp(15.,10.)(5.,10.)(0.,){/Straight}{0}
\FAProp(5.,10.)(5.,15.)(0.,){/Straight}{0}
\FAProp(10.,10.)(10.,15.)(0.,){/Straight}{0}
\FAProp(15.,10.)(15.,15.)(0.,){/Straight}{0}
\FAProp(5.,15.)(2.,15.)(0.,){/Straight}{0}
\FAProp(5.,15.)(5.,18.)(0.,){/Straight}{0}
\FAProp(15.,15.)(17.1213,17.1213)(0.,){/Straight}{0}
\FAProp(15.,10.)(17.1213,7.87868)(0.,){/Straight}{0}
\FAProp(5.,10.)(5.,7.)(0.,){/Straight}{0}
\FAProp(5.,10.)(2.,10.)(0.,){/Straight}{0}
\FAProp(10.,17.5)(10.,7.5)(0.6,){/ScalarDash}{0}
\FAProp(10.,17.5)(10.,7.5)(-0.6,){/ScalarDash}{0}
\FAProp(2.5,12.5)(17.5,12.5)(0.,){/ScalarDash}{0}
\FAVert(10.,17.5){0}
\FAVert(10.,7.5){0}
\FAVert(2.5,12.5){0}
\FAVert(17.5,12.5){0}
\FALabel(10.,3.)[]{$0 \times {\textstyle{\frac{1}{2}}}$}
\FADiagram{}
\FAProp(5.,15.)(15.,15.)(0.,){/Straight}{0}
\FAProp(15.,10.)(5.,10.)(0.,){/Straight}{0}
\FAProp(5.,10.)(5.,15.)(0.,){/Straight}{0}
\FAProp(10.,10.)(10.,15.)(0.,){/Straight}{0}
\FAProp(15.,10.)(15.,15.)(0.,){/Straight}{0}
\FAProp(5.,15.)(2.87868,17.1213)(0.,){/Straight}{0}
\FAProp(15.,15.)(18.,15.)(0.,){/Straight}{0}
\FAProp(15.,15.)(15.,18.)(0.,){/Straight}{0}
\FAProp(15.,10.)(17.1213,7.87868)(0.,){/Straight}{0}
\FAProp(5.,10.)(5.,7.)(0.,){/Straight}{0}
\FAProp(5.,10.)(2.,10.)(0.,){/Straight}{0}
\FAProp(10.,17.5)(10.,7.5)(0.6,){/ScalarDash}{0}
\FAProp(10.,17.5)(10.,7.5)(-0.6,){/ScalarDash}{0}
\FAProp(2.5,12.5)(17.5,12.5)(0.,){/ScalarDash}{0}
\FAVert(10.,17.5){0}
\FAVert(10.,7.5){0}
\FAVert(2.5,12.5){0}
\FAVert(17.5,12.5){0}
\FALabel(10.,3.)[]{$1 \times {\textstyle{\frac{1}{2}}}$}
\FADiagram{}
\FAProp(5.,15.)(15.,15.)(0.,){/Straight}{0}
\FAProp(15.,10.)(5.,10.)(0.,){/Straight}{0}
\FAProp(5.,10.)(5.,15.)(0.,){/Straight}{0}
\FAProp(10.,10.)(10.,15.)(0.,){/Straight}{0}
\FAProp(15.,10.)(15.,15.)(0.,){/Straight}{0}
\FAProp(5.,15.)(2.87868,17.1213)(0.,){/Straight}{0}
\FAProp(15.,15.)(18.,15.)(0.,){/Straight}{0}
\FAProp(15.,15.)(15.,18.)(0.,){/Straight}{0}
\FAProp(15.,10.)(17.1213,7.87868)(0.,){/Straight}{0}
\FAProp(5.,10.)(5.,7.)(0.,){/Straight}{0}
\FAProp(5.,10.)(2.,10.)(0.,){/Straight}{0}
\FAProp(10.,17.5)(10.,7.5)(-0.3,){/ScalarDash}{0}
\FAProp(2.5,12.5)(10.,7.5)(0.,){/ScalarDash}{0}
\FAProp(10.,17.5)(17.5,12.5)(0.,){/ScalarDash}{0}
\FAVert(10.,17.5){0}
\FAVert(10.,7.5){0}
\FAVert(2.5,12.5){0}
\FAVert(17.5,12.5){0}
\FALabel(10.,3.)[]{$-1 \times {\textstyle{\frac{1}{2}}}$}
\FADiagram{}
\FAProp(5.,15.)(15.,15.)(0.,){/Straight}{0}
\FAProp(15.,10.)(5.,10.)(0.,){/Straight}{0}
\FAProp(5.,10.)(5.,15.)(0.,){/Straight}{0}
\FAProp(10.,10.)(10.,15.)(0.,){/Straight}{0}
\FAProp(15.,10.)(15.,15.)(0.,){/Straight}{0}
\FAProp(5.,15.)(2.87868,17.1213)(0.,){/Straight}{0}
\FAProp(15.,15.)(17.1213,17.1213)(0.,){/Straight}{0}
\FAProp(15.,10.)(18.,10.)(0.,){/Straight}{0}
\FAProp(15.,10.)(15.,7.)(0.,){/Straight}{0}
\FAProp(5.,10.)(5.,7.)(0.,){/Straight}{0}
\FAProp(5.,10.)(2.,10.)(0.,){/Straight}{0}
\FAProp(10.,17.5)(10.,7.5)(0.6,){/ScalarDash}{0}
\FAProp(10.,17.5)(10.,7.5)(-0.6,){/ScalarDash}{0}
\FAProp(2.5,12.5)(17.5,12.5)(0.,){/ScalarDash}{0}
\FAVert(10.,17.5){0}
\FAVert(10.,7.5){0}
\FAVert(2.5,12.5){0}
\FAVert(17.5,12.5){0}
\FALabel(10.,3.)[]{$1 \times {\textstyle{\frac{1}{2}}}$}
\FADiagram{}
\FAProp(5.,15.)(15.,15.)(0.,){/Straight}{0}
\FAProp(15.,10.)(5.,10.)(0.,){/Straight}{0}
\FAProp(5.,10.)(5.,15.)(0.,){/Straight}{0}
\FAProp(10.,10.)(10.,15.)(0.,){/Straight}{0}
\FAProp(15.,10.)(15.,15.)(0.,){/Straight}{0}
\FAProp(5.,15.)(2.87868,17.1213)(0.,){/Straight}{0}
\FAProp(15.,15.)(17.1213,17.1213)(0.,){/Straight}{0}
\FAProp(15.,10.)(17.1213,7.87868)(0.,){/Straight}{0}
\FAProp(5.,10.)(5.,7.)(0.,){/Straight}{0}
\FAProp(5.,10.)(2.,10.)(0.,){/Straight}{0}
\FAProp(10.,15.)(10.,18.)(0.,){/Straight}{0}
\FAProp(2.5,12.5)(17.5,12.5)(0.,){/ScalarDash}{0}
\FAProp(10.,7.5)(12.5,17.5)(0.2,){/ScalarDash}{0}
\FAProp(10.,7.5)(7.5,17.5)(-0.2,){/ScalarDash}{0}
\FAVert(7.5,17.5){0}
\FAVert(12.5,17.5){0}
\FAVert(10.,7.5){0}
\FAVert(2.5,12.5){0}
\FAVert(17.5,12.5){0}
\FALabel(10.,3.)[]{$1 \times 1$}
\FADiagram{}
\FAProp(5.,15.)(15.,15.)(0.,){/Straight}{0}
\FAProp(15.,10.)(5.,10.)(0.,){/Straight}{0}
\FAProp(5.,10.)(5.,15.)(0.,){/Straight}{0}
\FAProp(10.,10.)(10.,15.)(0.,){/Straight}{0}
\FAProp(15.,10.)(15.,15.)(0.,){/Straight}{0}
\FAProp(5.,15.)(2.87868,17.1213)(0.,){/Straight}{0}
\FAProp(15.,15.)(17.1213,17.1213)(0.,){/Straight}{0}
\FAProp(15.,10.)(17.1213,7.87868)(0.,){/Straight}{0}
\FAProp(5.,10.)(5.,7.)(0.,){/Straight}{0}
\FAProp(5.,10.)(2.,10.)(0.,){/Straight}{0}
\FAProp(10.,15.)(10.,18.)(0.,){/Straight}{0}
\FAProp(2.5,12.5)(10.,7.5)(-0.2,){/ScalarDash}{0}
\FAProp(10.,7.5)(12.5,17.5)(0.2,){/ScalarDash}{0}
\FAProp(7.5,17.5)(17.5,12.5)(0.2,){/ScalarDash}{0}
\FAVert(7.5,17.5){0}
\FAVert(12.5,17.5){0}
\FAVert(10.,7.5){0}
\FAVert(2.5,12.5){0}
\FAVert(17.5,12.5){0}
\FALabel(10.,3.)[]{$-2 \times 1$}
\FADiagram{}
\FAProp(5.,15.)(15.,15.)(0.,){/Straight}{0}
\FAProp(15.,10.)(5.,10.)(0.,){/Straight}{0}
\FAProp(5.,10.)(5.,15.)(0.,){/Straight}{0}
\FAProp(10.,10.)(10.,15.)(0.,){/Straight}{0}
\FAProp(15.,10.)(15.,15.)(0.,){/Straight}{0}
\FAProp(5.,15.)(2.87868,17.1213)(0.,){/Straight}{0}
\FAProp(15.,15.)(17.1213,17.1213)(0.,){/Straight}{0}
\FAProp(15.,10.)(17.1213,7.87868)(0.,){/Straight}{0}
\FAProp(5.,10.)(5.,7.)(0.,){/Straight}{0}
\FAProp(5.,10.)(2.,10.)(0.,){/Straight}{0}
\FAProp(10.,10.)(10.,7.)(0.,){/Straight}{0}
\FAProp(2.5,12.5)(17.5,12.5)(0.,){/ScalarDash}{0}
\FAProp(10.,17.5)(12.5,7.5)(-0.2,){/ScalarDash}{0}
\FAProp(10.,17.5)(7.5,7.5)(0.2,){/ScalarDash}{0}
\FAVert(7.5,7.5){0}
\FAVert(12.5,7.5){0}
\FAVert(10.,17.5){0}
\FAVert(2.5,12.5){0}
\FAVert(17.5,12.5){0}
\FALabel(10.,3.)[]{$-1 \times 1$}
\FADiagram{}
\FAProp(5.,15.)(15.,15.)(0.,){/Straight}{0}
\FAProp(15.,10.)(5.,10.)(0.,){/Straight}{0}
\FAProp(5.,10.)(5.,15.)(0.,){/Straight}{0}
\FAProp(10.,10.)(10.,15.)(0.,){/Straight}{0}
\FAProp(15.,10.)(15.,15.)(0.,){/Straight}{0}
\FAProp(5.,15.)(2.87868,17.1213)(0.,){/Straight}{0}
\FAProp(15.,15.)(17.1213,17.1213)(0.,){/Straight}{0}
\FAProp(15.,10.)(17.1213,7.87868)(0.,){/Straight}{0}
\FAProp(5.,10.)(2.87868,7.87868)(0.,){/Straight}{0}
\FAProp(10.,10.)(10.,7.)(0.,){/Straight}{0}
\FAProp(10.,15.)(10.,18.)(0.,){/Straight}{0}
\FAProp(7.5,7.5)(7.5,17.5)(0.,){/ScalarDash}{0}
\FAProp(12.5,7.5)(12.5,17.5)(0.,){/ScalarDash}{0}
\FAProp(2.5,12.5)(17.5,12.5)(0.,){/ScalarDash}{0}
\FAVert(7.5,7.5){0}
\FAVert(12.5,7.5){0}
\FAVert(7.5,17.5){0}
\FAVert(12.5,17.5){0}
\FAVert(2.5,12.5){0}
\FAVert(17.5,12.5){0}
\FALabel(10.,3.)[]{$-4 \times {\textstyle{\frac{1}{4}}}$}
\FADiagram{}
\FAProp(5.,15.)(15.,15.)(0.,){/Straight}{0}
\FAProp(15.,10.)(5.,10.)(0.,){/Straight}{0}
\FAProp(5.,10.)(5.,15.)(0.,){/Straight}{0}
\FAProp(10.,10.)(10.,15.)(0.,){/Straight}{0}
\FAProp(15.,10.)(15.,15.)(0.,){/Straight}{0}
\FAProp(5.,15.)(2.87868,17.1213)(0.,){/Straight}{0}
\FAProp(15.,15.)(17.1213,17.1213)(0.,){/Straight}{0}
\FAProp(15.,10.)(17.1213,7.87868)(0.,){/Straight}{0}
\FAProp(5.,10.)(2.87868,7.87868)(0.,){/Straight}{0}
\FAProp(10.,10.)(10.,7.)(0.,){/Straight}{0}
\FAProp(10.,15.)(10.,18.)(0.,){/Straight}{0}
\FAProp(2.5,12.5)(12.5,17.5)(0.5,){/ScalarDash}{0}
\FAProp(7.5,7.5)(17.5,12.5)(-0.5,){/ScalarDash}{0}
\FAProp(7.5,17.5)(12.5,7.5)(0.,){/ScalarDash}{0}
\FAVert(7.5,7.5){0}
\FAVert(12.5,7.5){0}
\FAVert(7.5,17.5){0}
\FAVert(12.5,17.5){0}
\FAVert(2.5,12.5){0}
\FAVert(17.5,12.5){0}
\FALabel(10.,3.)[]{$-1 \times {\textstyle{\frac{1}{2}}}$}
\FADiagram{}
\FAProp(5.,15.)(15.,15.)(0.,){/Straight}{0}
\FAProp(15.,10.)(5.,10.)(0.,){/Straight}{0}
\FAProp(5.,10.)(5.,15.)(0.,){/Straight}{0}
\FAProp(10.,10.)(10.,15.)(0.,){/Straight}{0}
\FAProp(15.,10.)(15.,15.)(0.,){/Straight}{0}
\FAProp(5.,15.)(2.87868,17.1213)(0.,){/Straight}{0}
\FAProp(15.,15.)(17.1213,17.1213)(0.,){/Straight}{0}
\FAProp(15.,10.)(17.1213,7.87868)(0.,){/Straight}{0}
\FAProp(5.,10.)(2.87868,7.87868)(0.,){/Straight}{0}
\FAProp(10.,10.)(10.,7.)(0.,){/Straight}{0}
\FAProp(10.,15.)(10.,18.)(0.,){/Straight}{0}
\FAProp(7.5,7.5)(12.5,17.5)(0.,){/ScalarDash}{0}
\FAProp(12.5,7.5)(7.5,17.5)(0.,){/ScalarDash}{0}
\FAProp(2.5,12.5)(17.5,12.5)(0.,){/ScalarDash}{0}
\FAVert(7.5,7.5){0}
\FAVert(12.5,7.5){0}
\FAVert(7.5,17.5){0}
\FAVert(12.5,17.5){0}
\FAVert(2.5,12.5){0}
\FAVert(17.5,12.5){0}
\FALabel(10.,3.)[]{$2 \times {\textstyle{\frac{1}{4}}}$}
\FADiagram{}
\FAProp(5.,15.)(15.,15.)(0.,){/Straight}{0}
\FAProp(15.,10.)(5.,10.)(0.,){/Straight}{0}
\FAProp(5.,10.)(5.,15.)(0.,){/Straight}{0}
\FAProp(10.,10.)(10.,15.)(0.,){/Straight}{0}
\FAProp(15.,10.)(15.,15.)(0.,){/Straight}{0}
\FAProp(5.,15.)(2.87868,17.1213)(0.,){/Straight}{0}
\FAProp(15.,15.)(17.1213,17.1213)(0.,){/Straight}{0}
\FAProp(15.,10.)(17.1213,7.87868)(0.,){/Straight}{0}
\FAProp(5.,10.)(2.87868,7.87868)(0.,){/Straight}{0}
\FAProp(10.,10.)(8.5,7.4019)(0.,){/Straight}{0}
\FAProp(10.,10.)(11.5,7.4019)(0.,){/Straight}{0}
\FAProp(2.5,12.5)(17.5,12.5)(0.,){/ScalarDash}{0}
\FAProp(10.,17.5)(13.5,7.5)(-0.2,){/ScalarDash}{0}
\FAProp(10.,17.5)(6.5,7.5)(0.2,){/ScalarDash}{0}
\FAVert(6.5,7.5){0}
\FAVert(13.5,7.5){0}
\FAVert(10.,17.5){0}
\FAVert(2.5,12.5){0}
\FAVert(17.5,12.5){0}
\FALabel(10.,3.)[]{$0 \times {\textstyle{\frac{1}{2}}}$}
\FADiagram{}
\FAProp(3.6529,15.)(8.6529,15.)(0.,){/Straight}{0}
\FAProp(8.6529,15.)(8.6529,10.)(0.,){/Straight}{0}
\FAProp(8.6529,10.)(3.6529,10.)(0.,){/Straight}{0}
\FAProp(3.6529,10.)(3.6529,15.)(0.,){/Straight}{0}
\FAProp(8.6529,15.)(13.4082,16.5451)(0.,){/Straight}{0}
\FAProp(8.6529,10.)(13.4082,8.45492)(0.,){/Straight}{0}
\FAProp(13.4082,8.45492)(16.3471,12.5)(0.,){/Straight}{0}
\FAProp(13.4082,16.5451)(16.3471,12.5)(0.,){/Straight}{0}
\FAProp(3.6529,15.)(1.5316,17.1213)(0.,){/Straight}{0}
\FAProp(3.6529,10.)(1.5316,7.87868)(0.,){/Straight}{0}
\FAProp(16.3471,12.5)(19.3471,12.5)(0.,){/Straight}{0}
\FAProp(13.4082,8.45492)(14.1085,5.53781)(0.,){/Straight}{0}
\FAProp(13.4082,16.5451)(15.7396,18.4331)(0.,){/Straight}{0}
\FAProp(13.4082,16.5451)(11.7743,19.0611)(0.,){/Straight}{0}
\FAProp(6.1529,7.5)(6.1529,17.5)(0.,){/ScalarDash}{0}
\FAProp(1.1529,12.5)(12.5,12.5)(0.,){/ScalarDash}{0}
\FAProp(6.1529,17.5)(17.6349,8.7693)(-0.2,){/ScalarDash}{0}
\FAProp(6.1529,7.5)(17.6349,16.2307)(0.2,){/ScalarDash}{0}
\FAVert(12.5,12.5){0}
\FAVert(6.1529,7.5){0}
\FAVert(6.1529,17.5){0}
\FAVert(1.1529,12.5){0}
\FAVert(17.6349,8.7693){0}
\FAVert(17.6349,16.2307){0}
\FALabel(10.,3.)[]{$0 \times 1$}
\FADiagram{}
\FAProp(3.6529,15.)(8.6529,15.)(0.,){/Straight}{0}
\FAProp(8.6529,15.)(8.6529,10.)(0.,){/Straight}{0}
\FAProp(8.6529,10.)(3.6529,10.)(0.,){/Straight}{0}
\FAProp(3.6529,10.)(3.6529,15.)(0.,){/Straight}{0}
\FAProp(8.6529,15.)(13.4082,16.5451)(0.,){/Straight}{0}
\FAProp(8.6529,10.)(13.4082,8.45492)(0.,){/Straight}{0}
\FAProp(13.4082,8.45492)(16.3471,12.5)(0.,){/Straight}{0}
\FAProp(13.4082,16.5451)(16.3471,12.5)(0.,){/Straight}{0}
\FAProp(3.6529,15.)(1.5316,17.1213)(0.,){/Straight}{0}
\FAProp(3.6529,10.)(1.5316,7.87868)(0.,){/Straight}{0}
\FAProp(16.3471,12.5)(18.4684,14.6213)(0.,){/Straight}{0}
\FAProp(16.3471,12.5)(18.4684,10.3787)(0.,){/Straight}{0}
\FAProp(13.4082,8.45492)(14.1085,5.53781)(0.,){/Straight}{0}
\FAProp(13.4082,16.5451)(14.1085,19.4622)(0.,){/Straight}{0}
\FAProp(1.1529,12.5)(12.5,12.5)(0.,){/ScalarDash}{0}
\FAProp(6.1529,7.5)(6.1529,17.5)(0.,){/ScalarDash}{0}
\FAProp(6.1529,17.5)(17.6349,8.7693)(-0.2,){/ScalarDash}{0}
\FAProp(6.1529,7.5)(17.6349,16.2307)(0.2,){/ScalarDash}{0}
\FAVert(12.5,12.5){0}
\FAVert(6.1529,7.5){0}
\FAVert(6.1529,17.5){0}
\FAVert(1.1529,12.5){0}
\FAVert(17.6349,8.7693){0}
\FAVert(17.6349,16.2307){0}
\FALabel(10.,3.)[]{$2 \times {\textstyle{\frac{1}{2}}}$}
\FADiagram{}
\FAProp(3.6529,15.)(8.6529,15.)(0.,){/Straight}{0}
\FAProp(8.6529,15.)(8.6529,10.)(0.,){/Straight}{0}
\FAProp(8.6529,10.)(3.6529,10.)(0.,){/Straight}{0}
\FAProp(3.6529,10.)(3.6529,15.)(0.,){/Straight}{0}
\FAProp(8.6529,15.)(13.4082,16.5451)(0.,){/Straight}{0}
\FAProp(8.6529,10.)(13.4082,8.45492)(0.,){/Straight}{0}
\FAProp(13.4082,8.45492)(16.3471,12.5)(0.,){/Straight}{0}
\FAProp(13.4082,16.5451)(16.3471,12.5)(0.,){/Straight}{0}
\FAProp(3.6529,15.)(1.5316,17.1213)(0.,){/Straight}{0}
\FAProp(3.6529,10.)(1.5316,7.87868)(0.,){/Straight}{0}
\FAProp(16.3471,12.5)(18.4684,14.6213)(0.,){/Straight}{0}
\FAProp(16.3471,12.5)(18.4684,10.3787)(0.,){/Straight}{0}
\FAProp(13.4082,8.45492)(14.1085,5.53781)(0.,){/Straight}{0}
\FAProp(13.4082,16.5451)(14.1085,19.4622)(0.,){/Straight}{0}
\FAProp(1.1529,12.5)(12.5,12.5)(0.,){/ScalarDash}{0}
\FAProp(6.1529,7.5)(6.1529,17.5)(0.,){/ScalarDash}{0}
\FAProp(6.1529,7.5)(6.1529,17.5)(0.9,){/ScalarDash}{0}
\FAProp(17.6349,8.7693)(17.6349,16.2307)(-0.9,){/ScalarDash}{0}
\FAVert(12.5,12.5){0}
\FAVert(6.1529,7.5){0}
\FAVert(6.1529,17.5){0}
\FAVert(1.1529,12.5){0}
\FAVert(17.6349,8.7693){0}
\FAVert(17.6349,16.2307){0}
\FALabel(10.,3.)[]{$-2 \times {\textstyle{\frac{1}{2}}}$}
\FADiagram{}
\FAProp(3.6529,15.)(8.6529,15.)(0.,){/Straight}{0}
\FAProp(8.6529,15.)(8.6529,10.)(0.,){/Straight}{0}
\FAProp(8.6529,10.)(3.6529,10.)(0.,){/Straight}{0}
\FAProp(3.6529,10.)(3.6529,15.)(0.,){/Straight}{0}
\FAProp(8.6529,15.)(13.4082,16.5451)(0.,){/Straight}{0}
\FAProp(8.6529,10.)(13.4082,8.45492)(0.,){/Straight}{0}
\FAProp(13.4082,8.45492)(16.3471,12.5)(0.,){/Straight}{0}
\FAProp(13.4082,16.5451)(16.3471,12.5)(0.,){/Straight}{0}
\FAProp(3.6529,15.)(1.5316,17.1213)(0.,){/Straight}{0}
\FAProp(3.6529,10.)(0.6529,10.)(0.,){/Straight}{0}
\FAProp(3.6529,10.)(3.6529,7.)(0.,){/Straight}{0}
\FAProp(16.3471,12.5)(19.3471,12.5)(0.,){/Straight}{0}
\FAProp(13.4082,8.45492)(14.1085,5.53781)(0.,){/Straight}{0}
\FAProp(13.4082,16.5451)(14.1085,19.4622)(0.,){/Straight}{0}
\FAProp(6.1529,7.5)(6.1529,17.5)(0.,){/ScalarDash}{0}
\FAProp(1.1529,12.5)(12.5,12.5)(0.,){/ScalarDash}{0}
\FAProp(6.1529,17.5)(17.6349,8.7693)(-0.2,){/ScalarDash}{0}
\FAProp(6.1529,7.5)(17.6349,16.2307)(0.2,){/ScalarDash}{0}
\FAVert(12.5,12.5){0}
\FAVert(6.1529,7.5){0}
\FAVert(6.1529,17.5){0}
\FAVert(1.1529,12.5){0}
\FAVert(17.6349,8.7693){0}
\FAVert(17.6349,16.2307){0}
\FALabel(10.,3.)[]{$1 \times 1$}
\FADiagram{}
\FADiagram{}
\FAProp(3.6529,15.)(8.6529,15.)(0.,){/Straight}{0}
\FAProp(8.6529,15.)(8.6529,10.)(0.,){/Straight}{0}
\FAProp(8.6529,10.)(3.6529,10.)(0.,){/Straight}{0}
\FAProp(3.6529,10.)(3.6529,15.)(0.,){/Straight}{0}
\FAProp(8.6529,15.)(13.4082,16.5451)(0.,){/Straight}{0}
\FAProp(8.6529,10.)(13.4082,8.45492)(0.,){/Straight}{0}
\FAProp(13.4082,8.45492)(16.3471,12.5)(0.,){/Straight}{0}
\FAProp(13.4082,16.5451)(16.3471,12.5)(0.,){/Straight}{0}
\FAProp(3.6529,15.)(1.5316,17.1213)(0.,){/Straight}{0}
\FAProp(3.6529,10.)(1.5316,7.87868)(0.,){/Straight}{0}
\FAProp(16.3471,12.5)(19.3471,12.5)(0.,){/Straight}{0}
\FAProp(13.4082,8.45492)(14.1085,5.53781)(0.,){/Straight}{0}
\FAProp(13.4082,16.5451)(14.1085,19.4622)(0.,){/Straight}{0}
\FAProp(8.6529,10.)(7.9526,7.0829)(0.,){/Straight}{0}
\FAProp(1.1529,12.5)(12.5,12.5)(0.,){/ScalarDash}{0}
\FAProp(5.1529,17.5)(17.6349,8.7693)(-0.2,){/ScalarDash}{0}
\FAProp(5.1529,7.5)(5.1529,17.5)(0.2,){/ScalarDash}{0}
\FAProp(10.5386,6.4635)(17.6349,16.2307)(-0.1,){/ScalarDash}{0}
\FAVert(12.5,12.5){0}
\FAVert(5.1529,7.5){0}
\FAVert(5.1529,17.5){0}
\FAVert(1.1529,12.5){0}
\FAVert(17.6349,8.7693){0}
\FAVert(17.6349,16.2307){0}
\FAVert(10.5386,6.4635){0}
\FALabel(10.,3.)[]{$2 \times 1$}
\FADiagram{}
\FAProp(3.6529,15.)(8.6529,15.)(0.,){/Straight}{0}
\FAProp(8.6529,15.)(8.6529,10.)(0.,){/Straight}{0}
\FAProp(8.6529,10.)(3.6529,10.)(0.,){/Straight}{0}
\FAProp(3.6529,10.)(3.6529,15.)(0.,){/Straight}{0}
\FAProp(8.6529,15.)(13.4082,16.5451)(0.,){/Straight}{0}
\FAProp(8.6529,10.)(13.4082,8.45492)(0.,){/Straight}{0}
\FAProp(13.4082,8.45492)(16.3471,12.5)(0.,){/Straight}{0}
\FAProp(13.4082,16.5451)(16.3471,12.5)(0.,){/Straight}{0}
\FAProp(3.6529,15.)(1.5316,17.1213)(0.,){/Straight}{0}
\FAProp(3.6529,10.)(1.5316,7.87868)(0.,){/Straight}{0}
\FAProp(16.3471,12.5)(19.3471,12.5)(0.,){/Straight}{0}
\FAProp(13.4082,8.45492)(14.1085,5.53781)(0.,){/Straight}{0}
\FAProp(13.4082,16.5451)(14.1085,19.4622)(0.,){/Straight}{0}
\FAProp(8.6529,10.)(7.9526,7.0829)(0.,){/Straight}{0}
\FAProp(1.1529,12.5)(12.5,12.5)(0.,){/ScalarDash}{0}
\FAProp(5.1529,17.5)(10.5386,6.4635)(-0.2,){/ScalarDash}{0}
\FAProp(5.1529,17.5)(17.6349,8.7693)(-0.2,){/ScalarDash}{0}
\FAProp(5.1529,7.5)(17.6349,16.2307)(-0.2,){/ScalarDash}{0}
\FAVert(12.5,12.5){0}
\FAVert(5.1529,7.5){0}
\FAVert(5.1529,17.5){0}
\FAVert(1.1529,12.5){0}
\FAVert(17.6349,8.7693){0}
\FAVert(17.6349,16.2307){0}
\FAVert(10.5386,6.4635){0}
\FALabel(10.,3.)[]{$-1 \times 1$}
\FADiagram{}
\FAProp(3.6529,15.)(8.6529,15.)(0.,){/Straight}{0}
\FAProp(8.6529,15.)(8.6529,10.)(0.,){/Straight}{0}
\FAProp(8.6529,10.)(3.6529,10.)(0.,){/Straight}{0}
\FAProp(3.6529,10.)(3.6529,15.)(0.,){/Straight}{0}
\FAProp(8.6529,15.)(13.4082,16.5451)(0.,){/Straight}{0}
\FAProp(8.6529,10.)(13.4082,8.45492)(0.,){/Straight}{0}
\FAProp(13.4082,8.45492)(16.3471,12.5)(0.,){/Straight}{0}
\FAProp(13.4082,16.5451)(16.3471,12.5)(0.,){/Straight}{0}
\FAProp(3.6529,15.)(1.5316,17.1213)(0.,){/Straight}{0}
\FAProp(3.6529,10.)(1.5316,7.87868)(0.,){/Straight}{0}
\FAProp(16.3471,12.5)(19.3471,12.5)(0.,){/Straight}{0}
\FAProp(13.4082,8.45492)(14.1085,5.53781)(0.,){/Straight}{0}
\FAProp(13.4082,16.5451)(14.1085,19.4622)(0.,){/Straight}{0}
\FAProp(8.6529,10.)(7.9526,7.0829)(0.,){/Straight}{0}
\FAProp(12.5,12.5)(5.1529,7.5)(0.4,){/ScalarDash}{0}
\FAProp(5.1529,17.5)(10.5386,6.4635)(-0.2,){/ScalarDash}{0}
\FAProp(5.1529,17.5)(17.6349,8.7693)(-0.2,){/ScalarDash}{0}
\FAProp(1.1529,12.5)(17.6349,16.2307)(0.1,){/ScalarDash}{0}
\FAVert(12.5,12.5){0}
\FAVert(5.1529,7.5){0}
\FAVert(5.1529,17.5){0}
\FAVert(1.1529,12.5){0}
\FAVert(17.6349,8.7693){0}
\FAVert(17.6349,16.2307){0}
\FAVert(10.5386,6.4635){0}
\FALabel(10.,3.)[]{$1 \times 1$}
\FADiagram{}
\FADiagram{}
\FADiagram{}
\FAProp(10.,10.)(10.,15.)(0.,){/Straight}{0}
\FAProp(10.,15.)(14.7553,16.5451)(0.,){/Straight}{0}
\FAProp(10.,10.)(14.7553,8.45492)(0.,){/Straight}{0}
\FAProp(14.7553,8.45492)(17.6942,12.5)(0.,){/Straight}{0}
\FAProp(14.7553,16.5451)(17.6942,12.5)(0.,){/Straight}{0}
\FAProp(17.6942,12.5)(20.6942,12.5)(0.,){/Straight}{0}
\FAProp(14.7553,8.45492)(15.4556,5.53781)(0.,){/Straight}{0}
\FAProp(14.7553,16.5451)(15.4556,19.4622)(0.,){/Straight}{0}
\FAProp(10.,15.)(5.2447,16.5451)(0.,){/Straight}{0}
\FAProp(10.,10.)(5.2447,8.45492)(0.,){/Straight}{0}
\FAProp(5.2447,8.45492)(2.3058,12.5)(0.,){/Straight}{0}
\FAProp(5.2447,16.5451)(2.3058,12.5)(0.,){/Straight}{0}
\FAProp(5.2447,8.45492)(4.5444,5.53781)(0.,){/Straight}{0}
\FAProp(5.2447,16.5451)(4.5444,19.4622)(0.,){/Straight}{0}
\FAProp(2.3058,12.5)(-0.6942,12.5)(0.,){/Straight}{0}
\FAProp(13.8471,12.5)(0.9910,8.7693)(0.,){/ScalarDash}{0}
\FAProp(6.1529,12.5)(18.9820,16.2307)(0.,){/ScalarDash}{0}
\FAProp(10.,18.8471)(10.,6.1529)(0.3,){/ScalarDash}{0}
\FAProp(10.,18.8471)(10.,6.1529)(-0.3,){/ScalarDash}{0}
\FAProp(0.9910,16.2307)(18.9820,8.7693)(0.,){/ScalarDash}{0}
\FAVert(13.8471,12.5){0}
\FAVert(6.1529,12.5){0}
\FAVert(10.,18.8471){0}
\FAVert(10.,6.1529){0}
\FAVert(18.9820,16.2307){0}
\FAVert(18.9820,8.7693){0}
\FAVert(0.9910,8.7693){0}
\FAVert(0.9910,16.2307){0}
\FALabel(10.,3.)[]{$1 \times {\textstyle{\frac{1}{2}}}$}
\FADiagram{}
\FAProp(10.,10.)(10.,15.)(0.,){/Straight}{0}
\FAProp(10.,15.)(14.7553,16.5451)(0.,){/Straight}{0}
\FAProp(10.,10.)(14.7553,8.45492)(0.,){/Straight}{0}
\FAProp(14.7553,8.45492)(17.6942,12.5)(0.,){/Straight}{0}
\FAProp(14.7553,16.5451)(17.6942,12.5)(0.,){/Straight}{0}
\FAProp(17.6942,12.5)(20.6942,12.5)(0.,){/Straight}{0}
\FAProp(14.7553,8.45492)(15.4556,5.53781)(0.,){/Straight}{0}
\FAProp(14.7553,16.5451)(15.4556,19.4622)(0.,){/Straight}{0}
\FAProp(10.,15.)(5.2447,16.5451)(0.,){/Straight}{0}
\FAProp(10.,10.)(5.2447,8.45492)(0.,){/Straight}{0}
\FAProp(5.2447,8.45492)(2.3058,12.5)(0.,){/Straight}{0}
\FAProp(5.2447,16.5451)(2.3058,12.5)(0.,){/Straight}{0}
\FAProp(5.2447,8.45492)(4.5444,5.53781)(0.,){/Straight}{0}
\FAProp(5.2447,16.5451)(4.5444,19.4622)(0.,){/Straight}{0}
\FAProp(2.3058,12.5)(-0.6942,12.5)(0.,){/Straight}{0}
\FAProp(13.8471,12.5)(0.9910,8.7693)(0.,){/ScalarDash}{0}
\FAProp(6.1529,12.5)(18.9820,16.2307)(0.,){/ScalarDash}{0}
\FAProp(0.9910,16.2307)(10.,6.1529)(0.,){/ScalarDash}{0}
\FAProp(10.,18.8471)(18.9820,8.7693)(0.,){/ScalarDash}{0}
\FAProp(10.,18.8471)(10.,6.1529)(-0.3,){/ScalarDash}{0}
\FAVert(13.8471,12.5){0}
\FAVert(6.1529,12.5){0}
\FAVert(10.,18.8471){0}
\FAVert(10.,6.1529){0}
\FAVert(18.9820,16.2307){0}
\FAVert(18.9820,8.7693){0}
\FAVert(0.9910,8.7693){0}
\FAVert(0.9910,16.2307){0}
\FALabel(10.,3.)[]{$-1 \times {\textstyle{\frac{1}{2}}}$}
\FADiagram{}
\FAProp(10.,10.)(10.,15.)(0.,){/Straight}{0}
\FAProp(10.,15.)(14.7553,16.5451)(0.,){/Straight}{0}
\FAProp(10.,10.)(14.7553,8.45492)(0.,){/Straight}{0}
\FAProp(14.7553,8.45492)(17.6942,12.5)(0.,){/Straight}{0}
\FAProp(14.7553,16.5451)(17.6942,12.5)(0.,){/Straight}{0}
\FAProp(17.6942,12.5)(20.6942,12.5)(0.,){/Straight}{0}
\FAProp(14.7553,8.45492)(15.4556,5.53781)(0.,){/Straight}{0}
\FAProp(14.7553,16.5451)(15.4556,19.4622)(0.,){/Straight}{0}
\FAProp(10.,15.)(5.2447,16.5451)(0.,){/Straight}{0}
\FAProp(10.,10.)(5.2447,8.45492)(0.,){/Straight}{0}
\FAProp(5.2447,8.45492)(2.3058,12.5)(0.,){/Straight}{0}
\FAProp(5.2447,16.5451)(2.3058,12.5)(0.,){/Straight}{0}
\FAProp(5.2447,8.45492)(4.5444,5.53781)(0.,){/Straight}{0}
\FAProp(5.2447,16.5451)(4.5444,19.4622)(0.,){/Straight}{0}
\FAProp(2.3058,12.5)(-0.6942,12.5)(0.,){/Straight}{0}
\FAProp(13.8471,12.5)(0.9910,16.2307)(0.,){/ScalarDash}{0}
\FAProp(6.1529,12.5)(18.9820,16.2307)(0.,){/ScalarDash}{0}
\FAProp(10.,18.8471)(10.,6.1529)(0.3,){/ScalarDash}{0}
\FAProp(10.,18.8471)(10.,6.1529)(-0.3,){/ScalarDash}{0}
\FAProp(0.9910,8.7693)(18.9820,8.7693)(-0.3,){/ScalarDash}{0}
\FAVert(13.8471,12.5){0}
\FAVert(6.1529,12.5){0}
\FAVert(10.,18.8471){0}
\FAVert(10.,6.1529){0}
\FAVert(18.9820,16.2307){0}
\FAVert(18.9820,8.7693){0}
\FAVert(0.9910,8.7693){0}
\FAVert(0.9910,16.2307){0}
\FALabel(10.,3.)[]{$-1 \times {\textstyle{\frac{1}{2}}}$}
\end{feynartspicture}
\vskip +0.6125 cm 
\caption{
The 26 different integrals which are allowed, by
the hypothesis of dual conformal symmetry, to contribute to the
amplitude $M_6^{(2),D=4}$.  Beneath each diagram is the
coefficient with which the corresponding integral, defined
according to the rules reviewed in \fig{ConformalIntegralsFigure},
enters into our result for $M_6^{(2),D=4}$.
An overall factor of $1/16$ is suppressed and
it is understood that one should sum over the 12 cyclic 
and reflection permutations of the external legs.
In each coefficient, the second factor is a symmetry factor
that accounts for overcounting in this sum.}
\label{MotherFigure}
\end{figure}

As with the odd part of $M_5^{(2)}$, the dual conformal properties
of the two integrals in $M_6^{(2), \mu}$, $I^{(14)}$ and $I^{(15)}$, 
are not apparent.
It is possible that they can be re-expressed as a linear combination
of integrals with manifest properties.  However,
we will argue in the next section that they cancel against
the analogous one-loop piece $M_6^{(1),\mu}$,
in the remainder function~(\ref{def_remainder}) for the ABDK/BDS ansatz.
Hence we focus on the surviving piece $M_6^{(2),D=4}$.
We find that this piece can indeed be written
as a linear combination of the 26 independent pseudo-conformal
integrals, as exhibited in \fig{MotherFigure}.

The integrals appearing in the four-point amplitude have a number of
striking features, partly explained by heuristics such
as the rung rule~\cite{BRY} and box substitution rule~\cite{FiveLoop}. Three
interesting features are observed in the four-point amplitude through five
loops:
\begin{itemize}
\item All pseudo-conformal integrals
appear with relative weights of $\pm 1$ or
$0$~\cite{BRY,ABDK,BCDKS,FiveLoop}.

\item
Moreover,
an integral appears with coefficient zero if 
and only if the integral is ill-defined (unregulated)
after taking its external legs off shell
and taking $\ep \rightarrow 0$~\cite{DrummondVanishing}.

\item
Finally, it has been proposed that the signs $\pm 1$ of the
contributing integrals can be understood by the requirement
of cancelling unphysical singularities~\cite{CachazoSkinner}.
\end{itemize}
It is clear from \fig{MotherFigure} that the first of these does not
hold for our representation of the six-point amplitude; in particular,
some relative weights are $\pm2$, and there is one weight of $-4$.  An
examination of the integrals in \fig{NonContributingIntegralsFigure},
shows that the second observation also requires some
modification. This should not be too surprising since at six points we
expect that some of the well-defined integrals appear in non-MHV
amplitudes but not in the MHV ones.  It would certainly be very
interesting to determine whether any of these considerations could be
generalized or modified to explain the pattern of coefficients
appearing in \fig{MotherFigure}.


\section{Results}
\label{ResultsSection}

Because the two-loop iteration formula (\ref{TwoloopOneloop})
incorporates the known infrared singularities, it must
hold for infrared-singular terms.  We have evaluated the integrals in
\figs{ContributingIntegralsFigure}{NonContributingIntegralsFigure}
through $\Ord(\ep^{-2})$ analytically in terms of ordinary polylogarithms.  In
\app{IntegralsAnalyticAppendix} we have collected their values though
$\Ord(\ep^{-3})$; we refrain from presenting the much lengthier
$\Ord(\ep^{-2})$ contributions.  By inserting the values of the
two-loop integrals into the assembly equation~(\ref{TwoLoopAssembly})
we find that \eqn{TwoloopOneloop} holds analytically through
$\Ord(\ep^{-2})$.
This provides a non-trivial check on our cut construction, evaluation
of integrals, and assembly of contributions.
The reader may check agreement through
$\Ord(\ep^{-3})$ using the values of the two-loop integrals in
\app{IntegralsAnalyticAppendix} and the one-loop
amplitude~(\ref{OneloopMHVAmplitude}) with $n=6$.

Beyond $\Ord(\ep^{-2})$, we resort to numerical integration.  We
first constructed Mellin-Barnes representations, 
in order to make use of the package {\tt MB}~\cite{MB}.  The
package {\tt AMBRE}~\cite{AMBRE} provides a simple means for obtaining
Mellin-Barnes representations that can be integrated using {\tt MB}.
One must treat the most complicated double-pentagon integrals, $I^{(12)}$ and
$I^{(13)}$, with some care to produce a numerically suitable
representation, so we give Mellin-Barnes representations for these 
two integrals in \app{MBAppendix}.

In four dimensions, there are at most four linearly independent 
momenta.  For the six-point amplitude, therefore, the Gram determinant
of any five external momenta must vanish,
\begin{equation}
\det(k_i \cdot k_j) = 0\,, \qquad\quad i,j = 1,2,3,4,5.
\label{GramDetConstraint}
\end{equation}
This constraint turns out not to be relevant for any of our checks,
but it is important to choose at least a few kinematic points
satisfying this constraint, in order to ensure that any deviation 
from the BDS ansatz does not arise from choosing momentum invariants that
cannot be realized in four dimensions.

\subsection{The $\mu$-Dependent Terms}
\label{ResultsSectionMu}

In section~\ref{IntegrandSection} we split both the one-
and two-loop amplitudes into a $D=4$ part and a part containing
explicit dependence on $\mu$, the ($-2\e$)-dependent part of the loop
momentum, according to \eqns{OneLoopAssembly}{TwoLoopAssembly}.
The $\mu$-dependent part of the one-loop amplitude, 
$M_6^{(1), \mu} (\epsilon)$, vanishes as $\e\to0$.  It only contributes
to the remainder function $R_6^{(2)}$, defined in \eqn{def_remainder},
because it appears in $(M_n^{(1)}(\e))^2$ multiplied by the singular 
terms in the one-loop amplitude, which are given in \eqn{OneloopMHVAmplitude}.
Thus the contribution of the one- and two-loop $\mu$-dependent terms
to $R_6^{(2)}$ is
\begin{eqnarray}
\Remainder^{(2),\mu}_6 &=& \lim_{\epsilon \to 0}
 \left[ M_6^{\twoloop,\mu}(\e) 
 - \left( - {1 \over 2}{1\over \e^2} \sum_{i=1}^6 (-s_{i,i+1})^{-\e}
    M_6^{\oneloop,\mu}(\e) \right)
\right] \,,
\label{def_remainder_mu}
\end{eqnarray}
where we set the dimensional regularization scale $\mu\to1$ here
to avoid confusion with $\mu_p$.

Integrals containing numerator factors of $\mu_p$ and $\mu_q$ can
be computed by differentiating integrands for scalar integrals
with respect to Schwinger parameters~\cite{NeqOneTwoLoop}.
This result holds because
the dependence of the integrals on the $(-2\e)$ components
of the loop momenta is very simple.  At two loops, it is given by,
\begin{equation}
\int d^{-2\e} \mu_p \ d^{-2\e} \mu_q 
\exp\Bigl[ - \mu_p^2 \, T_p - \mu_q^2 \, T_q 
           - \mu_{p+q}^2 \, T_{pq} \Bigr] 
\ \propto\ \Delta^\e \,,
\label{extraDIntegral}
\end{equation}
where $T_p$, $T_q$ and $T_{pq}$ are the sums of Schwinger parameters for
propagators carrying loop momenta $p$, $q$ and $p+q$, respectively,
and $\Delta = T_p T_q + T_p T_{pq} + T_q T_{pq}$.
Differentiation leads to the parameter insertions (see eq.~(4.26) of
ref.~\cite{NeqOneTwoLoop}),
\begin{eqnarray} 
 \mu_p^2 &\to&  -\e {T_q+T_{pq} \over \Delta} \,, \label{mupsqinsert} \\ 
 \mu_p\cdot\mu_q 
&\to&  {-\e\over2} {(T_p+T_q)-(T_q+T_{pq})-(T_p+T_{pq}) \over \Delta}
= \e {T_{pq} \over \Delta}  \,. \label{mupqinsert}
\end{eqnarray} 
At one loop, $\mu_p^2 \to -\e/T$, where $T$ is the sum of all the Schwinger
parameters.  

At one loop, the insertion of $1/T$ shifts the dimension of the
integral from $D=4-2\e$ to $D=6-2\e$, which makes the integral
infrared finite (and it remains ultraviolet finite).  Thus $I^{\rm
hex}(\epsilon)$, and hence $M_6^{(1),\mu} (\epsilon)$, vanish as
$\e\to0$.  At two loops, the factor of $1/\Delta$ in
\eqns{mupsqinsert}{mupqinsert} also shifts the dimension to $D=6-2\e$.
However, the Schwinger parameters in the numerator lead to doubled
propagators (see, for example, the discussion in
ref.~\cite{NeqOneTwoLoop}), which can cause infrared divergences, even
near $D=6$.

In the case of the double pentagon integral $I^{(14)}$, \eqn{mupqinsert}
shows that the doubled propagator is the central one.  Because this
propagator does not touch any on-shell external legs, doubling it
is ``safe'', and we expect that $I^{(14)} = \Ord(\e)$. 
We have confirmed this expectation, both numerically, and by checking
analytically the analogous planar double box integral for massless
four-point kinematics.

In the case of the ``hexabox'' integral, $I^{(15)}$, \eqn{mupsqinsert}
leads to doubled propagators on the box loop, which create infrared
divergences.  However, the hexagon loop remains infrared safe.  Reasoning
by analogy to the factorization of soft and collinear singularities at
the amplitude level~\cite{KnownIR}, the $\mu_p^2$-hexagon inside the hexabox
can be thought of as a hard process.  Thus one can shrink it to zero size
in space-time, and decorate it by a one-loop scalar triangle integral
representing the infrared-divergent contributions.  Thus we expect,
\begin{equation}
I^{(15)}(\e) =
-{1\over\e^2} (-s_{61})^{-1-\e} \, I^{\rm hex}(\e)\,+\,\Ord(\e)\,.
\label{hexaboxsing}
\end{equation}
The reason the equation is valid to $\Ord(\e)$ rather than $\Ord(\e^0)$
is simply because the ``hard'' part $I^{\rm hex}$ is itself $\Ord(\e)$.  
Once again, we have checked \eqn{hexaboxsing} numerically.  
We have also checked analytically that the same relation holds for 
the analogous planar double box integral.  Inserting \eqn{hexaboxsing}
into the explicit expressions for the one- and two-loop $\mu$-dependent
contributions $M_6^{(1),\mu}$ and $M_6^{(2),\mu}$ in
\eqn{def_remainder_mu}, we see that $R_6^{(2),\mu}$ vanishes.

\subsection{Evaluation of Remainder Function}

Explicit computations~\cite{ABDK,TwoLoopFiveA,TwoLoopFiveB}
have demonstrated that the remainder function $R_n^{(2)}$ defined
in \eqn{def_remainder} vanishes for $n=4,5$.
In this section we shall evaluate $R^{(2)}_6$ numerically at a few kinematic
points, and find that it is nonzero and nonconstant.

We choose Euclidean kinematics for all points, as this 
simplifies the numerical evaluation.  A particularly convenient 
kinematic point is
\def\hs{\hskip .4 cm}
\begin{equation}
K^{(0)} :  s_{i,i+1} = -1\,, \hs s_{i,i+1,i+2} = -2 \,, \nn \\
\label{KinematicStandard}
\end{equation}
which we take to be our standard reference point.  This point has several
advantages. Firstly, because it is symmetric under cyclic relabeling
$i \rightarrow i+1$ and under the reflection $i \rightarrow 6-i+1$, we do not
need to evaluate any relabelings of the integrals to obtain the
amplitude.  We have also exploited $s_{i,i+1}=-1$ to simplify the
Mellin-Barnes representations.  Moreover, this kinematic point
satisfies the Gram determinant constraint~(\ref{GramDetConstraint}).

The numerical values of all the integrals in
\fig{ContributingIntegralsFigure} at the standard kinematic point are
given in \app{IntegralsNumericalAppendix}.  (For completeness we also
give the values of the non-contributing integrals in
\fig{NonContributingIntegralsFigure}.)  Inserting these values into
the amplitude~(\ref{TwoLoopAssemblyA}), we obtain,
\begin{equation}
M_6^{(2),D=4} = {9\over 2 \ep^4} -
 {12.2457 \over \ep^2} - {21.99 \over \ep} - 20.8534 \pm 0.0057 + 
\Ord(\ep)\,,
\label{AmplKin2Value}
\end{equation}
at the standard kinematic point $K^{(0)}$.
The values of the $\mu$-dependent contributions (\ref{TwoLoopAssemblyB}) are,
\begin{equation}
M_6^{(2), \mu} = {2.3510 \over \ep } + 8.6024 \pm  0.0010  + \Ord(\ep)\,.
\label{AmplMuKin2Value}
\end{equation}
We have included estimated errors from the numerical integration reported 
by {\tt CUBA}~\cite{CUBA}, added in quadrature. In many cases the errors 
appear to be overestimated.  However, there can be correlations between 
different subintegrals in which the amplitude is expanded, 
because the same random seed is used.  The reported errors do
appear to give us a reliable measure of how many digits are
trustworthy.

The values~(\ref{AmplKin2Value}) and (\ref{AmplMuKin2Value}) 
may be compared to those for the ABDK/BDS ansatz~(\ref{TwoloopOneloop}). 
Separating it out, in analogy to~(\ref{TwoLoopAssembly}), as
\begin{equation}
M_6^{\rm BDS}=M_6^{{\rm BDS,}\,D=4}+M_6^{\rm BDS, \mu} \,,
\end{equation}
where the second term arises from $I^{\rm hex}$ in $M_6^{\oneloop,\mu}$,
we find that
\begin{equation}
M_6^{{\rm BDS,} \, D=4} =  {9\over 2 \ep^4} - {12.2457\over \ep^2} 
                        - {21.995\over \ep} - 21.9471 + \Ord(\ep) \,,
\label{BDSNonMuKin2Value}
\end{equation}
and that the $\mu$ pieces 
are given by
\begin{equation}
M_6^{\rm BDS, \mu} = { 2.3510 \over \ep} + 8.6017  + \Ord(\ep) \,.
\label{BDSMuKin2Value}
\end{equation}
Since these formul\ae{} involve computing only one-loop integrals, the
numerical integration errors are much smaller and do not affect the
answer to the quoted precision.

By comparing \eqns{AmplMuKin2Value}{BDSMuKin2Value}, we see that the
$\mu$ terms agree.  This result is in accord with the general vanishing of
$R_6^{(2),\mu}$ described in section~\ref{ResultsSectionMu}.
However, there is a difference in the $D=4$ terms, between our explicit 
calculation of the amplitude and the BDS ansatz.  Defining it as
\begin{equation}
\Remainder_A \equiv \Remainder^{\twoloop}_6 
= M_6^\twoloop -  M_6^{\rm BDS} \,,
\label{defRA}
\end{equation}
we find that at our standard kinematic point (\ref{KinematicStandard})
it equals
\begin{equation}
R_A^{0}\equiv R_A(K^{(0)}) = 1.0937 \pm 0.0057 \,.
\end{equation}
Although the remainder is only $5$ percent of the finite term in \eqn{AmplKin2Value},
it is nonzero at very high confidence level, demonstrating that the
ABDK/BDS ansatz needs to be modified.

Besides our standard kinematic point, we also evaluated $R_A$ 
at various other kinematic points,
\def\hs{\hskip .4 cm}
\begin{eqnarray}
%
%
K^{(1)} &:&  s_{12} = -0.723 6200, \hs  s_{23} = -0.921 3500, \hs  
 s_{34} = -0.272 3200, \hs  s_{45} = -0.358 2300, \hs \nn \\
&&  
 s_{56} = -0.423 5500, \hs  s_{61} = -0.321 8573,  \hs 
 s_{123} = -2.148 6192,  \hs  s_{234} = -0.726 4904, \hs \nn \\
&&
 s_{345} =  -0.482 5841,  \nn\\
%
%
K^{(2)} &:& s_{12} =-0.322 3100, \hs s_{23} = -0.2323220,
  \hs s_{34} = -0.523 8300, \hs s_{45} = -0.823 7640,   \hs \nn\\
&&  
  s_{56} = -0.532 3200, \hs s_{61} = -0.923 7600, \hs
  s_{123} = -0.732 2000,  \hs s_{234} = -0.828 6700, \hs \nn\\
&&
  s_{345} = -0.662 6116, \nn\\
%
%
K^{(3)} &:& s_{i,i+1} = -1,  \hs  s_{123} = -1/2,  \hs s_{234} = -5/8, \hs 
             s_{345} = -17/14, \nn \\
%
%
K^{(4)} &:&  s_{i,i+1} = -1,\hs   s_{i,i+1,i+2} = -3 , \nn \\
%
K^{(5)} &:&  s_{i,i+1} = -1,\hs   s_{i,i+1,i+2} = -9/2.
\label{KinematicPoints}
\end{eqnarray}

With six-point kinematics we have sufficient freedom to construct
three nontrivial conformal cross ratios:
\begin{eqnarray}
u_1&=&\frac{x_{13}^2x_{46}^2}{x_{14}^2x_{36}^2}=
      \frac{s_{12}s_{45}}{s_{123}s_{345}}\,, \nn\\
u_2&=&\frac{x_{24}^2x_{51}^2}{x_{25}^2x_{41}^2}=
      \frac{s_{23}s_{56}}{s_{234}s_{123}} \,, \label{SixPtConformalCrossRatios} \\
u_3&=&\frac{x_{35}^2x_{62}^2}{x_{36}^2x_{52}^2}=
      \frac{s_{34}s_{61}}{s_{345}s_{234}} \,.
\nn
\end{eqnarray}
The three conformal cross ratios~(\ref{SixPtConformalCrossRatios}) for
these kinematic points are given in the second column of
Table~\ref{RemainderTable}. Here $K^{(1)}$, $K^{(2)}$ and $K^{(3)}$
satisfy the Gram determinant constraint~(\ref{GramDetConstraint})
while $K^{(4)}$ and $K^{(5)}$ do not.  

\begin{table}
\caption{\label{RemainderTable} The numerical remainder compared
with the ABDK ansatz~(\ref{TwoloopOneloop}) for various kinematic
points. The second column gives the conformal cross ratios defined in
\eqn{SixPtConformalCrossRatios}.}

\vskip .4 cm

\begin{tabular}{||c|c||c||}
\hline
\hline
kinematic point & $(u_1, u_2, u_3)$ & $\Remainder_{A}$    \\
\hline
\hline
$K^{(0)}$ & $(1/4, 1/4, 1/4)$  & $ 1.0937 \pm  0.0057$   \\
\hline
$K^{(1)}$ &$(1/4, 1/4, 1/4)$  & $  1.076 \pm 0.022$  \\
\hline
$K^{(2)}$ &  $\, (0.547253,\, 0.203822,\, 0.881270) \,$ 
                            & $ -1.659 \pm  0.014$   \\
\hline
$K^{(3)}$ &($28/17, 16/5, 112/85)$  & $ -3.6508 \pm 0.0032\,$  \\
\hline
$K^{(4)}$ & $(1/9, 1/9, 1/9)$ & $  5.21  \pm   0.10$  \\
\hline
$K^{(5)}$ & $(4/81, 4/81, 4/81)$ &  $ 11.09 \pm 0.50 $ \\
\hline
\end{tabular}
\end{table}
 
The point $K^{(1)}$ is chosen so that the conformal cross ratios $u_i
= 1/4$ are identical to the cross ratios for our reference kinematic
point $K^{(0)}$.  The agreement, within the errors, between the
remainder functions for these two kinematic points, suggests that
$R_A$ is a function of only the cross ratios, {\it i.e.} is invariant
under dual conformal transformations.


\subsection{Comparison with Wilson Loop}
\label{WLcompare}

Drummond, Henn, Korchemsky and Sokatchev have already 
shown~\cite{HexagonWilson} that the Wilson loop
expectation value $\langle W_6^{(2)}\rangle$ corresponding to the 
two-loop six-point MHV amplitude, is not equal to that suggested
by the ABDK/BDS ansatz for amplitudes.  That is, they found 
a nonvanishing remainder function,
\begin{equation}
\Remainder_{W} \equiv \langle W_6^{(2)}\rangle - W_6^{\rm BDS}\,.
\label{RWdef}
\end{equation}
Here $W_6^{BDS}$ is the Wilson loop analog of $M_6^{BDS}$ defined in 
\eqn{BDSAnsatz}.  It involves the same function $M_6^{(1)}(\e)$, 
but different constants appear~\cite{HexagonWilson,WilsonValues}
than the ones for amplitudes (which are captured by $f^{(l)}(\e)$ and 
$C^{(l)}$).  We are motivated by the correspondence 
between MHV amplitudes and Wilson loops at one 
loop~\cite{DrummondVanishing,BrandhuberWilson},
to ask how these two remainder functions, $R_A$ and $R_W$, compare.

\begin{table}
\caption{\label{ComparisonTable} The comparison between the remainder
functions $R_A$ and $R_W$ for the MHV amplitude and the Wilson loop.  
To account for various constants of the kinematics, we subtract from 
the remainders their values at the standard kinematic point $K^{(0)}$, 
denoted by $R_A^0$ and $R_W^0$.
The third column contains the difference of remainders for the amplitude, 
while the fourth column has the corresponding difference for the Wilson
loop.  The numerical agreement between the third and fourth columns 
provides strong evidence that the finite remainder for the Wilson loop
is identical to that for the MHV amplitude. }

\vskip .4 cm

\begin{tabular}{||c|c||c|c||}
\hline
\hline
kinematic point & $(u_1, u_2, u_3)$ & $\Remainder_{A} - \Remainder_A^{0}$ & 
    $\Remainder_{W} - \Remainder_{W}^{0}$ \\
\hline
\hline
$K^{(1)}$ &$(1/4, 1/4, 1/4)$ & $-0.018 \pm  0.023 $ &  $ <10^{-5}$  \\
\hline
$K^{(2)}$ & $\,(0.547253,\, 0.203822,\, 0.881270)\,$  & $-2.753 \pm 0.015$ 
 & $ -2.7553$ \\
\hline
$K^{(3)}$ & $(28/17, 16/5, 112/85)$ & $\, -4.7445 \pm  0.0075\, $ & $ -4.7446$ \\
\hline
$K^{(4)}$ & $(1/9, 1/9, 1/9)$  & $ 4.12  \pm  0.10$  & $   4.0914$  \\
\hline
$K^{(5)}$ & $(4/81, 4/81, 4/81)$  & $ 10.00 \pm  0.50$  & $  9.7255$  \\
\hline
\end{tabular}
\end{table}

As explained in ref.~\cite{HexagonWilson}, in the collinear limits,
corresponding to $u_1 = 0, u_3 = 1-u_2$, the Wilson loop remainder
function $\Remainder_{W}$ becomes a constant, corresponding to
eq.~(17) that paper.  On the other hand, as explained in
\sect{RemainderPropertiesSection}, the MHV amplitude remainder
function must vanish in the collinear limits in order to be consistent
with collinear factorization.  This suggests a simple relation
between the two remainders,
\begin{equation}
\Remainder_A = \Remainder_{W} - c_W\,,
\label{RemainderWA}
\end{equation}
which we test numerically.
{}From DHKS~\cite{WilsonValues},
the constant $c_W$ takes on the value,
\begin{equation}
c_W = 12.1756\,,
\label{cvalue}
\end{equation}
with a precision of $\sim 10^{-3}$.

The numerical determination of the constant $c_W$ from the collinear
limits of the Wilson loop leads to some loss of precision, so instead
in Table~\ref{ComparisonTable} we compare the Wilson loop and
MHV-amplitude remainder functions by considering a ``difference of
differences''.  That is, for both the Wilson loop and MHV amplitude
the remainders $R_A$ and $R_W$ are found by subtracting the value of the
appropriate ABDK/BDS formula at that point.
{}From $R_A$ and $R_W$ we subtract the corresponding values $R_A^0$ and
$R_W^0$ at the standard kinematic point $K^{(0)}$. From DHKS the remainder
at the standard point is~\cite{WilsonValues} $R_W^0 = 13.26530$.
This subtraction eliminates any dependence on $c_W$.
The Wilson loop results are obtained from ref.~\cite{WilsonValues}.
In general, the Wilson loop results have
much smaller errors than those of the amplitude.  This is due to the
much simpler integral representations appearing in the Wilson loop
computation~\cite{WilsonValues}.

Using the value of $c_W$ from \eqn{cvalue}, we can also compare $R_A$ and
$R_W$ directly, albeit at lower precision.  We find that
within errors \eqn{RemainderWA} is satisfied for all six kinematic
points in \eqns{KinematicStandard} {KinematicPoints}. 
As mentioned above, we tested dual conformal invariance directly
at one point.  However, this invariance was tested much more extensively
(also numerically) for the Wilson loop~\cite{HexagonWilson,WilsonValues}.  
Thus our numerical agreement with the Wilson loop remainder, displayed in 
Table~\ref{ComparisonTable}, obviously provides considerable
additional evidence that $R_A$ possesses dual conformal invariance.

\section{The Remainder Function}
\label{RemainderPropertiesSection}

In the previous section we found a numerical difference between 
the ABDK/BDS ansatz for the two-loop six point amplitude and 
the explicit calculation.  In this section we constrain the analytic
form of the remainder and point out that its functional form can 
be determined from triple-collinear limits.

\subsection{Constraints on the Remainder Function}

The form of the ABDK/BDS ansatz is tightly constrained by
factorization properties and also exhibits dual conformal
invariance. As discussed in the previous section, numerical evidence
confirms that, while it departs from the ansatz, the even part
of the two-loop six-point amplitude is invariant under dual conformal
transformations. We can therefore discuss further constraints imposed
by this symmetry on the remainder function. We will discuss the case
of $n$-particle amplitudes and specialize to $n=6$ at the end. 

DHKS~\cite{ConformalWard} argue that MHV amplitudes (like Wilson loops),
should obey anomalous dual conformal Ward identities, the anomaly due
to infrared (ultraviolet) divergences.  
The fact that the BDS ansatz accounts for all infrared divergences 
of MHV amplitudes to all loop orders implies that $\Remainder_n$ is 
finite and thus independent of the regulator.  It also means that 
$\Remainder_n$ should satisfy {\it non}-anomalous Ward identities;
that is, it must actually be invariant under dual conformal
transformations and thus depend only on conformally-invariant 
cross ratios:
\begin{eqnarray}
\Remainder_n =
\Remainder_n(\{ u_{i j k l} \})\,.
\end{eqnarray} 
Here $(ijkl)$ denote the allowed quartets of the external legs leading
to well-defined cross ratios (\ref{CrossRatios}).  The symmetry
properties of the remainder function $\Remainder_n$ under the
permutation of its arguments follow from the reflection and cyclic
identities obeyed by the rescaled MHV scattering amplitudes $M_n^{(L)}$.

Further restrictions on $R_n$ come from the fact that the BDS ansatz
correctly captures the two-particle collinear factorization of MHV
amplitudes. In general, the $L$-loop rescaled
planar amplitudes $M_n^{(L)}(1, 2, \ldots, n)$ satisfy simple
relations as the momenta of two color-adjacent legs $k_i$, $k_{i+1}$
become collinear,~\cite{TreeReview,NeqFourOneLoop,OneLoopSplitting,
KosowerAllOrder},
\begin{eqnarray}
 &&M_n^\Lloop(\ldots,i^{\lambda_i},(i+1)^{\lambda_{i+1}},\ldots)
  \, \longrightarrow
\sum_{l = 0}^L \sum_{\lambda=\pm}
  r^\lloop_{-\lambda}(z;i^{\lambda_i}\kern-1pt,(i+1)^{\lambda_{i+1}})
  M_{n-1}^{(L-l)}(\ldots,P^\lambda \kern-5pt,\ldots) \,. \hskip 1 cm
\label{LoopSplit}
\end{eqnarray}
The index $l$ sums over the different loop orders of the rescaled
splitting amplitudes,
\begin{equation}
r_{-\lambda}^\lloop(z;i^{\lambda_i}\kern-1pt,(i+1)^{\lambda_{i+1}})
\equiv { \Split^\lloop_{-\lambda}(z;i^{\lambda_i}\kern-1pt,(i+1)^{\lambda_{i+1}})
   \over \Split^{\tree}_{-\lambda}(z;i^{\lambda_i}\kern-1pt,(i+1)^{\lambda_{i+1}}) }
 \,,
\label{r2particledef}
\end{equation}
while $\lambda$ sums over
the helicities of the intermediate leg $k_P=(k_i + k_{i+1})$, and $z$
is the longitudinal momentum fraction of $k_i$, $k_i \approx z k_P$.

The relevant two-loop splitting amplitudes were calculated in
refs.~\cite{ABDK,BDKTwoLoopSplit}. If we assume that dual conformal
symmetry holds to all orders, then the five-point amplitudes are fully
determined by this symmetry. By taking the collinear limit, the all-loop 
splitting amplitude,
\begin{equation}
r^{\rm full}
\equiv 1 + \sum_{L=1}^\infty a^L r^{(L)} \,,
\label{rfulldef}
\end{equation}
must have the form
\begin{equation}
\ln r^{\rm full}
= \sum_{l=1}^\infty a^l \, f^{(l)}(\e) \, r^{(1)}(l \e) + \Ord(\e)\,.
\label{BDSAnsatzSplit}
\end{equation}
As discussed in ref.~\cite{BDS}, this iterative structure yields the
correct collinear behavior to all loop orders.

Since the BDS ansatz accounts for collinear factorization, to all
orders in perturbation theory, the
remainder functions must have a trivial behavior under collinear
factorization.  The $n$- and $(n-1)$-point remainder functions must be
related by
\begin{equation}
\label{doublecollinear_consistency}
\lim_{x_{i,i+2}^2\rightarrow 0}\Remainder_n(\{u_{i_1, i_2, i_3, i_4}\})=
\Remainder_{n-1}(\{u_{i_1, i_2, i_3, i_4}\}^\prime)
\,,
\end{equation}
for any two-particle Mandelstam invariant $x_{i,i+2}^2=s_{i,i+1}$.
The arguments of $\Remainder_{n-1}$ are the subset of the $n$-point
conformal cross ratios that are non-vanishing and well-defined in 
the collinear limit $s_{i,i+1}~\rightarrow~0$.

For five-point amplitudes, no conformal cross ratio with the required
properties may be constructed.  Thus no remainder function can exist consistent
with collinear factorization and the requirement of dual conformal
invariance. A constant remainder is ruled out by collinear
factorization.

Nontrivial conformal cross ratios can be first constructed with
six-particle kinematics, $u_1$, $u_2$ and $u_3$ 
in \eqn{SixPtConformalCrossRatios}. 
The cyclic and reflection symmetries imply that $\Remainder_6(u_1,u_2,u_3)$
is a totally symmetric function of its arguments.
Because the remainder function $R_5^{(2)}$ for the five-point
amplitude vanishes~\cite{TwoLoopFiveA,TwoLoopFiveB},
$\Remainder_6^{(2)}(u_1,u_2,u_3)$ must vanish in all collinear limits.
If dual conformal symmetry is valid to all loop orders then $R_5$ vanishes
exactly and therefore $R_6$ must vanish in all collinear limits to
all loop orders.
DHKS~\cite{HexagonWilson} reached a similar conclusion in
their analysis of the two-loop six-sided Wilson loop, namely that
there is a remainder $\hat f$ above and beyond the ABDK/BDS ansatz
with the properties described above.


\subsection{Remainder Function from Triple-Collinear Limits}

Planar color-ordered scattering amplitudes exhibit singularities when 
several adjacent momenta become collinear.  
General all-order factorization properties of scattering amplitudes 
have been discussed in ref.~\cite{KosowerAllOrder}.
The most familiar of these limits are when just two particles become 
collinear.  However, the limits when more particles become collinear
simultaneously can provide additional constraints%
\footnote{We thank Gregory Korchemsky and Emery Sokatchev for discussion on 
the triple-collinear limits.}.
(Multi-collinear configurations should not be confused with multi-particle 
factorization limits, in which amplitudes factorize into products 
of lower-point, non-degenerate scattering amplitudes. The latter limits
are trivial for MHV amplitudes in supersymmetric theories.)

As discussed previously, the ABDK/BDS ansatz incorporates the correct
two-particle collinear factorization of MHV amplitudes. It also makes definite
predictions, which remain to be tested, for the multi-collinear
factorization of MHV amplitudes in ${\cal N}=4$ super-Yang-Mills theory.
These tests amount to constraints on the remainder functions $R_n$.
For the six-gluon amplitude we only have the triple-collinear limit.
As we will see, it is possible to completely determine the remainder 
function $\Remainder_6$ from this limit.

Let us consider three adjacent momenta $k_{a,b,c}$ in the limit that
they become collinear and introduce the three momentum fractions
\be
k_a=z_1 P\,,
\qquad
k_b=z_2 P\,,
\qquad
k_c=z_3 P\,,
\qquad
z_1+z_2+z_3=1\,,
\qquad 0\le z_i\le 1\,,
\qquad
P^2\rightarrow 0\,.
\label{momentumfractions}
\ee
An $n$-point amplitude at $l$ loops factorizes as follows:
\bea
A_n^{(l)}(k_1,\dots , k_{n-2},k_{n-1},k_{n}) &\mapsto&
\sum_{\lambda=\pm}\sum_{s=0}^l
A_{n-2}^{(l-s)}(k_1,\dots,k_{n-3},P^{\lambda})\;
\Split^{(s)}_{-\lambda}(k_{n-2}k_{n-1}k_{n}; P)\,.
\nonumber\\
{~}
\label{all_loop_factorization}
\eea

Taking into account parity and reflection symmetries,
there are six independent triple-collinear splitting amplitudes:
\bea
&&\Split_{+}(k_a^{+}k_b^{+}k_c^{+}; P),
~~~~~~~~
\label{lambdasum4}\\
&&\Split_{-\lambda_P}(k_a^{\lambda_a}k_b^{\lambda_b}k_c^{\lambda_c}; P),
~~~~~~~~
\lambda_a+\lambda_b+\lambda_c-\lambda_P=2\,,
\label{lambdasum2}\\
&&\Split_{-\lambda_P}(k_a^{\lambda_a}k_b^{\lambda_b}k_c^{\lambda_c}; P),
~~~~~~~~
\lambda_a+\lambda_b+\lambda_c-\lambda_P=0\,.
\label{lambdasum0}
\eea
The first one~(\ref{lambdasum4}) vanishes in any supersymmetric theory.
The three triple-collinear splitting amplitudes of the second 
type~(\ref{lambdasum2}), an example of which is 
$\lambda_a=\lambda_b=\lambda_c=\lambda_P=1$,
appear in limits of MHV amplitudes. 
The ${\cal N}=4$ supersymmetry Ward identities for MHV amplitudes imply
that their rescaled forms\footnote{%
We omit a trivial dimensional dependence on $s_{abc}$ from the
argument list of $r_S^{(l)}$.}
are all equal,
\bea
\frac{\Split^{(l)}_\mp(k_a^\pm k_b^+k_c^+; P)}
{\Split^{(0)}_\mp(k_a^\pm k_b^+k_c^+; P)}=
\frac{\Split^{(l)}_\mp(k_a^+ k_b^\pm k_c^+; P)}
{\Split^{(0)}_\mp(k_a^+ k_b^\pm k_c^+; P)}=
r_S^{(l)}({\textstyle{\frac{s_{ab}}{s_{abc}}}}, 
{\textstyle{\frac{s_{bc}}{s_{abc}}}}, z_1, z_3)\,.
\label{RescaledSplitting}
\eea
The two splitting amplitudes of the third kind~(\ref{lambdasum0}) arise 
only in limits of non-MHV amplitudes and do not have a simple factorized
form similar to (\ref{RescaledSplitting}).%
\footnote{The spin-averaged absolute values squared of tree-level
triple-collinear splitting amplitudes have been computed in
ref.~\cite{Campbell}; without spin-averaging they have been
computed in refs.~\cite{CataniTripleCollinear}.  The tree-level 
triple (and higher) collinear splitting amplitudes themselves
have been computed in ref.~\cite{DFM}, and in ref.~\cite{Birthwright}
using the MHV rules~\cite{CSW}.  The one-loop correction to the 
$q\rightarrow q{\bar Q}Q$ triple-collinear splitting amplitude in QCD
was computed in ref.~\cite{CdFR}.
}

On general grounds, the six-gluon amplitude exhibits a nontrivial
triple-collinear limit. In the limit~(\ref{momentumfractions}), 
for $a=4$, $b=5$, $c=6$, the three conformal cross 
ratios~(\ref{SixPtConformalCrossRatios}) are all
nonvanishing and arbitrary:
\be
{\bar u}_1=\frac{s_{45}}{s_{456}}\,\frac{1}{1-z_3}\,,
\qquad
{\bar u}_2=\frac{s_{56}}{s_{456}}\,\frac{1}{1-z_1}\,,
\qquad
{\bar u}_3=\frac{z_1z_3}{(1-z_1)(1-z_3)}\,.
\ee
The remainder function therefore survives the triple-collinear
limit, and is evaluated at $({\bar u}_1,{\bar u}_2,{\bar u}_3)$. 
Thus, assuming dual conformal invariance, finding the remainder function 
in this limit is equivalent to finding it for generic six-particle kinematics.

We can expose the two-loop remainder function by comparing
\eqn{all_loop_factorization} for the rescaled MHV six-point amplitude,
\bea
\lim _{4||5||6}M_6^{(2)}=
 M_4^{(2)} 
+ M_4^{(1)} r^{(1)}_{S}({\textstyle{\frac{s_{45}}{s_{456}}}}, 
{\textstyle{\frac{s_{56}}{s_{456}}}}, z_1, z_3,\epsilon)
+ r^{(2)}_{S}({\textstyle{\frac{s_{45}}{s_{456}}}}, 
{\textstyle{\frac{s_{56}}{s_{456}}}}, z_1, z_3,\epsilon)\,,
\label{M6triple}
\eea
with the triple-collinear limit of \eqn{def_remainder}.
The second term in \eqn{M6triple}
is already determined by one-loop calculations, and it is incorporated
in the ABDK/BDS ansatz.  Therefore $\Remainder_6^{(2)}$ enters only in the 
two-loop splitting amplitude $r^{(2)}_{S}$, as the deviation from the 
ABDK/BDS prediction $r^{(2)\,{\rm BDS}}_{S}$:
\bea
r^{(2)}_{S}({\textstyle{\frac{s_{45}}{s_{456}}}}, 
{\textstyle{\frac{s_{56}}{s_{456}}}}, z_1, z_3,\epsilon)=
r^{(2)\,{\rm BDS}}_{S}({\textstyle{\frac{s_{45}}{s_{456}}}}, 
{\textstyle{\frac{s_{56}}{s_{456}}}}, z_1, z_3,\epsilon)
+\Remainder_6^{(2)}({\bar u_1}, {\bar u_2},{\bar u_3})\,,
\eea
with
\bea
&&r^{(2)\,{\rm BDS}}_{S}({\textstyle{\frac{s_{45}}{s_{456}}}}, 
{\textstyle{\frac{s_{56}}{s_{456}}}}, z_1,
z_3,\epsilon)=
\cr
&& \hskip 2 cm 
\frac{1}{2}
\left(r^{(1)}_{S}({\textstyle{\frac{s_{45}}{s_{456}}}}, 
{\textstyle{\frac{s_{56}}{s_{456}}}}, z_1, z_3,\epsilon)\right)^2
+f^{(2)}(\epsilon)\;r^{(1)}_{S}({\textstyle{\frac{s_{45}}{s_{456}}}}, 
{\textstyle{\frac{s_{56}}{s_{456}}}}, z_1, z_3,\epsilon)
\,.
\eea
Thus, the two-loop remainder function is {\it completely determined}
by the two-loop triple-collinear splitting amplitude, {\it e.g} for
the helicity configuration~(\ref{lambdasum2}).  While {\it a priori} it
may depend on all four arguments of the splitting amplitude, dual
conformal invariance requires that it depend only on the three
cross ratios.

The triple-collinear splitting amplitudes may be computed using the
unitarity method following the strategy in
refs.~\cite{KosowerAllOrder,BDKTwoLoopSplit}.  It is important
to understand whether they satisfy an iteration relation generalizing
that of the double-collinear splitting amplitude~\cite{ABDK}.  If such
an iteration relation exists, then it should be straightforward to
construct an all-order iteration relation for the six-point gluon
amplitude. This would allow us to add in a correction term to the BDS
ansatz, at least for the six-point case.

The remainder function beyond two loops can also be extracted from
the triple-collinear splitting amplitude, though they are no
longer equal. Instead, they are related iteratively via,
\begin{eqnarray}
\Remainder_6^{(l)}({\bar u}_1,{\bar u}_2,{\bar u}_3)=
         \sum_{s=2}^l M_4^{(l-s)}(\ep)\,
\Bigl[ r_S^{(s)}({\textstyle{\frac{s_{45}}{s_{456}}}}, 
                 {\textstyle{\frac{s_{56}}{s_{456}}}}, z_1, z_3,\epsilon)
     - r_S^{(s)\,{\rm BDS}}({\textstyle{\frac{s_{45}}{s_{456}}}},
                 {\textstyle{\frac{s_{56}}{s_{456}}}}, z_1, z_3,\epsilon)
\Bigr] \,. \hskip .7 cm 
\label{allloopR6}
\end{eqnarray}

For amplitudes with additional external legs, it is unclear whether
the triple-collinear limits suffice to constrain the remainder
functions completely.  If these limits do not suffice, we can
formulate additional constraints along the lines above.  In
particular, it is easy to see that the remainder function of the
two-loop $n$-point MHV amplitude is completely determined by the difference
between the two-loop $(n-3)$-point splitting amplitude and the
iteration of the one-loop $(n-3)$-point splitting amplitude. Also, the
consideration of the $m$-particle collinear limit with $m\le n-4$
leads to consistency conditions analogous to
\eqn{doublecollinear_consistency}.


\section{Summary and Conclusions}
\label{ConclusionSection}

The $\NeqFour$ supersymmetric gauge theory has proven an important
laboratory and testing ground for inquiry into the properties of gauge
theories, both at weak and strong coupling.  The BDS ansatz for planar
MHV scattering amplitudes~\cite{BDS} in this theory, along with the BES
integral equation for the cusp anomalous dimension~\cite{BES} and the
strong-coupling calculation of Alday and
Maldacena~\cite{AldayMaldacena}, point to the possibility of
computing planar amplitudes for {\it any\/} value of the coupling.

In this paper we have checked the BDS ansatz directly by computing the
parity-even parts of the leading-color part of the planar two-loop six-point
amplitude.  Using the unitarity method, we have obtained an integral
representation for it.  Numerical evaluation of this representation
shows that there is a remainder beyond the ABDK/BDS prediction for
this amplitude.  Strikingly, the remainder agrees with the 
corresponding remainder for the hexagonal Wilson
loop~\cite{HexagonWilson,WilsonValues}.

This remainder must vanish in any limit where two color-adjacent 
momenta become collinear, because the ABDK/BDS construction accounts
for all terms with collinear singularities.
As we showed, it should be possible to fully reconstruct 
the remainder function for the six-point amplitude
by evaluating triple-collinear splitting amplitudes. 

There are a number of interesting open issues which remain to be
clarified. The origin of the dual conformal symmetry remains
mysterious.  In the context of the AdS/CFT correspondence it has been
suggested~\cite{AMTrouble} that it is related to symmetries of the
space defined by the coordinates $y^\mu$ introduced in \eqn{winding}.
Based on this, one might wonder whether dual conformal symmetry
can be found in the planar amplitudes of all four-dimensional CFTs with a
string-theory dual.

We have found that the integrals appearing in the amplitude either
vanish as the loop momenta are taken to be four dimensional or are
pseudo-conformal (with the external momenta taken
off shell to make them infrared finite).  Contributions containing 
the $(-2\eps)$-dimensional components of loop momenta in the numerator 
factors satisfy the BDS ansatz and drop out of the remainder function. 
It seems reasonable to expect that this pattern continues to all loop orders.  
It would therefore be very useful to have a set of rules for writing 
the coefficients of all pseudo-conformal integrals directly, without 
resorting to evaluations of the cuts.

So far dual conformal transformations have only been discussed in the
context of their action on Wilson loops or planar MHV amplitudes 
after dividing out the tree amplitude prefactor.  Can we extend
this to non-MHV amplitudes?  At least at one loop the integrals 
appearing in non-MHV amplitudes are pseudo-conformal scalar box 
integrals~\cite{NeqFourOneLoop}, hinting that dual conformal invariance 
might be a general property of the planar limit of the theory.  However,
in this case the tree amplitude does not factor 
out~\cite{Fusing,OneloopTwistorB}, leaving the question of how the dual 
conformal symmetry might act on the spinor products that enter the relative 
factors of different pseudo-conformal integrals.
Related to this is the question of whether the dual conformal symmetry 
can be extended to the Lagrangian of ${\cal N}=4$ super-Yang-Mills theory.
It is not clear how it should act on the Lagrangian, given that it is 
only understood at present for planar MHV amplitudes.

What other properties can we employ to constrain the scattering
amplitudes?  Integrability~\cite{Integrability,ES,BES} of the
dilatation operator in planar ${\cal N} = 4$ super-Yang-Mills theory
has not yet been used.  Similarly, we expect the amplitudes to have
simple structures in twistor space~\cite{TwistorStructure}.  At one
loop the coefficients in front of each integral have been shown to lie
on simple curves in twistor space, though the precise structure of
complete loop amplitudes has not been
determined~\cite{OneloopTwistorA, OneloopTwistorB,
HolomorphicAnomaly}.  It would be very interesting to explore the
structure at higher loops.

In order to shed light on the remainder function it is very
important to find its analytic form for the two-loop six-point amplitude. 
This form would be extremely useful 
for understanding the missing terms in the BDS ansatz at higher loops.
It would also be useful for analytic continuation into the physical high
energy or Regge limits of $2\to4$ and $3\to3$ scattering that have
been discussed recently~\cite{BNST,Lipatov}.
In \sect{RemainderPropertiesSection} we proposed a possible means for
constructing it at six points from triple-collinear limits.  
The equality of the six-point remainder function and the corresponding
Wilson loop quantity~\cite{WilsonValues} is
rather surprising.  It would obviously be very desirable to understand
whether this equality holds to all loop orders, as well as at 
strong coupling.

In summary, our computation demonstrates that the BDS ansatz requires
modification for amplitudes with six or more external legs.  The
surprising equality of the Wilson loop and MHV amplitude remainders,
however, points to an additional structure in the theory which
constrains its form.  This in turn provides hope of determining the
remainder function analytically, first at two loops, and eventually 
to all loop orders.

\section*{Acknowledgments}

We are grateful to Luis Fernando Alday, John Joseph Carrasco, Harald
Ita, Henrik Johansson, Daniel Ma\^{\i}tre, Juan Maldacena, Chung-I
Tan, and Arkady Tseytlin for very stimulating discussions and
correspondence.  We are indebted to James Drummond, Johannes Henn,
Gregory Korchemsky and Emery Sokatchev for equally stimulating
discussions, and especially for sharing their new results allowing a
comparison to Wilson loops~\cite{WilsonValues}.  M.~S. and A.~V. are
grateful to Freddy Cachazo for collaboration in the early stages of
this work. L.~J.~D. thanks the Ecole Normale Sup\'erieure for
hospitality. Part of this work was done while R.~R. was a participant
of the programme ``Strong Fields, Integrability and Strings'' at the
Isaac Newton Institute in Cambridge, U.K.  This work was supported in
part by the US Department of Energy under contracts DE-FG03-91ER40662
(Z.~B.), DE-AC02-76SF00515 (L.~J.~D.) and DE-FG02-91ER40688
(M.~S. (OJI) and A.~V.); the Agence Nationale de la Recherce of France
under grant ANR--05--BLAN--0073--01 (D.~A.~K.); the US National
Science Foundation under grants PHY-0455649 and PHY-0608114 (R.~R.),
PHY-0610259 (M.~S.) and PHY-0643150 CAREER (A.~V.); and the
A.~P.~Sloan Foundation (R.~R.).  We also thank Academic Technology
Services at UCLA for computer support.  Several of the figures were
generated using Jaxodraw~\cite{Jaxo} (based on Axodraw~\cite{Axo}) or
FeynEdit~\cite{FeynEdit} (based on FeynArts~\cite{FeynArts}).

\appendix

\section{Analytic Values of Integrals}
\label{IntegralsAnalyticAppendix}

\subsection{One-loop integrals}

In the Euclidean region with all $s_{i,i+1}$ and $s_{i,i+1,i+2}$ negative,
the one-mass box, $I^{1\rm m}$ in \fig{OneLoopIntegralsFigure}, is
given through $\Ord(\ep^0)$ by
\begin{eqnarray}
I^{1\rm m}(s_{45}, s_{56}, s_{123})
&=&
{2 \over s_{45} s_{56}}
\Biggl[\frac{1}{\ep^2}\,
\Bigl( (-s_{45})^{-\ep} + (-s_{56})^{-\ep} - (-s_{123})^{-\ep} \Bigr)
- {1\over 2} \ln^2\Bigg( {s_{45}\over s_{56}}\Biggr)
\nn\\
&&\null
-\Li_2\Biggl(1-{s_{123}\over s_{45}}\Biggr)
-\Li_2\Biggl(1-{s_{123}\over s_{56}}\Biggr)
-\frac{\pi^2}{4}
\Biggr]+\Ord(\ep)\,.
\label{IntOneloopAnalytic1m}
\end{eqnarray}
Similarly, the two-mass ``easy'' box, $I^{{\rm 2 m} e}$, is
\begin{eqnarray}
I^{2{\rm m}e}(s_{123},s_{345},s_{12},s_{45})  &=& 
{2\over (s_{12} s_{45} - s_{123} s_{345})}  \Biggl[ {1\over \ep^2} 
\Bigl( (-s_{12})^{-\ep} + (-s_{45})^{-\ep} \nn \\
&& \null - (-s_{123})^{-\ep} - (-s_{345})^{-\ep} \Bigr)
 + {1\over 2} \ln^2\Biggl( {s_{123}\over s_{345}}\Biggr) \nn \\
&& \null 
 + \Li_2 \biggl(1 - {s_{12} \over s_{123}} \biggr) 
 + \Li_2 \biggl(1 - {s_{45} \over s_{123}} \biggr) 
 + \Li_2 \biggl(1 - {s_{12} \over s_{345}} \biggr) \nn \\
&& \null 
 + \Li_2 \biggl(1 - {s_{45} \over s_{345}} \biggr) 
 - \Li_2 \biggl( 1 - {s_{12} s_{45} \over s_{123} s_{345}} \biggr)
  \Biggr]
 + \Ord(\ep) \,.
\label{IntOneloopAnalytic2me}
\end{eqnarray}

\subsection{Two-loop integrals}

We have computed analytic expressions for the two-loop
integrals appearing in \fig{ContributingIntegralsFigure} through
$\Ord(\ep^{-2})$.  The $\Ord(\ep^{-2})$ expressions are rather
cumbersome, so we display here only the results through $\Ord(\ep^{-3})$,
omitting $I^{(1)}$ as it is given by a product of two one-loop
integrals of type $I^{1\rm m}$:
\begin{eqnarray} 
I^{(2)} & = &
- {1 \over (-s_{12})^{1+2 \eps} \, s_{23}^2} \Biggl[
    {1\over \ep^4}  + {2\over \eps^3} \,
            \ln\biggl({s_{123} \over s_{23}}  \biggr)
             \Biggr] 
    + \Ord(\ep^{-2}) \,,\nn\\
I^{(3)} & = & 
\Ord(\ep^{-2}) \,, \nn\\
%
I^{(4)} & = &  
-{1\over (-s_{12})^{1+2\ep} \, s_{234}^2}  \Biggl[
{1\over 4 \ep^4} + {1\over 2 \ep^3} \ln\biggl({s_{3 4} s_{5 6} \over
                        s_{2 3 4}^2} \biggr)  \Biggr]
   + \Ord(\ep^{-2})\,, \nn\\  
I^{(5)} & = & 
 {3 \over 2 \ep^3}\,  {1 \over 
        s_{3 4}  (s_{2 3} s_{5 6}  - s_{1 2 3} s_{2 3 4}) } \,
      \ln\biggl({s_{1 2 3} s_{2 3 4} \over s_{2 3} s_{5 6}} \biggr)
 + \Ord(\ep^{-2})\,, \nn\\  
I^{(6)} &=&  
-{1\over (-s_{1 2})^{1+2\ep} \, s_{2 3} s_{2 3 4}} 
 \Biggl[ {3 \over 4 \ep^4} + {1\over 2 \ep^3}
   \ln\biggl( {s_{3 4} s_{5 6}^3 \over s_{2 3 4}^4} \biggr)\Biggr]  
 + \Ord(\ep^{-2})\,, \nn\\  
I^{(7)} &=& 
-{1\over s_{61} s_{3 4} (-s_{1 2 3})^{1+2\ep}}
 \Biggl[ {1 \over \ep^4} + {1\over \ep^3}
       \ln\biggl({s_{12} s_{23} s_{45} s_{56} 
                    \over s_{61}s_{3 4} s_{123}^2} \biggr) \Biggr]
+ \Ord(\ep^{-2})\,, \nn\\  
I^{(8)} &=& 
 {1\over s_{61} (s_{61} s_{34} - s_{2 3 4} s_{345})} 
               {3\over \ep^3}
      \ln\biggl( {s_{234} s_{345} \over s_{61} s_{34}} \biggr)
    + \Ord(\ep^{-2})\,, \nn\\
I^{(9)} &=& 
-{1\over (-s_{2 3})^{1+2\ep}\, s_{3 4} s_{2 3 4}}
\Biggl[{1\over 2 \ep^4} + {1\over \ep^3}
   \ln\biggl({s_{23} s_{56}\over s_{1 2} s_{234}}\biggr) \nn \\
  && \null \hskip 1.8 cm
   + {3\over 2\ep^3} {s_{123} s_{234}  \over  s_{123} s_{234} - s_{23} s_{56} }
     \ln\biggl({ s_{23} s_{56} \over s_{123} s_{234}} \biggr) 
  \Biggr] 
   + \Ord(\ep^{-2})\,, \nn\\  
I^{(10)} &=& 
-{1\over (-s_{61})^{1+2\ep} s_{2 3} s_{3 4}} \Biggl[
    {5\over 2 \ep^4} + {1\over 2 \ep^3}
      \ln\biggl( {s_{2 3 4}^5  s_{4 5}^2 s_{3 4 5}^3  s_{61}^5 \over
             s_{1 2}^4 s_{2 3}^2 
                s_{3 4}^5  s_{1 2 3}^4 } \biggr)  
     \Biggr] 
   + \Ord(\ep^{-2})\,, \nn\\ 
I^{(11)} &=& 
-  {1\over \ep^4 s_{23}} \Biggl[
  {s_{45} \over s_{61} s_{34} (-s_{123})^{1+2\ep} } + 
  {3 \over 4 s_{12} (-s_{234})^{1+2\ep} } + 
   {3 s_{345} \over  2 (-s_{12})^{1+2\ep}\, s_{61} s_{34} } \Biggr] \nn \\
 && \null 
  + {1\over 2 \ep^3  s_{23}} \Biggl[
      {2 s_{45} \over s_{61} s_{34} s_{123}} 
      \ln\biggl({s_{12} s_{23} s_{45} s_{56} \over
         s_{61} s_{34}  s_{123}^2} \biggr)
  + {1\over s_{12} s_{234}}  
   \ln\biggl({s_{34}^3 s_{123} \over s_{12}^3 s_{23}}\biggr) \nn \\
 && \null \hskip 1.8 cm 
 + {s_{61} s_{23} s_{34} s_{56} + 2 s_{12} s_{45} s_{234}^2 
     \over  s_{12} s_{61} s_{34} s_{234} 
    (s_{23} s_{56} -  s_{123} s_{234} )}
    \ln\biggl( {s_{23} s_{56} \over s_{123} s_{234}} \biggr) \nn \\
 && \null \hskip 1.8 cm 
  + {s_{345} \over s_{12} s_{61} s_{34} }
   \ln\biggl({s_{123}^2 s_{234}^3 s_{345}^3 \over
                      s_{61}^3 s_{23}^2 s_{34}^3 } 
                  \biggr)   \Biggr]
   + \Ord(\ep^{-2})\,, \nn\\ 
I^{(12)} &=& 
-  {1\over \ep^4} \Biggl[
    {3 s_{123}\over (-s_{12})^{1+2\ep} \, s_{61} s_{34} s_{45} }
   + {s_{23} s_{56} \over s_{12} s_{61} s_{34} s_{45} (-s_{234})^{1+2\ep} }
   + {1\over s_{61} s_{34} (-s_{345})^{1+2\ep} } \Biggr]  \nn \\
&& \null 
+ {1\over\ep^3} \Biggl[
{s_{123} \over s_{12} s_{61} s_{34} s_{45}}
 \ln\biggl({ s_{234}^2 s_{345}^6 \over 
       s_{23} s_{34}^3 s_{45}^3 s_{56}} \biggr)
+ {s_{23} s_{56} \over s_{12} s_{61} s_{34} s_{45} s_{234}}
    \ln\biggl({s_{23} s_{56} s_{345}^2 \over 
              s_{12}^2 s_{34} s_{45} } \biggr) \nn \\
  && \null \hskip 1.2 cm 
+ {1\over s_{61} s_{34} s_{345}}
 \ln\biggl({s_{45} s_{234} s_{345} \over s_{23} s_{34} s_{56} }\biggr) \nn\\
  && \null \hskip 1.2 cm 
+ {1\over s_{61} s_{34} - s_{234} s_{345}}
   \,  { s_{45} s_{234} s_{12} + 2 s_{345} s_{23} s_{56} \over 
          s_{45} s_{234} s_{12} s_{345}      }\,  
  \ln\biggl({s_{61} s_{34} \over s_{234} s_{345}} \biggr) \nn \\
  && \null \hskip 1.2 cm 
+ { s_{12} s_{45} s_{234} + 
    (s_{23} s_{56} + 3 s_{123} s_{234}) s_{345} \over 
    s_{12} s_{61} s_{34} s_{45} s_{234} s_{345} } 
     \ln\biggl({s_{12} \over s_{61}} \biggr)
  \Biggr]
   + \Ord(\ep^{-2})\,, \nn\\ 
I^{(13)} &=&  
-{1\over \ep^4} \Biggl[
  {3 s_{23} \over s_{12} s_{61} s_{34} (-s_{45})^{1+2\ep} } 
 + {s_{123} s_{234} \over s_{12} s_{34} s_{45} (-s_{56})^{1+2\ep}\,  s_{61} } 
 + {1\over 4 (-s_{56})^{1+2\ep} \, s_{345}^2}\nn  \\
&& \null \hskip 4 cm 
 + {s_{123} \over 2 s_{12} s_{45} s_{56} (-s_{345})^{1+2\ep} } 
 + {s_{234} \over 2 s_{61} s_{34} (-s_{56})^{1+2\ep} s_{345} } \Biggr] \nn \\
&& \null 
+ {1\over \ep^3} \Biggl[ 
 {s_{23} \over s_{12} s_{61} s_{34} s_{45}}
   \ln\biggl({s_{345}^6 s_{56}^2 s_{45}^3\over s_{12}^3 s_{61}^3 s_{34}^3 
             s_{123} s_{234}} \biggr)
+   {s_{123} s_{234} \over s_{12} s_{34} s_{45} s_{56} s_{61}}
      \ln\biggl({ s_{123} s_{234} s_{345}^2 \over 
                  s_{12} s_{61} s_{34} s_{45}} \biggr)  \nn \\
 && \null 
 + {1\over 2}  {1\over s_{56} s_{345}^2 }
      \ln\biggl({s_{12}  s_{34} \over s_{345}^2} \biggr) 
 +   {s_{23} \over s_{61} s_{34} s_{123} s_{345} }
    \ln\biggl( {s_{12} s_{45} \over  s_{123} s_{345}} \biggr) \nn \\
 && \null 
 +   {s_{123} \over  s_{12} s_{45} s_{56} s_{345}}
    \ln\biggl( {s_{34}\over s_{56} }\biggr) 
 + {1\over 2} {s_{234} \over s_{61} s_{34} s_{56} s_{345} }
   \ln \biggl( { s_{12}^2 s_{234} \over 
         s_{61} s_{34} s_{345} } \biggr)  \nn \\
 && \null 
+ 2 {s_{23} \over s_{12} s_{61} s_{34} s_{45} - s_{61} s_{34} s_{123} s_{345}}
      \ln\biggl({s_{12} s_{45} \over s_{123} s_{345}} \biggr)  \nn \\
 && \null 
+ {1\over 2} {s_{123}^2 \over 
     -s_{12}^2 s_{45}^2 s_{56} + s_{12} s_{45} s_{56} s_{123} s_{345} }
      \ln\biggl( {s_{123} s_{345} \over s_{12} s_{45}}\biggr)  \nn \\
 && \null 
+ {s_{23} \over 
       s_{12} s_{61} s_{34} s_{45} - s_{12} s_{45} s_{234} s_{345} }
     \ln\biggl( {s_{61} s_{34} \over s_{234} s_{345}} \biggr)    \nn \\
 && \null 
 -  {s_{12} s_{23} s_{45} \over
        s_{12} s_{61} s_{34} s_{45} s_{123} s_{345} - 
           s_{61} s_{34} s_{123}^2 s_{345}^2 } 
    \ln\biggl({s_{12} s_{45} \over s_{123} s_{345} } \biggr)  \nn \\
 && \null 
 + {1\over 2}
   {s_{234} \over s_{61} s_{34} s_{56} s_{345} - s_{56} s_{234} s_{345}^2}
   \ln\biggl({ s_{61} s_{34} \over s_{234} s_{345} }\biggr) 
         \Biggr]  \nn \\
%
&& \null  + \Ord(\ep^{-2})\,, \nn\\ 
I^{(14)} &=& \Ord(\ep) \,, \nn\\
I^{(15)} &=& \Ord(\ep^{-1}) \,, \nn\\
I^{(16)} & = &
-{1\over s_{34}^2 (-s_{123})^{1+2\ep}} \Biggl[
 {1\over 4 \ep^4} + {1\over 2 \ep^3} \ln\biggl({s_{12} s_{5 6} \over
          s_{3 4} s_{1 2 3} }\biggr) \Biggr]
   + \Ord(\ep^{-2})\,, \nn\\
I^{(17)} &=&
-{1\over (-s_{12})^{1+2\ep}\, s_{61} s_{23}}
 \Biggl[ {9\over 4 \ep^4} + {3\over 2 \ep^3}
  \ln\biggl({s_{1 2 3} s_{3 4 5} \over s_{6 1} s_{2 3}} \biggr)
      \Biggr]
 + \Ord(\ep^{-2})\,, \nn\\
I^{(18)} &=&
 -{1\over s_{61} (-s_{45})^{1+2\ep}\,  s_{234}}
\Biggl[ {3 \over 2 \ep^4} +
 {1\over \ep^3} \ln\biggl({s_{23} s_{45}^2\over s_{5 6}^2 s_{2 3 4}}\biggr)
       \Biggr]
   + \Ord(\ep^{-2})\,.
\end{eqnarray}

\section{Numerical Value of Integrals}
\label{IntegralsNumericalAppendix}

In this appendix we give numerical values of the one- and two-loop 
integrals at the symmetric kinematic point $K^{(0)}$.

The numerical values of the one-loop integrals~(\ref{IntOneloopAnalytic1m}) 
and (\ref{IntOneloopAnalytic2me}) are needed through $\Ord(\ep^2)$,
\begin{eqnarray}
I^{1\rm m}(\ep) &=&
{2\over \ep^2} + {2 \ln 2\over \ep} - 2.125387 - 
              6.638772 \ep - 9.871006 \ep^2 
 + \Ord(\ep^3)\,, \nn\\
I^{2{\rm m}e}(\ep) &=&
 - {4 \ln 2 \over 3 \ep} - 0.580026 + 1.033726 \ep 
                       + 3.0089373 \ep^2
 + \Ord(\ep^3) \,. 
\end{eqnarray}
The one-loop hexagon in \fig{OneLoopIntegralsFigure} is
\begin{equation}
I^{\rm hex}(\ep) =
 - 1.56735 \ep -  ( 5.73447 \pm{0.00067} ) \ep^2 \,.
\end{equation}

The numerical values of the two-loop integrals at $K^{(0)}$ are,
through $\Ord(\eps^0)$,
\begin{eqnarray}
I^{(1)}(\ep) &=& 
   {4 \over \ep^4} + {8 \ln{2} \over \ep^3} -
      {6.57974 \over \ep^2} - {32.4479 \over \ep}
      -53.373339 
+ \Ord(\ep) \,, \nn \\
I^{(2)}(\ep) &=& 
- {1\over \ep^4} - {2 \ln{2} \over \ep^3} - 
       {0.822467\over \ep^2} - {0.121225 \over \ep}
        +  3.655417 
+ \Ord(\ep) \,, \nn\\
%
I^{(3)}(\ep) &=&  
 - {0.64646\over \ep^2} - {0.55042\over \ep} +  1.767276 
+ \Ord(\ep) \,, \nn\\
%
I^{(4)}(\ep) &=& 
   - {1\over 16 \ep^4} +  {\ln{2} \over 4 \ep^3} + 
      { 0.27382 \over \ep^2} + {0.57483 \over \ep} + 
       0.511132 \pm{0.000013}
 + \Ord(\ep)\,,  \nn\\
%
I^{(5)}(\ep) &=& 
 {\ln{2} \over \ep^3} - {0.53467 \over \ep^2} 
        - {4.06579\over \ep}  -2.47430 \pm{0.00070}
 + \Ord(\ep) \,, \nn\\
%
I^{(6)}(\ep) &=& 
 - {3 \over 8\ep^4} + {\ln{2} \over \ep^3} 
      + {3.11575 \over \ep^2} + {5.60293 \over \ep}
      +  6.7248 
+ \Ord(\ep)  \,, \nn\\
I^{(7)}(\ep) &=&
     - {1\over 2 \ep^4} + {2 \ln{2}  \over \ep^3} 
     + {2.48496 \over \ep^2} - {5.9524 \over \ep} 
       - 28.8375 
+ \Ord(\ep) \,, \nn\\
%
I^{(8)}(\ep) &=& 
{2 \ln{2} \over \ep^3} + {3.45587 \over \ep^2} 
       + {2.62502 \over \ep}  -12.847078
 + \Ord(\ep) \,,  \nn \\
I^{(9)}(\ep) &=& 
 - {1\over 4 \ep^4} + {5 \ln{2} \over 2 \ep^3} 
       + {2.46847 \over \ep^2} - {3.82945 \over \ep}
         -7.799036
 + \Ord(\ep)  \,,  \nn \\
I^{(10)}(\ep) &=& 
 - {5\over 2\ep^4} - {2 \ln{2}\over \ep^3} + 
      {7.89144 \over \ep^2} + {17.0922 \over \ep }
      +  8.592639
 + \Ord(\ep)   \,,  \nn \\
I^{(11)}(\ep) &=&
     - {31 \over 8 \ep^4} - {7 \ln{2} \over \ep^3} + {10.37440 \over \ep^2} 
     + {49.447\over \ep} +   107.1558\pm{0.0027}
 + \Ord(\ep) \,,  \nn \\
%
I^{(12)}(\ep) &=&
   - {7\over \ep^4} - {17 \ln{2} \over \ep^3} + 
       {9.99506 \over \ep^2} + {70.50 \over \ep} +  148.0118 \pm{0.0013}
      + \Ord(\ep) \,, \nn \\
%
I^{(13)}(\ep) &=& 
- {129 \over 16\ep^4} - {87 \ln{2}\over 4 \ep^3} 
      +  {9.74788 \over \ep^2} + {94.31 \over \ep}
      + 236.1222 \pm {0.0016}
+ \Ord(\ep)   \,, \nn \\
I^{(14)}(\ep) &=& \Ord(\ep) \,, \nn \\
%
I^{(15)}(\ep) &=& 
 { 1.56735 \over \ep} + 5.73494  \pm{0.00069}
      + \Ord(\ep) \,, \nn \\
%
I^{(16)}(\ep) &=& 
 - {1\over 8 \ep^4} + {\ln{2}\over 2 \ep^3} -
        {0.68607 \over \ep^2} - {2.83047\over \ep}
       +  2.218047
  + \Ord(\ep)\,, \nn\\
%
I^{(17)}(\ep) &=& 
- {9\over 4 \ep^4} - {3 \ln{2} \over \ep^3}
         + {8.61834 \over \ep^2} + {36.07160 \over \ep}
         +  78.922647  
+ \Ord(\ep) \,, \nn \\
%
I^{(18)}(\ep) &=&  
- {3 \over 4 \ep^4} + {\ln{2} \over 2 \ep^3}
       + {3.76700 \over \ep^2} + {7.50556 \over \ep}
       +  7.57613 
 + \Ord(\ep)\,.
\end{eqnarray}
If no errors are quoted, the integration errors are smaller than the
quoted precision.

\section{Mellin-Barnes Representations}
\label{MBAppendix} 

In this appendix we present Mellin-Barnes representations of the most
complicated integrals of \fig{ContributingIntegralsFigure}, 
$I^{(12)}$ and $I^{(13)}$.  In both cases it is necessary
to introduce an auxiliary parameter $\eta$ in order to render the
integral well-defined.  The value of the integral in the desired
limit $\eta \to 0$ is obtained by analytic continuation.
Using the notation $z_{i,j,\ldots} = z_i + z_j + \cdots$ we have:
\begin{eqnarray}
I^{(12)} &=& \frac{(-1)^{1+2 \eta} e^{2 \eps \gamma}}
{\Gamma(-1 - 2 \eps - \eta) \Gamma(\eta)}
\int_{-i \infty}^{+i \infty} \cdots \int_{-i \infty}^{+i \infty}
\prod_{j=1}^{18} \frac{dz_j}{2 \pi i} \Gamma(-z_j)
\frac{\Gamma(3 + \eps+\eta+
z_{1,2,3,4,5,6,7,8,9,10})}{\Gamma(4+\eps+\eta + 
z_{1,2,3,4,5,6,7,8,9,10})}
\cr
&&\times
(-s_{12})^{z_{8,13}}
(-s_{23})^{z_{14}}
(-s_{34})^{z_{1,18}}
(-s_{45})^{z_{3,15}}
(-s_{61})^{z_{11}}
(-s_{123})^{z_{9,16}}
(-s_{234})^{z_{17}}
(-s_{345})^{z_{2,12}}
\cr
&&\times
(-s_{56})^{-5 - 2 \eps - 2 \eta -
z_{1, 2, 3, 8, 9, 11, 12, 13, 14, 15, 16, 17, 18}}
\cr
&&\times
\frac{\Gamma(-3-\eps-z_{1, 2, 3, 4, 5, 6, 7}) \Gamma(5 + 2 \eps+2 
\eta + z_{1, 2, 3, 8, 9, 10, 11, 12, 13, 14, 15, 16, 17, 
18})}{\Gamma(1-z_4) \Gamma(\eta-z_5) \Gamma(-z_6) \Gamma(1-z_7)
\Gamma(-3-3 \eps-2\eta - z_{1,2,3,8,9,10})}
\cr
&&\times
\Gamma(-5 - 2 \eps - 2 \eta - z_{1, 2, 3, 6, 8, 9, 10, 11, 12, 13, 14, 
15, 16})
\Gamma(-1-\eps-\eta+z_{4,5,6,7} -z_{11,12,14,15,17,18})
\cr
&&\times
\Gamma(-2-\eps-\eta-z_{1,2,3,8,9,10})
\Gamma(\eta-z_5 + z_{14,15,16}) \Gamma(1-z_4 + z_{12,13,18})
\Gamma(1+z_{1,2,4,8})
\cr
&&\times
\Gamma(1-z_7+z_{11,12,15})
\Gamma(1+z_{1,6,10}) \Gamma(1+z_{2,3,7}) \Gamma(1 + z_{3,5,9})
\Gamma(1+z_{11,14,17})\,,
\end{eqnarray}
\begin{eqnarray}
I^{(13)} &=& \frac{(-1)^{1+2 \eta} e^{2 \eps \gamma}}
{\Gamma(-1 - 2 \eps - \eta) \Gamma(\eta)}
\int_{-i \infty}^{+i \infty} \cdots \int_{-i \infty}^{+i \infty}
\prod_{j=1}^{18} \frac{dz_j}{2 \pi i} \Gamma(-z_j)
\frac{\Gamma(3+\eps+\eta
+ z_{1, 2, 3, 4, 5, 6, 7, 8, 9, 10})}{\Gamma(4+\eps+\eta+z_{1, 2, 3, 4, 
5, 6, 7, 8, 9, 10})}
\cr
&&\times
(-s_{12})^{z_{3,17}}
(-s_{23})^{z_{8,14}}
(-s_{34})^{z_{11}}
(-s_{45})^{z_{9,13}}
(-s_{61})^{z_{1,16}}
(-s_{123})^{z_{10,18}}
(-s_{234})^{z_{15}}
(-s_{345})^{z_{2,12}}
\cr
&&\times (-s_{56})^{-5-2\eps-2\eta - z_{1, 2, 3, 8, 9, 10, 11, 12, 13, 
14, 15, 16, 17, 18}}
\Gamma(\eta-z_4 + z_{14,15,16})
\cr
&&\times
\frac{
\Gamma(-3-\eps-z_{1, 2, 3, 4, 5, 6, 7})
\Gamma(5+2 \eps+2\eta + z_{1, 2, 3, 8, 9, 10, 11, 12, 13, 14, 15, 16, 
17, 18}) \Gamma(1 + z_{1,2,6,9})
}{\Gamma(\eta-z_4) \Gamma(1-z_5) \Gamma(1-z_6) \Gamma(-z_7)
\Gamma(-3-3 \eps-2\eta  -z_{1, 2, 3, 8, 9, 10})}
\cr
&&\times
\Gamma(-5 - 2 \eps - 2 \eta - z_{1, 2, 3, 7, 8, 9, 10, 11, 12, 13, 14, 
15, 16})
\Gamma(4+\eps+\eta+z_{1, 2, 3, 4, 5, 6, 7, 8, 9, 10, 13, 14, 18})
\cr
&&\times
\Gamma(-4-2 \eps-2 \eta - z_{1, 2, 3, 8, 9, 10, 12, 13, 14, 16, 17, 
18})
\Gamma(-2\eps-\eta-z_{1, 2, 3, 8, 9, 10})
\Gamma(1+z_{2,3,5})
\cr
&&\times
\Gamma(1-z_5 + z_{11,12,17})
\Gamma(1-z_6 + z_{12,13,16})
\Gamma(1+z_{1,4,8})
\Gamma(1+z_{3,7,10})\,.
\end{eqnarray}


\end{document}